\documentclass{jfm}
\usepackage{graphicx}
\usepackage{epstopdf, epsfig}

\usepackage{amsmath}
\usepackage{mathrsfs}
\usepackage{overpic}
\usepackage{color}
\usepackage[final]{changes}
\definechangesauthor[name={Anatoliy Khait},color=blue]{AK}
\definechangesauthor[name={Zhihua Ma},color=red]{ZM}

\newcommand{\myhighlight}[1]{{\color{black}{#1}}}

\shorttitle{Phase shifting in deep water wave breaking}
\shortauthor{A. Khait and Z. Ma}

\title{Phase shifting, dispersion variation and defocusing suppression
       in wave breaking}

\author{Anatoliy Khait 
        \corresp{\email{haitanatoliy@gmail.com}}
        \and
        Zhihua Ma 
        \corresp{\email{z.ma@mmu.ac.uk}}}

\affiliation{
	 Centre for Mathematical Modelling and Flow Analysis,
Department of Computing and Mathematics,
Manchester Metropolitan University,
Chester Street, Manchester M1 5GD, UK}

\begin{document}

\maketitle

\begin{abstract}
We present an investigation of the fundamental physical processes involved in deep water wave breaking. Our motivation is to identify the underlying reason causing the deficiency of the eddy viscosity breaking model (EVBM) in predicting surface elevation for strongly nonlinear waves. Owing to the limitation of experimental methods in the provision of high-resolution flow information, we propose a numerical methodology by developing an EVBM enclosed standalone fully-nonlinear quasi-potential (FNP) flow model and a coupled FNP plus Navier-Stokes flow model. The numerical models were firstly verified with a wave train subject to modulational instability, then used to simulate a series of broad-banded focusing wave trains under non-, moderate- and strong-breaking conditions. A systematic analysis was carried out to investigate the discrepancies of numerical solutions produced by the two models in surface elevation and other important physical properties. It is found that EVBM predicts accurately the energy dissipated by breaking and the amplitude spectrum of free waves in terms of magnitude, but fails to capture accurately breaking induced phase shifting. The shift of phase grows with breaking intensity and is especially strong for high wavenumber components. This is identified as a cause of the upshift of wave dispersion relation, which increases the frequencies of large wavenumber components. Such a variation drives large-wavenumber components to propagate at nearly the same speed, which is significantly higher than the linear dispersion levels. This suppresses the instant dispersive spreading of harmonics after the focal point, prolonging the lifespan of focused waves and expanding their propagation space.
\end{abstract}

\begin{keywords}
surface gravity waves, wave breaking, non-potential effects, phase shift 
\end{keywords}

\section{Introduction}
\added[id=ZM]{Wave breaking is a highly important and challenging topic in
  engineering and environmental science. Violent breaking waves can produce
  destructive loads causing severe damage to, and even complete failure/loss
  of naval, coastal and marine structures including ships, breakwaters,
  seawalls, and oil and gas platforms \citep{BabaninBook2011, Ma2014, Ma2016}.}
\added[id=ZM]{More broadly, wave breaking plays a crucial role in the
  planetary-scale atmosphere-ocean system by enhancing the
  exchange of mass, momentum and heat across the air-sea interface, thus
  influencing the earth's climate and weather \citep{Duncan2012, Veron2015}.
} \deleted[id=ZM]{Impact of high waves is dangerous for coastal
  and open-sea structures.  Inevitably breaking extremely high waves release
  significantly more destruction energy as compared to stable waves. Violent
  impulsive breaking wave loads may destroy the coastal defence walls, break up
  the oil platforms and dismantle large ships.  Aeration of the ocean that is
  much intensified by wave breaking in harsh weather conditions induces shock
  wave loading leading to high frequency vibrations of structures.  Wave
  breaking is also responsible for intensification of the green water slumming
  onto the coastal lines and decks possibly causing flood and destruction of the
  facilities
  Wave breaking plays a key role in the environmental oceanography being a part
  of nonlinear processes in the ocean-atmosphere system.  It has a direct impact
  on the dissolved oxygen and other gases in the ocean and, therefore, aquatic
  life. It is an important source of ocean turbulence and currents, as well as a
  conductor of the ocean-atmosphere interactions that potentially influence the
  harsh weather.} \replaced[id=ZM]{Breaking is also a major mechanism to dissipate wave energy,}{Wave
  breaking is the main sink of \deleted[id=ZM]{the} ocean wave energy}
\added[id=ZM]{preventing the endless growth
  of wind waves \citep{Paper-1996-Melville-279}.} \deleted[id=ZM]{in oceans that balances the energy input due
  to winds and consequently prevents waves from endless growth providing limit
  for the energy concentration.}
Breaking wave models, in particular the accurate estimate of energy dissipated through the process, constitute \replaced[id=ZM]{a key}{an important} part
of \replaced[id=ZM]{numerical}{the} ocean \added[id=ZM]{wave} forecast,
\deleted[id=ZM]{software} \replaced[id=ZM]{which is essential to the safety of
  maritime activities including, but not limited to,}{that provides an essential
  information for maritime safety} \added[id=ZM]{fishery,} ship navigation and
\replaced[id=ZM]{offshore construction and operation}{scheduling on-site operations and constructions in open seas}.

Despite \replaced[id=ZM]{the vast amount of theoretical, experimental and
  numerical work reported in the past}{its crucial importance},
\replaced[id=ZM]{the fundamental mechanism of wave breaking has not been
  understood thoroughly yet due to its extraordinary complexity}{the wave
  breaking is not well understood}.  Wave breaking is a
\added[id=ZM]{multi-scale and} \replaced[id=ZM]{multi}{two}-phase problem,
\replaced[id=ZM]{which involves multiple orders of scales ranging from the large
  orbital motions induced by surface water waves to the small air bubbles
  entrained into water mass and spray ejected into the atmosphere.}{involving
  air entertainment into the water as well as water spray production. Only
  two-phase viscous Navier-Stokes equations are applicable to study such
  phenomenon.  Breaking-induced fluid flows incorporate multiple orders of
  scales ranging from orbital motions induced by the long surface water waves to
  the small bubbles of air entrapped during the breaking} \added[id=ZM]{To
  fully resolve the transient flow features of breaking waves in numerical
  simulations,} \replaced{extremely}{Very} fine meshes and small time steps are
needed. \deleted{for accurate numerical treatment of all scales within the
  Navier-Stokes framework require exceptionally significant computational
  effort.}  \added[id=ZM]{However, this will place a prohibitive burden on
  computing resources}, \replaced{restricting the computational domain
  of high-fidelity numerical models to}{Limited computational capabilities restrict the possible
  sizes of the domains to} only several representative wave lengths for 3D problems
 \citep{Iafrati2009, Paper-2015-Lubin-364, Melville2017}.

\replaced[id=ZM]{It is}{On the other hand,
it is} known that \replaced[id=ZM]{the onset of breaking}{breaking development} and \added[id=ZM]{the} post-breaking evolution of \deleted[id=ZM]{the} wave
field \replaced[id=ZM]{are}{is} dependent on the breaking crest formation process, which is usually
highly nonlinear \citep{Khait2018}. \replaced[id=ZM]{The development of
breaking wave crest involves significantly large}{Breaking crest development is seen at
significant} temporal and spatial scales \replaced[id=ZM]{that cannot be efficiently handled by high-fidelity numerical models alone yet}{beyond capabilities of the two-phase
Navier-Stokes models} \citep{Iafrati2018, Iafrati2019}. \replaced[id=ZM]{To effectively deal with
these large scales, it is necessary to use low-fidelity models such as the potential 
model which assumes the flow to be inviscid and irrotational. 
 Under such an approximation, the flow velocity can be calculated as the gradient of the potential function. 
}
{Necessity of mathematical
simplification in order to extend the computational domain to larger sizes
brings an assumption of potentiality of the velocity field:}
\deleted[id=ZM]{
\begin{eqnarray}
&& \vec{U} = - \vec{\nabla} \varphi 
\label{eq:potent}
\end{eqnarray}
The potential $\varphi$ is governed by the Laplace equation with
appropriate nonlinear boundary conditions.
}%
Although \added[id=ZM]{the} potential approximation allows a substantial simplification
of the problem, it disregards the \replaced[id=ZM]{crucial physical effects such as fluid viscosity,}{viscous effects,} flow vorticity and two-phase features \replaced[id=ZM]{for}{,all of them are essential
in} wave breaking \added[id=ZM]{problems}. Empirical closures are thus needed to take into
account these important effects in the evolution of wave field subject to breaking.

\citet{Chalikov2005} noticed that \deleted[id=ZM]{at the breaking inception }the free
surface close to \replaced[id=ZM]{an}{the} unstable crest became nearly vertical \added[id=ZM] upon the inception of breaking, \replaced[id=ZM]{and}{leading to formation of} high wavenumber spectral harmonics, accompanied by the
nonlinear flux of energy from low to high wavenumbers, were generated.
Damping the high wavenumber components of the spectrum
and therefore dissipating the associated energy \replaced[id=ZM]{can help}{allows} to \replaced[id=ZM]{stabilise}{stabilize}
the computation \citep{Chalikov2005, Chalikov2014}.
The damping process \replaced[id=ZM]{was actually accomplished by introducing}{is actually modelled by addition of} \deleted[id=ZM]{the} empirical
terms to the free surface boundary conditions.
Similar approaches can be found in \deleted[id=ZM]{some formulations of} the \added[id=ZM]{extended} High-Order
Spectral Models \added[id=ZM]{of} \cite{Ducrozet2012, Ducrozet2016}.
\replaced[id=ZM]{For spectral ocean forecasting
	models, the nonlinear evolution of waves is considered as an energy cascade that transfers energy between different frequency harmonics.}{The nonlinear wave evolution is seen in the spectral ocean forecasting
models as a cascade transferring energy between the frequency harmonics
in the energy spectrum.} It allows to take into account the
spectral energy dissipation due to breaking by \replaced[id=ZM]{using}{introduction of}
observation-based empirical source terms \replaced[id=ZM]{to parameterise}{parametrising} the reduction
of spectral components \deleted[id=ZM]{associated with breaking effects}
\citep{Babanin2011, Shrira2018}.

Although\deleted[id=ZM]{the methods of the} damping \deleted[id=ZM]{the} high frequency spectral
components \replaced[id=ZM]{is computationally}{are often} efficient, \replaced[id=ZM]{this kind of method has}{they have} inherent restrictions.
\replaced[id=ZM]{One significant difficulty is that wave breaking}{The wave breaking in physical space is a strongly localized event that} cannot be adequately described in the Fourier space because it 
is strongly localised in the physical space.  \added[id=ZM]{An} alternative
semi-empirical closure for \deleted[id=ZM]{the} wave
breaking\deleted[id=ZM]{that is free of the aforementioned limitations,} was
proposed by \citet{Perlin2010, Perlin2012}.
\deleted[id=ZM]
{
 \replaced[id=ZM]{This model has been applied and/or extended}{and further investigated} by others
\citep{Ducrozet2017, Ducrozet2018, Sriram2019, Craciunescu2020}.
\replaced[id=ZM]{These approaches are}{Their approach is} based on the \deleted[id=ZM]{theory of} weakly-damped quasi-potential
small-amplitude \replaced[id=ZM]{wave}{waves} \added[id=ZM]{theory proposed} by \citet{Ruvinsky1991}.
The potential velocity field (\ref{eq:potent}) is complemented by \replaced[id=ZM]{a}{the}
small rotational component:
\begin{eqnarray}
&& \vec{U} = - \vec{\nabla} \varphi - \vec{\nabla} \times \vec{\Psi},
\label{eq:quasipotent}
\end{eqnarray}
imposing $\left| \vec{\nabla} \times \vec{\Psi} \right| \ll 
\left| \vec{\nabla} \varphi \right|$.
Substituting (\ref{eq:quasipotent}) into Navier-Stokes equations,
assuming \deleted[id=ZM]{the concentration of} the rotational flow \added{to be concentrated} in the narrow layer
$\delta \ll 1$ near the free surface and \replaced[id=ZM]{applying}{involving} several additional
simplifications, \citet{Ruvinsky1991} and \citet{Dias2008}
reduced the equations governing the stream function
$\vec{\Psi}$ to a simple free surface boundary condition.
\cite{Perlin2010, Perlin2012} \replaced[id=ZM]{extended}{have complemented}
the model of \cite{Ruvinsky1991} \added{by proposing} an \added[id=ZM]{eddy viscosity based} empirical
closure, \replaced[id=ZM]{which was calibrated by} {for the eddy viscosity formulated based on the} numerous laboratory \replaced[id=ZM]{measurement of breaking wave energy dissipation}{experiments on energy dissipation due to wave breaking}.
}
\replaced[id=ZM]{This model}{The eddy viscosity weakly-potential wave breaking
  model} contains only one empirical constant as opposed to numerous fitting
parameters \added[id=ZM]{used} in \replaced[id=ZM]{other}{alternative} breaking
approximations. Therefore, it is preferable for studying \deleted[id=ZM]{the}
breaking processes at large spatial and temporal scales.  \replaced[id=ZM]{A
  number of researchers have demonstrated this model's}{The model has shown a
  high} accuracy and robustness in \added[id=ZM]{the} prediction of
\deleted[id=ZM]{the} energy flux reduction \replaced[id=ZM]{for}{in both}
spilling and plunging \replaced[id=ZM]{breakers}{breaker cases}
\citep{Perlin2010, Perlin2012, Ducrozet2017, Ducrozet2018, Sriram2019,
  Craciunescu2020}.  However, \deleted[id=ZM]{a} notable disagreement{s} with
experimental measurements in terms of the surface elevation for
\deleted[id=ZM]{several particular}
strongly-nonlinear wave trains \replaced[id=ZM]{have also}{has} been reported in
\replaced[id=ZM]{the literature}{multiple studies} \citep{Perlin2012,
  Ducrozet2018}\deleted[id=ZM]{, while other wave trains agreed well with
  measurements}.  \replaced[id=ZM]{The underlying reason causing the
  discrepancy between the eddy viscosity model and laboratory experiments
  remains unclear.}{The physical reasons for such deviations are still not clear
  and maybe of fundamental nature, thus requiring a closer investigation.}


\added[id=ZM]{This requests further investigation to identify the actual cause by producing a series of realistic wave breaking scenarios and analysing a considerable amount of detailed flow data.}
\added[id=ZM]{However, state-of-the-art} laboratory facilities, in particular
measurement equipment, \replaced[id=ZM]{are still deficient in the provision of
  needed large amount of high-resolution spatial and temporal flow
  information.}{are significantly limited in spatial and temporal resolutions.}
To circumvent these restrictions, a hybrid low- and high-fidelity numerical
model is developed and applied in the present work. We coupled a boundary
element method (BEM) based fully nonlinear quasi-potential (FNP) model with a
volume-of-fluid method (VOF) based two-phase incompressible Navier-Stokes (NS)
model to formulate the hybrid FNP-NS model. This new approach is used to deal with
the generation, propagation and breaking of deep water waves and their
post-breaking evolution.  The accuracy of the numerical model is carefully
assessed through a breaking wave train subject to modulational instability. The computed results are compared against the laboratory measurements and
other numerical solutions reported in the literature. The FNP-NS model is then
used to simulate the evolution of six broad-banded wave trains under non-, moderate- and strong-breaking conditions.  The standalone FNP model, incorporated with the eddy viscosity
enclosure, is also applied to compute these wave trains. This allows us to
perform a comprehensive comparative study of breaking waves with the standalone
FNP model and the hybrid FNP-NS model to quantify the deviations in
surface elevation, energy dissipation and other important physical properties. Close
attention is paid to the phases of free waves, the dispersion relationship
between wavenumber and frequency, the trajectory of wave trains and their height before, during and after focusing. Detailed analyses of these important
properties are carried out to examine the fundamental physical processes involved
in breaking.


\replaced[id=ZM]{The remainder of the}{This} paper is \replaced[id=ZM]{organised}{organized} as follows. The numerical methodology is \replaced[id=ZM]{described}{formulated}
in Section~\ref{sec:method} and carefully validated against wave flume
experiments in Section~\ref{sec:valid}. A detailed analysis of the wave train
evolution observed in numerical simulations is presented in
Sections~\ref{sec:elevat} and \ref{sec:dissip}. \deleted{in terms of the free
surface shape and the wave energy dissipation.}
The breaking induced phase shifting phenomenon and variation
of wave train dispersion are \replaced{discussed}{considered} in Sections~\ref{sec:shift}
and \ref{sec:disper}. Section~\ref{sec:trajec} is devoted to the discussion of the
suppression of dispersive defocusing and associated processes.
Conclusions are drawn and practical implications are discussed in
Section \ref{sec:concl}.


\section{\label{sec:method} \replaced[id=ZM]{Methodology}{Methods}}

\begin{figure}
    \centering
    \includegraphics[width=0.9\textwidth]{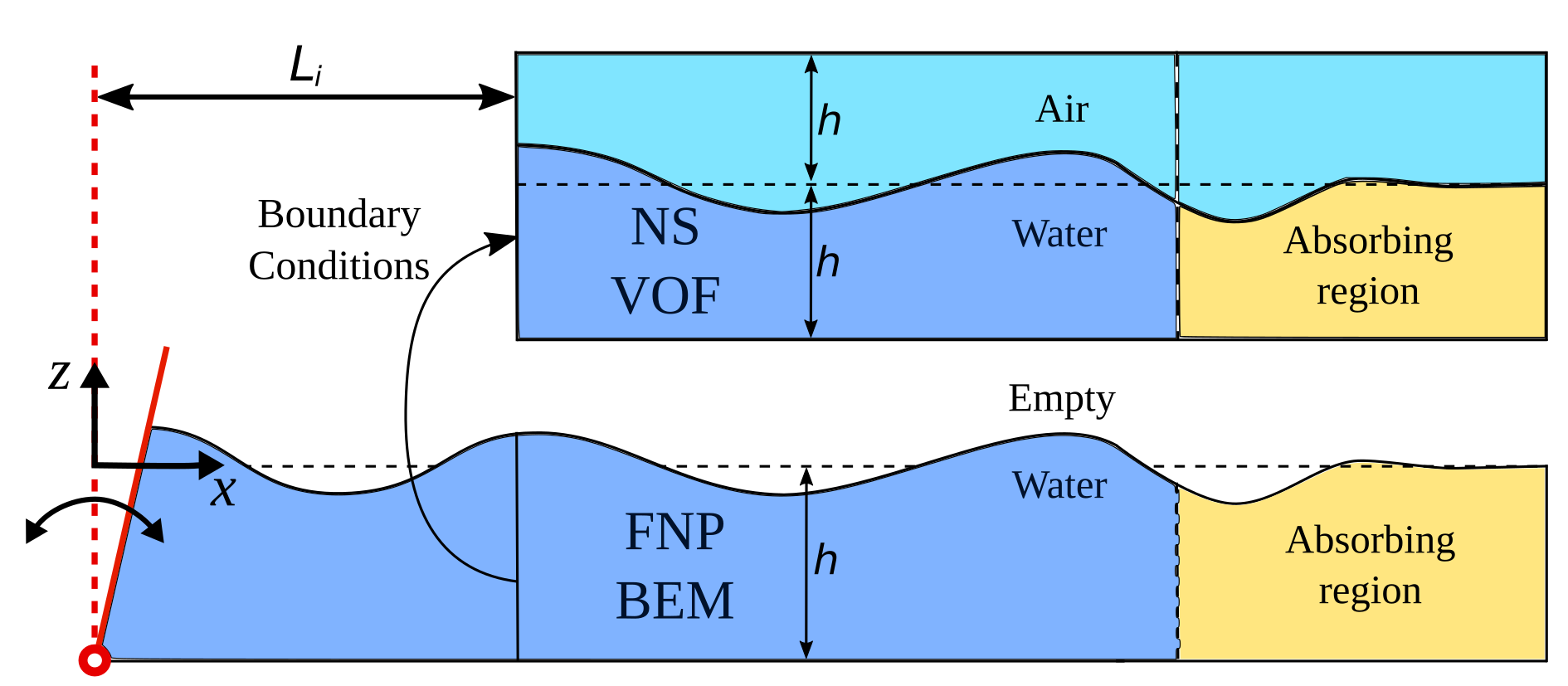}
    \caption{Schematic of the coupled FNP-NS model. {Boundary Element Method
      (BEM) is used discretise the FNP domain. Volume-of-Fluid (VOF) method is used to discretise the NS domain. In the present work we also use `BEM' and `VOF' to refer to the low- and high-fidelity flow models, respectively.}}
    \label{fig:domain}
\end{figure}

\added[id=ZM]{To reproduce realistically the generation, propagation and
  breaking of surface waves as well as their afterwards evolution, a number of
  numerical approaches are applied in the present work. This includes
  a fully-nonlinear potential flow (FNP) model and a two-phase incompressible
  viscous Navier-Stokes (NS) model.  A hybrid FNP-NS approach is developed
  by connecting the FNP and NS models through a one-way coupling strategy (see
  Figure~\ref{fig:domain}). Firstly waves are generated and propagated in the
  domain of the FNP model. The NS model is initialised after a certain amount
  of time when the waves arrive at its inlet boundary. Free
  surface elevation and flow velocity computed by the FNP model at the coupling
  boundary are then transferred to the NS model (see more details in Section \ref{sec:coupling-bem-vof}).}

\deleted[id=ZM]{We attempt to investigate the breaking wave field seen at
  different levels of mathematical simplification.  The two-phase incompressible
  viscous Navier-Stokes model (NS) is assumed to provide the high-fidelity
  results that closely reflect the experimental observations. On the other hand,
  the fully-nonlinear potential model (FNP) assumes the flow to be irrotational.
  The need of quantitative comparison between these two models requires an
  accurate match between corresponding initial and boundary conditions used in
  NS and FNP. To this end, a one-way coupling was
utilized such that the wave generation and
propagation in FNP model was considered first. Simultaneously, at a certain time
the initialization of NS model is performed using the results of the FNP
computations. The boundary conditions used at the inlet of the NS model are also
derived from the FNP computations.  The schematic of the considered FNP-NS
coupling is presented in figure~\ref{fig:domain}.  In view of extreme
computational effort needed to perform the high-resolution NS simulations, both
numerical models were restricted to 2D case. }

\added[id=ZM]{As mentioned above, the FNP model is efficient for
  large-scale problems but ignores important physical effects. On the contrary,
  the NS model includes important physical effects and provides more detailed
  flow information at the cost of computational efficiency. The coupled FNP-NS
  model provides a way to balance low- and high-fidelity computations. Another
  benefit of using the coupled model is that it can produce more realistic
  scenarios without introducing artificial conditions for wave breaking
  \citep{Paper-2015-Lubin-364}.}

{In the present work we focused on two-dimensional wave breaking problems. A
  series of numerical computations of breaking events under different wave
  conditions were carried out by using the standalone FNP model and the coupled
  FNP-NS model, respectively. Detailed descriptions of the FNP and NS models are
  given in the following.}

\subsection{Fully-nonlinear quasi-potential model (BEM)}

\replaced[id=ZM]{Under the}{Within} potential approximation, the velocity field
$\mathbf{U} = \{u, w\}$ is \replaced[id=ZM]{given by the gradient of}{expressed
  through} the hydrodynamic potential $\varphi$, i.e. ${\mathbf{U} = -
  \mathbf{\nabla} \varphi}$.  The Boundary Element Method (BEM)  is used to determine the distribution of $\varphi$ across the FNP
domain \citep{Grilli1989, Grilli1990}. In this approach, the fluid flow is governed by the Green's second
identity:
\begin{eqnarray}
&& \alpha \varphi(\mathbf{r_s}) = 
   \oint_{\Gamma} \left(
                  \frac{\partial \varphi}{\partial n}(\mathbf{r})
                                 \Phi(\mathbf{r}, \mathbf{r_s})
                 -\varphi(\mathbf{r}) \frac{\partial \Phi}{\partial n}
                                     (\mathbf{r}, \mathbf{r_s})
                  \right) \,d\Gamma
\label{eq:green}
\end{eqnarray}
Here $\Phi(\mathbf{r},\mathbf{r_s}) = -1/(2\pi) \text{ln} |\mathbf{r} -
\mathbf{r_s}|$ is the fundamental solution that represents
the potential flow at point $\mathbf{r}$ due to a source located
at $\mathbf{r_s}$; $\Gamma$ is the closed boundary of the domain;
$\alpha = \pi$ for regular nodes, and $\alpha = \pi / 2$ for corner nodes;
$n$ is the outward normal direction to $\Gamma$.
\replaced[id=ZM]{The solution}{Solution} of (\ref{eq:green}) provides the value of
$\partial \varphi / \partial n$ or $\varphi$ at the point
$\mathbf{r}$ located on $\Gamma$.
The $3^{\text{rd}}$-order MII local interpolation technique
\citep{Grilli1996} was used to \replaced[id=ZM]{discretise}{discretize} the boundary $\Gamma$ for
numerical solution of (\ref{eq:green}).

The free surface\deleted[id=ZM]{boundary} is subject to the dynamic and kinematic
boundary conditions determining the time variation of its shape.
Since we investigate strongly breaking wave field, \replaced[id=ZM]{the inclusion}{involvement} of
empirical closures in the FNP model \replaced[id=ZM]{is required to stabilise the computation}{are required to perform FNP simulations}.
For weakly-damped waves \citep{Ruvinsky1991}, \replaced[id=ZM]{assuming}{assumed} the flow to be
quasi-potential with \added[id=ZM]{small} vortical velocity components\deleted[id=ZM]{being small}, \replaced[id=ZM]{we can obtain}{and thus
derived} the modified boundary conditions \citep{Dias2008, Shrira2020} given by:
\begin{eqnarray}
&& \frac{D \mathbf{r}}{D t} = - \mathbf{\nabla} \varphi
   - \underbrace{\mathbf{\nabla} \times \mathbf{\Psi}}
   _{\text{wave breaking}} 
\label{eq:freebc1}
\\
&& \frac{D \varphi}{D t} = g z -\frac{1}{2} | \nabla \varphi |^2
   - \underbrace{\widetilde{p}_d \sqrt{g h} \frac{\partial \varphi}{\partial n} b_f(x)}
   _{\text{wave absorption}}
   + \underbrace{2 \nu_{eddy} \frac{\partial^2 \varphi}{\partial s^2}}
   _{\text{wave breaking}}
\label{eq:freebc2}
\end{eqnarray}
The vector stream function $\mathbf{\Psi} = (0, 0, \psi)$ contains
only a vortical part of the flow; $\nu_{eddy}$ is the closure constant
for the wave breaking model; $s$ is the direction tangential to the
free surface. The value of the stream function at the free surface
is governed by the vorticity equation. The exact form of this equation
cannot be satisfied for \deleted[id=ZM]{the} potential \replaced[id=ZM]{flows}{flow}, therefore
its approximate version is used \citep{Ruvinsky1991, Shrira2020}:
\begin{eqnarray}
&& \frac{\partial}{\partial t} \frac{\partial \psi}{\partial s}
   = 2 \nu_{eddy} \frac{\partial^2}{\partial s^2}
       \frac{\partial \varphi}{\partial n},
\label{eq:psi}
\end{eqnarray}
where $\partial \varphi / \partial n$ is the solution of (\ref{eq:green}).
Equation (\ref{eq:freebc2}) is also responsible for \replaced[id=ZM]{wave}{waves} absorption
at the end of the domain\deleted[id=ZM]{to prevent their reflection},
see Figure \ref{fig:domain}.
The dimensionless constant $\widetilde{p}_d$ \replaced[id=ZM]{characterises}{characterizes} the strength of the
wave damping; function $b_f(x)$ determines the location
of the absorption region and the gradual increase of the
damping strength in the beginning of the region.
The most effective absorption occurs when $\widetilde{p}_d = 2$
\citep{Khait2018, Khait2019b}.

The non-penetration\deleted[id=ZM]{boundary} condition $\partial \varphi / \partial n = 0$
is \replaced[id=ZM]{applied}{used} at the bottom and\deleted[id=ZM]{at the} right\deleted[id=ZM]{lateral} boundaries.
Moving boundary with \deleted[id=ZM]{the} specified velocity \deleted[id=ZM]{field} is introduced
at the left side of the domain to replicate the motion of \replaced[id=ZM]{a wave paddle}{the paddle
wavemaker}, see Figure~\ref{fig:domain}.
The domain size and grid resolution used in the simulations are \replaced[id=ZM]{summarised}{summarized}
in Table~\ref{tab:bem}.
Grid \replaced[id=ZM]{convergence}{independence} study is presented in Appendix~\ref{app:A}.
Integration time step was taken to satisfy the numerical stability
criterion defined by the Courant number $CFL \leq 0.1$
\citep{Grilli1990}.

\begin{table}
\centering
\caption{\label{tab:bem} Parameters of the BEM model. Details on the
wave trains studied are presented in Section~\ref{sec:trains};
$\lambda_0$ is the carrier wave length.}
\begin{tabular}{ l c c }
 \hline
  & Gaussian wave train \;\; & \;\; Modulated wave train \\
 \hline
 Length of the domain & $26 \lambda_0$ & $31 \lambda_0$ \\
 Still water depth $h$ & 0.6 m & 3.0 m \\
 Length of the absorbing region & \multicolumn{2}{c}{$7 \lambda_0$} \\
 Free surface grid resolution & 
                       \multicolumn{2}{c}{30 nodes per $\lambda_0$} \\
 Bottom boundary resolution &
                       \multicolumn{2}{c}{20 nodes per $\lambda_0$} \\
 \hline
\end{tabular}
\end{table}

Two approaches \replaced[id=ZM]{namely regridding and empirical eddy viscosity for approximating}{in approximation of} the energy dissipation due to \deleted[id=ZM]{the}
wave breaking are considered in the paper. First, it was found \deleted[id=ZM]{out} that
\deleted[id=ZM]{the} regridding \deleted[id=ZM]{of} the free surface mesh at the instant of breaking
inception \replaced[id=ZM]{can stabilise}{leads to stabilization of} the numerical simulation \deleted[id=ZM]{, which
allows pass-through computation} without \replaced[id=ZM]{using}{involvement of} the eddy viscosity
closure, i.e. $\nu_{eddy} = 0$ in (\ref{eq:freebc1}) and
(\ref{eq:freebc2}). For convenience, this model is designated as `BEMr'
in the following.

\begin{figure}
    \centering
    \includegraphics[width=0.6\textwidth]{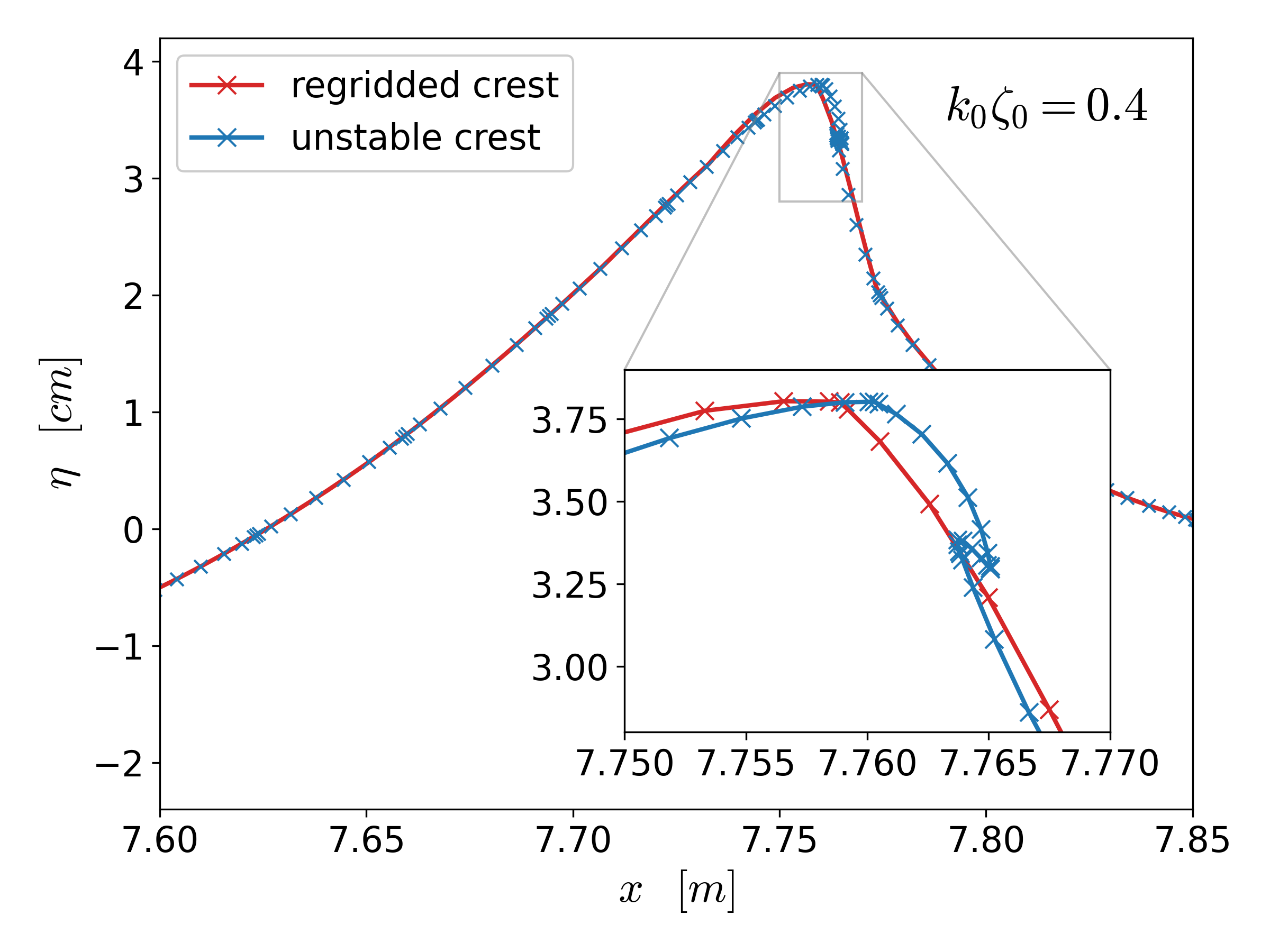}
    \caption{Shape of the pre-breaking wave crest before and after 
             re-meshing. Cross markers show the mesh nodes on the free surface.}
    \label{fig:regrid}
\end{figure}

The grid nodes at the free surface represent the floating Lagrangian
markers having all degrees of freedom. \replaced[id=ZM]{The distance}{Distance} between \replaced[id=ZM]{two}{the}
\replaced[id=ZM]{neighbouring}{neighbor} nodes is constantly varying due to the propagation of \deleted[id=ZM]{the}
nonlinear wave{s}. The regridding method developed by \citet{Grilli1994}
establishes equal lengths of the arcs between all neighbour nodes as
measured along the boundary elements.
At the instance of breaking inception, the distance between the nodes
at the pre-breaking crest becomes critically small leading to
consequent crest overturning and loss of computation stability.
The shape of the pre-breaking crest with and without regridding
is shown in Figure~\ref{fig:regrid}.
It can be \replaced[id=ZM]{seen}{said} that the regridding \added[id=ZM]{method} smoothens the shape of the
pre-breaking crest and removes the unstable overturning part.

A more advanced approximation for \deleted[id=ZM]{the} wave breaking energy dissipation
is based on the eddy viscosity empirical closure suggested by
\citet{Perlin2010, Perlin2012}.
According to this method, the location of the breaking crest
is established \added[id=ZM]{by} using the geometrical criterion $S_b \geq S_c$;
where $S_b$ is the local free surface slope,
while its threshold value is $S_c = 0.95$.
Once the location of the breaking is determined, the eddy viscosity
value is calculated \added[id=ZM]{by} using the empirical relation
\citep{Perlin2010, Perlin2012}:
\begin{eqnarray}
&& \nu_{eddy} = \alpha \frac{H_b L_b}{T_b}
\label{eq:eddy}
\end{eqnarray}
Here $H_b$ and $L_b$ are the characteristic vertical and horizontal
scales of the breaking event, respectively, while $T_b$ is the
characteristic time scale. All these values are determined by using
the empirical relations:
\begin{eqnarray}
\begin{aligned}
&& L_b = \frac{24.3 S_b - 1.5}{k_b} \\
&& T_b = \frac{18.4 S_b + 1.4}{\omega_b} \\
&& H_b = \frac{0.87 R_b - 0.3}{k_b}
\end{aligned}
\label{eq:HLT}
\end{eqnarray}
Here $k_b$ and $\omega_b$ are the local wave parameters; $R_b$ is the
geometrical factor showing vertical asymmetry of the breaking wave crest.
The value of the proportionality constant in (\ref{eq:eddy}) is
$\alpha = 0.02$. The determined eddy viscosity $\nu_{eddy}$
is applied in the region of the expected energy dissipation having
length $L_b$. The duration for the eddy viscosity impact is $T_b$;
afterwards the wave breaking is assumed to be finished 
implying $\nu_{eddy} = 0$. The methodology for \replaced[id=ZM]{determining all the}{determination of all}
required empirical constants is detailed in \added[id=ZM] the work of \cite{Perlin2012}.
Further, we designate the given eddy viscosity type of the BEM model
as `BEM$\nu$', see Table~\ref{tab:breaking}. In the following part of the paper, we use `BEM' and `FNP' interchangeably when referring to the potential flow model.

\begin{table}
\centering
\caption{\label{tab:breaking} Types of the breaking approximations}
\begin{tabular}{ l c c }
 \hline
 Model type & BEMr &  BEM$\nu$ \\
 \hline
 Eddy viscosity & $\nu_{eddy} = 0$ & $\nu_{eddy} \neq 0$ \\
 Regridding & on & off \\
 \hline
\end{tabular}
\end{table}

\subsection{\added[id=ZM]{Two-phase} Navier-Stokes model (\replaced{VOF}{FOAM})}
\label{sec:FOAM}

\added[id=ZM]{A volume-of-fluid (VOF) based two-phase incompressible Navier-Stokes flow
  solver namely \emph{interFoam}, available in the open source library OpenFOAM,
  is used in the present work to develop the coupled FNP-NS model. The
  underlying NS model has been tested extensively for a series of wave-structure
  interaction problems including dam break, water entry, wave propagation and
  breaking wave impacting with fixed and moving structures, and the computed
  results have been verified against analytical solutions, laboratory
  experiments and other numerical results reported in the literature
  \citep{Ma2016, Ferrer2016, Larsen2019}.}
%
%
The governing equations of the NS model represent momentum and mass conservation
laws supplemented with the transport equation for the volumetric
fraction of the water phase:
\begin{eqnarray}
   && \frac{\partial \rho \mathbf{U}}{\partial t}
      + \mathbf{\nabla} \cdot \left( \rho \mathbf{U} \mathbf{U} \right)
      = \mathbf{\nabla} \cdot \left( \mu \nabla \mathbf{U} \right)
      + \sigma  \kappa \mathbf{\nabla} \alpha
      - \mathbf{g} \cdot \mathbf{r} \mathbf{\nabla} \rho
      - \mathbf{p_d}
   \label{eq:ns1}\\
   && \mathbf{\nabla} \cdot \mathbf{U} = 0
   \label{eq:ns2}\\
   &&
   \frac{\partial \beta}{\partial t}
      + \mathbf{\nabla} \cdot \left( \beta \mathbf{U} \right) = 0
   \label{eq:ns3}
\end{eqnarray}
Density of the mixture $\rho$ is determined by using the water volumetric
fraction $\beta$ as follows: ${\rho = \beta \rho_w + (1-\beta) \rho_a}$, 
where $\rho_w$ and $\rho_a$ are the densities of water and air respectively.
Similar expression is used to determine the dynamic viscosity
of the mixture $\mu$. Equation (\ref{eq:ns1}) involves the dynamic
pressure $\mathbf{p_d} = \mathbf{p} - \rho \mathbf{g} \cdot \mathbf{r}$,
where $\mathbf{r}$ is the radius-vector in Cartesian coordinates
and $\mathbf{g}$ is the acceleration of gravity.
Surface tension \deleted[id=ZM]{force} is taken into account by the coefficient
$\sigma$ and the local interface curvature $\kappa$.
\replaced[id=ZM]{
The VOF based NS model (\ref{eq:ns1})-(\ref{eq:ns3})
are discretised
by a finite volume method on collocated grids and the transient
flow problem is solved by the PISO method  \citep{Issa2001}.}%
{
Note that the system (\ref{eq:ns1})-(\ref{eq:ns3})
cannot be explicitly solved with respect to $p_d$ ensuring conservation
of mass according to (\ref{eq:ns2}).
To overcome this numerical problem, the so-called PISO algorithm
is implemented in the OpenFOAM$^{\text{\textregistered}}$ framework
suggesting iterative solution of the equations \citep{Issa2001}.}
In the following part of the paper, we use `VOF' and `NS' interchangeably when referring to the two-phase incompressible viscous flow model.

\replaced{To be consistent with the BEM model's 2D domain}{In consistency with BEM model}, the VOF domain was \replaced[id=ZM]{discretised by}{discretized using}
\replaced[id=ZM]{cuboid mesh}{cubic} cells \replaced[id=ZM]{with one}{placed in} single layer in {the} $y$ direction,
\replaced[id=ZM]{thus generating a}{the so-called} pseudo-2D domain, see Figure~\ref{fig:domain}.
The left boundary of the VOF domain was displaced by $L_i$ with respect
to \deleted[id=ZM]{that of} the BEM domain as shown in Figure~\ref{fig:domain}.
The solution of the BEM model was thus used to determine both the initial
and the left boundary condition for the VOF model \replaced[id=ZM]{to establish}{establishing} a
one-way coupling between them. The numerical absorption of \deleted[id=ZM]{the} waves
at the end of the domains was performed independently in both BEM and
VOF models in order to avoid any interplay between them that may affect
the wave train evolution process.
The velocity field damping using the effective viscosity \replaced[id=ZM]{was}{has been}
implemented near the far-end boundary of the VOF domain.

The Reynolds number for the wave trains considered in the research is
\citep{Iafrati2009}:
\begin{eqnarray}
   && Re = \frac{\rho_w g^{1/2} \lambda_0^{3/2}}{\mu_w} > 10^5
   \label{eq:re}
\end{eqnarray}
It suggests that non-breaking wave trains may produce turbulence, as \deleted[id=ZM]{was}
demonstrated by \citet{Babanin2012}.
Even for a two-dimensional problem, the numerical simulation of flows
at such high Reynolds number requires enormous computational effort
to resolve all \replaced[id=ZM]{the scales involved in}{scales appearing as a result of the} wave breaking.
Assuming that the nearly laminar flow due to the surface gravity wave
is dominant in the problems considered, it is expected to have
the grid convergence in terms of free surface elevation.
In course of the study it was established that the grid resolution
of 256 cells per carrier wave length ($\lambda_0$) \replaced[id=ZM]{is sufficient to produce
converged solutions}{provides
the converging solution} while balancing the computational efficiency
and the capability to resolve \replaced[id=ZM]{key}{the small scale} flow features.
The details on the grid independence are given in Appendix~\ref{app:A},
while the domain configuration is \replaced[id=ZM]{summarised}{summarized} in Table~\ref{tab:foam}.

\begin{table}
\centering
\caption{\label{tab:foam}
Parameters of the VOF model. Details on the
wave trains studied are presented in Section~\ref{sec:trains};
$\lambda_0$ is the carrier wave length.}
\begin{tabular}{ l c c }
 \hline
  & Gaussian wave train \;\; & \;\; Modulated wave train \\
 \hline
 Coordinate of the left boundary $L_i$ & 3.0 m & 30 m \\
 Length of the domain & $24 \lambda_0$ & $16 \lambda_0$ \\
 Total height of the domain ($2 h$) & \multicolumn{2}{c}{1.2 m} \\
 Length of the absorbing region & \multicolumn{2}{c}{$8 \lambda_0$} \\
 Grid resolution & \multicolumn{2}{c}{256 cells per $\lambda_0$} \\
 \hline
\end{tabular}
\end{table}

The spatial and temporal numerical schemes for solution of the
equations (\ref{eq:ns1}) - (\ref{eq:ns3}) were selected \added[id=ZM]{by} following
the recommendations for the \emph{interFoam} solver \citep{Larsen2019}.
\replaced[id=ZM]{Adaptive}{Similarly, the integration} time step was \replaced[id=ZM]{used}{selected to be adaptive
maintaining} \added[id=ZM]{and} the stability criterion \added[id=ZM]{was set as} $CFL \leq 0.65$.

\subsection{Coupling of the FNP and NS models}
\label{sec:coupling-bem-vof}

The coupling \added{of the FNP and NS models} was achieved \replaced{through}{in} the following steps.
Firstly, the velocity field $\mathbf{U}$ \replaced{was}{is} constructed in the interior
area of the BEM domain using the known boundary values of $\varphi$
and $\partial \varphi / \partial n$.
The values of $\mathbf{U}$ in the BEM domain were calculated at the
coordinates corresponding to the cell centres of the VOF mesh.
Several numerical techniques of evaluation of
${\mathbf{U} \equiv \{u, w\} = -\mathbf{\nabla}\varphi}$
were examined in terms of \deleted[id=ZM]{the} accuracy and \deleted[id=ZM]{the} computational efficiency.
It was found that the simplest central differencing scheme provides
a reasonable accuracy, while keeping the process computationally
efficient:
\begin{eqnarray}
   \begin{aligned}
   && u = -\frac{\varphi(x_{cell}+\Delta x,z_{cell})
        - \varphi(x_{cell}-\Delta x,z_{cell})}{2\Delta x}, \\
   && w = -\frac{\varphi(x_{cell},z_{cell}+\Delta z)
        - \varphi(x_{cell},z_{cell}-\Delta z)}{2\Delta z},
   \end{aligned}
   \label{eq:Uphi}
\end{eqnarray}
where $x_{cell}$ and $z_{cell}$ are coordinates of the mesh cells
of the VOF domain.
The resolution of the scheme $\Delta x = \Delta z$ was taken equal to
$1/10$ of the size of the VOF cell.
Other resolutions, i.e. 1/6 and 1/16, were also tested.
The values of the potentials $\varphi(x,z)$ in (\ref{eq:Uphi})
were calculated in the BEM solver by selecting the location of the
source at the coordinates $\mathbf{r_s} = (x,z)$ and performing
integration of (\ref{eq:green}).
Secondly, the BEM velocity field and the free surface profile were used to derive the appropriate boundary and initial conditions for the VOF model.

\subsection{\label{sec:trains} Wave train generation}

Two types of \deleted[id=ZM]{the} wave train{s} are considered in this study. 
\replaced[id=ZM]{To validate the proposed hybrid}{For validation of the suggested coupled} BEM-VOF model against the
experiments of \citet{Perlin2012} and \citet{Ducrozet2018},
we investigate the wave breaking appearing in \replaced[id=ZM]{a}{the}
narrow-banded wave train subjected to modulational instability.
The surface elevation at the wavemaker location is:
\begin{eqnarray}
   && \eta(t) = a_0 \cos(\omega_0 t)
              + b \cos \left( \omega_1 t - \frac{\pi}{4} \right)
              + b \cos \left( \omega_2 t - \frac{\pi}{4} \right),
   \label{eq:mi}
\end{eqnarray}
where $a_0$ and $\omega_0$ are the amplitude and angular frequency
of the carrier wave; frequencies of the sideband perturbations are
$\omega_1 = \omega_0 - \Delta \omega / 2$ and
$\omega_2 = \omega_0 + \Delta \omega / 2$.
The following parameters were adopted from the case MI0719 of
\cite{Ducrozet2018} study:
$\omega_0 = 4.398 \; \text{s}^{-1}$, $\Delta \omega = 0.317$,
$b / a_0 = 0.5$. The carrier angular frequency is related to the
corresponding wavenumber $k_0(\omega_0)$
by the linear dispersion relation:
\begin{eqnarray}
   && \omega^2 = \left( g k + \frac{\sigma}{\rho} k^3 \right) \tanh (k h),
   \label{eq:disprel}
\end{eqnarray}
For the range of the wavenumbers considered in the paper, \replaced[id=ZM]{capillary effect}{the effect of
capillarity} is not significant \replaced[id=ZM]{so we set}{allowing assumption} $\sigma = 0$.
The initial steepness of the wave train was $k_0 a_0 = 0.19$.

\begin{figure}
\centering
\begin{overpic}[width=0.45\linewidth]{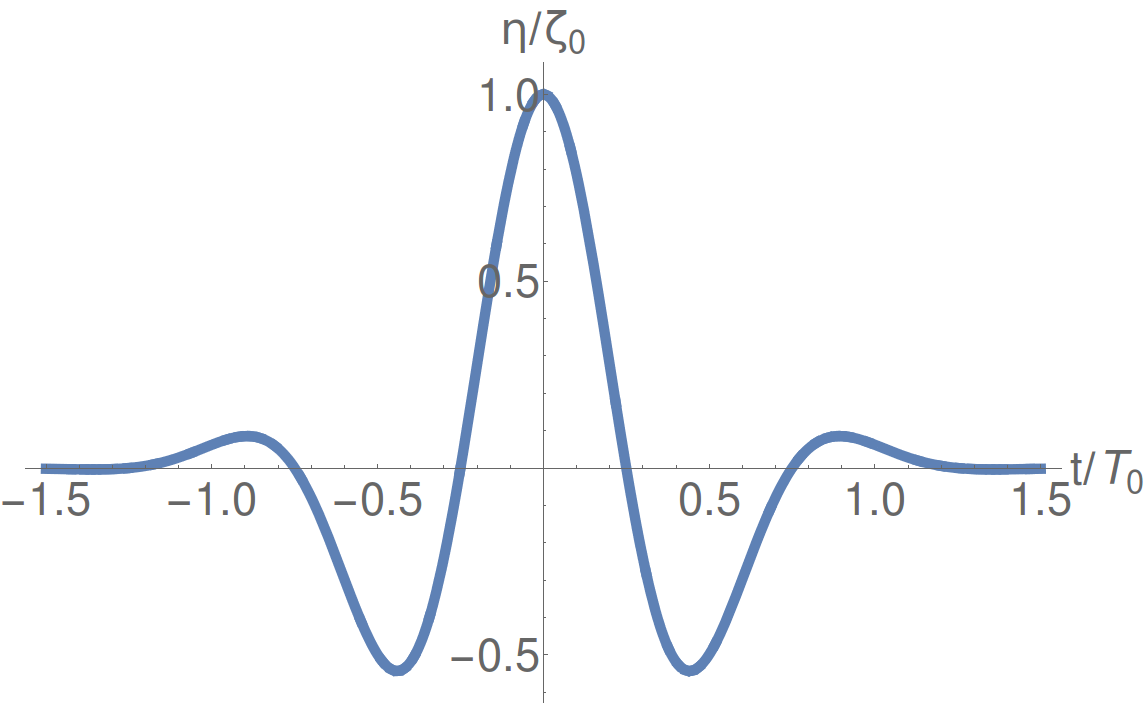}
    \put(30,54){\textbf{(a)}}
\end{overpic}
\begin{overpic}[width=0.45\linewidth]{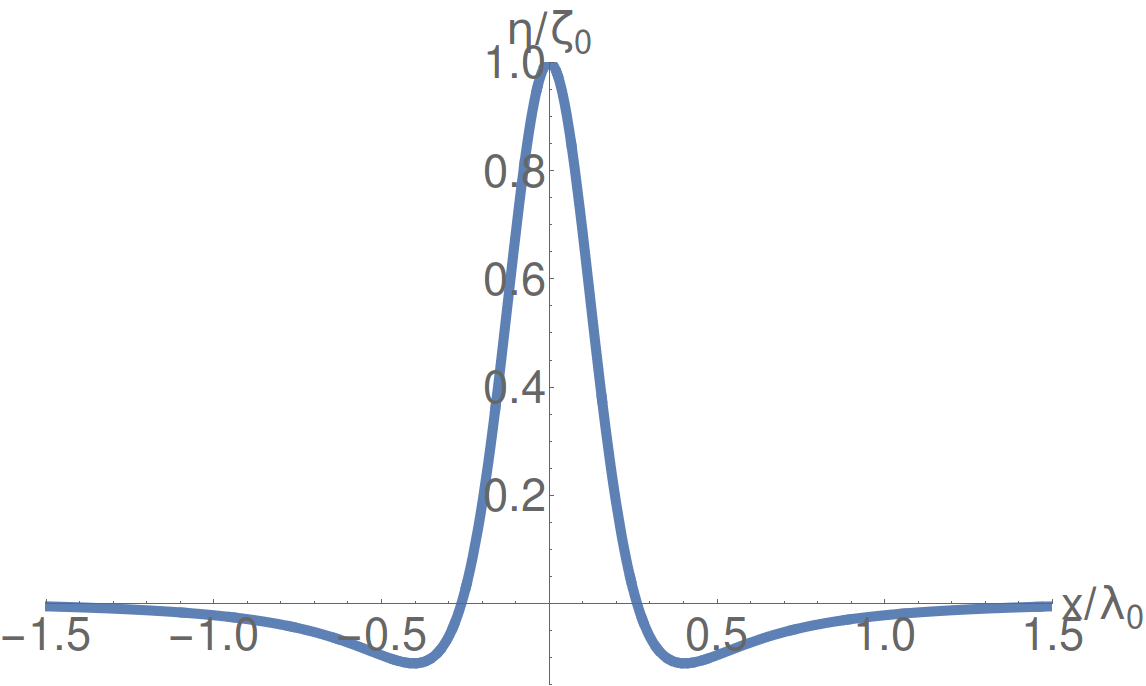}
    \put(30,54){\textbf{(b)}}
\end{overpic}\\
\begin{overpic}[width=0.45\linewidth]{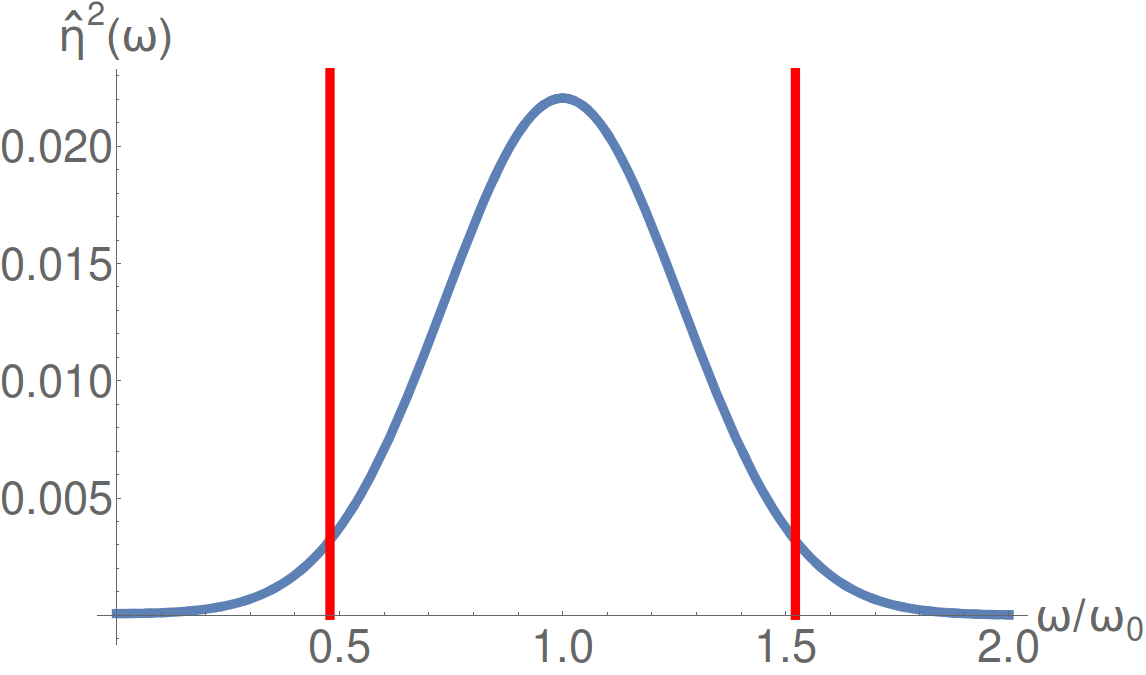}
    \put(45,54){\textbf{(c)}}
\end{overpic}
\begin{overpic}[width=0.45\linewidth]{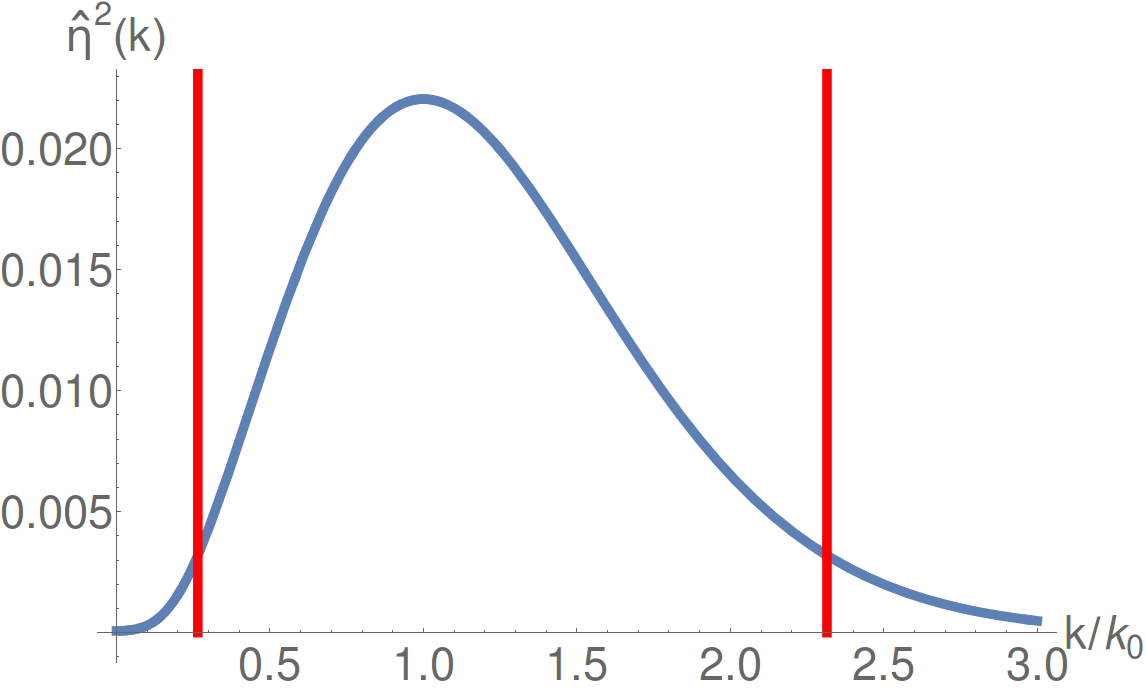}
    \put(45,54){\textbf{(d)}}
\end{overpic}\\
\caption{Broad-banded Gaussian-shaped focusing wave train considered
         in the study: 
\replaced[id=ZM]{(a) free surface elevation at the focal point $x_f=8.5$ m;
(b) wave profile at the focal time $t_f=0$;
(c) frequency and (d) wavenumber power spectra.}
{(a) and (b) temporal and spatial variation of the surface elevation;
         (c) and (d) frequency and wavenumber power spectra,
         respectively.}%
 Vertical red lines designate the range of
         frequencies and the wavenumbers containing $95 \%$
         of the spectral energy.}
\label{fig:math}
\end{figure}

A broad-banded Gaussian-shaped focusing wave train was selected
for further investigation of the wave breaking phenomenon.
This wave train implies the spatial periodicity of the free surface elevation
if the domain is sufficiently long, which is critically required
for the accurate post-processing of the simulation results.
The strongest wave breaking in this case is expected in the vicinity
of the focus location whose coordinate relative to the wave generating
boundary in the BEM domain ($x = 0$) was selected as $x_f = 8.5\;\text{m}$.
The surface elevation variation with time at $x_f$ is:
\begin{eqnarray}
   && \eta(t, x=x_f) = \zeta_0 \exp \left\lbrace 
                    - \left( \frac{t}{m T_0} \right)^2
                      \right\rbrace \cos(\omega_0 t)
   \label{eq:gauss}
\end{eqnarray}
The parameter $m = 0.6$ determines the broad-banded wave train;
the carrier wave period and angular frequency are $T_0 = 0.7$ s and
$\omega_0 = 2 \pi / T_0$, respectively. According to the
linear dispersion relation (\ref{eq:disprel}) the carrier wave length
is $\lambda_0 = 2 \pi / k_0 = 0.765$ m. The dimensionless water depth
corresponds to deep water condition \citep{D&D1991},
i.e. $k_0 h = 4.93 > \pi$.
The plot of $\eta(t)$ at the focal point $x_f$ is shown in
Figure~\ref{fig:math}(a); the spatial surface elevation $\eta(x)$
at the instant of focusing $t_f$ is plotted in Figure~\ref{fig:math}(b).

In this study we investigate the evolution of the free waves, and this requires 
to exclude the bound waves from the BEM and VOF results.  Since bound waves appear predominantly at high and low frequencies with respect to the carrier frequency $\omega_0$, it is possible to partially
avoid their influence by band-pass filtering those regions.  Consider the Fourier
transform of surface elevation (\ref{eq:gauss}):
%
\begin{equation}
      \hat{\eta}(\omega) = \mathscr{F}\{\eta(t)\} =
      \frac{\zeta_0}{2\sqrt{2}} m T_0 
      \exp \left\{ - \pi^2 m^2
           \left(1 + \frac{\omega}{\omega_0} \right)^2 \right\}
      \left( 1 + \exp \left\{ 4 \pi^2 m^2 \frac{\omega}{\omega_0}
                      \right\} \right)
   \label{eq:spec}
\end{equation}
%
The wavenumber spectrum for deep water waves can be obtained by expressing $\omega$
from the linear dispersion relation (\ref{eq:disprel}) and substituting it
into (\ref{eq:spec}).
The energy fraction $\delta e$ contained in the frequency range
$\left[ \omega_0 - \Delta \omega , \omega_0 + \Delta \omega \right]$ is:
\begin{eqnarray}
   && \delta e = \frac{\int_{\omega_0-\Delta\omega}^{\omega_0+\Delta\omega}
                     {\hat{\eta}^2 d \omega}}
               {\int_{0}^{+\infty}{\hat{\eta}^2 d \omega}}
      \approx 2 \sqrt{2 \pi} m
      \frac{e^{-6 \pi^2 m^2} \left( 1 + e^{4 \pi^2 m^2} \right)^2}
           {1 + e^{2 \pi^2 m^2}}
      \frac{\Delta \omega}{\omega_0}
   \label{eq:energy}
\end{eqnarray}
Solution of (\ref{eq:energy}) with respect to $\Delta \omega$
assuming $\delta e = 0.95$ allows to find the frequency band
containing $95 \%$ of the spectral energy.
Following this procedure, the frequency and the wavenumber ranges
that will be considered in the further analysis were estimated
as follows: $ 0.48 \leq \omega / \omega_0 \leq 1.52$ ;
$ 0.27 \leq k / k_0 \leq 2.31$.
The power spectra with the corresponding frequency and the wavenumber
bounds are depicted in Figures~\ref{fig:math}~(c)~and~(d).

Assuming deep water dispersion $k = \omega^2 / g$, the spatio-temporal
variation of the surface elevation can be obtained from
(\ref{eq:gauss}) and (\ref{eq:spec}) by using the linear approximation
for \deleted{the gravity} water waves:
\begin{eqnarray}
   && \eta(x, t) = \mathscr{F}^{-1} \left\{ \hat{\eta}(\omega) 
                 \exp \left[ i k (\omega) x \right] \right\}
             = \mathscr{F}^{-1} \left\{ \hat{\eta}(\omega) 
             \exp \left[ i \frac{\omega |\omega|}{g} x \right]
             \right\},
   \label{eq:spatial}
\end{eqnarray}
where $\mathscr{F}^{-1}$ is the inverse Fourier transform.
The linear approximation for the steepness variation in the course of
the wave train temporal evolution is:
\begin{eqnarray}
   && \varepsilon_{max} (t) = \underset{-\infty<x<+\infty}{\max}
      \left| \frac{\partial \eta(x, t)}{\partial x} \right|
   \label{eq:steepness}
\end{eqnarray}
It is known that waves break when their steepness satisfy the \deleted{approximate}
condition ${\varepsilon > 0.3}$. Increasing the wave steepness beyond
this value may lead to a single or multiple breaking events. The strength
of breaking is also dependent on the value of $\varepsilon$.
Varying the value of the constant $\zeta_0$ in (\ref{eq:gauss}),
(\ref{eq:spec}) and (\ref{eq:spatial}), six wave trains of different
steepness $k_0 \zeta_0 = 0.2,\; 0.3, \; 0.4, \; 0.6, \; 0.8\; \text{and} \; 1.0$
are taken for investigation.

\begin{figure}
    \centering
    \includegraphics[width=0.6\textwidth]{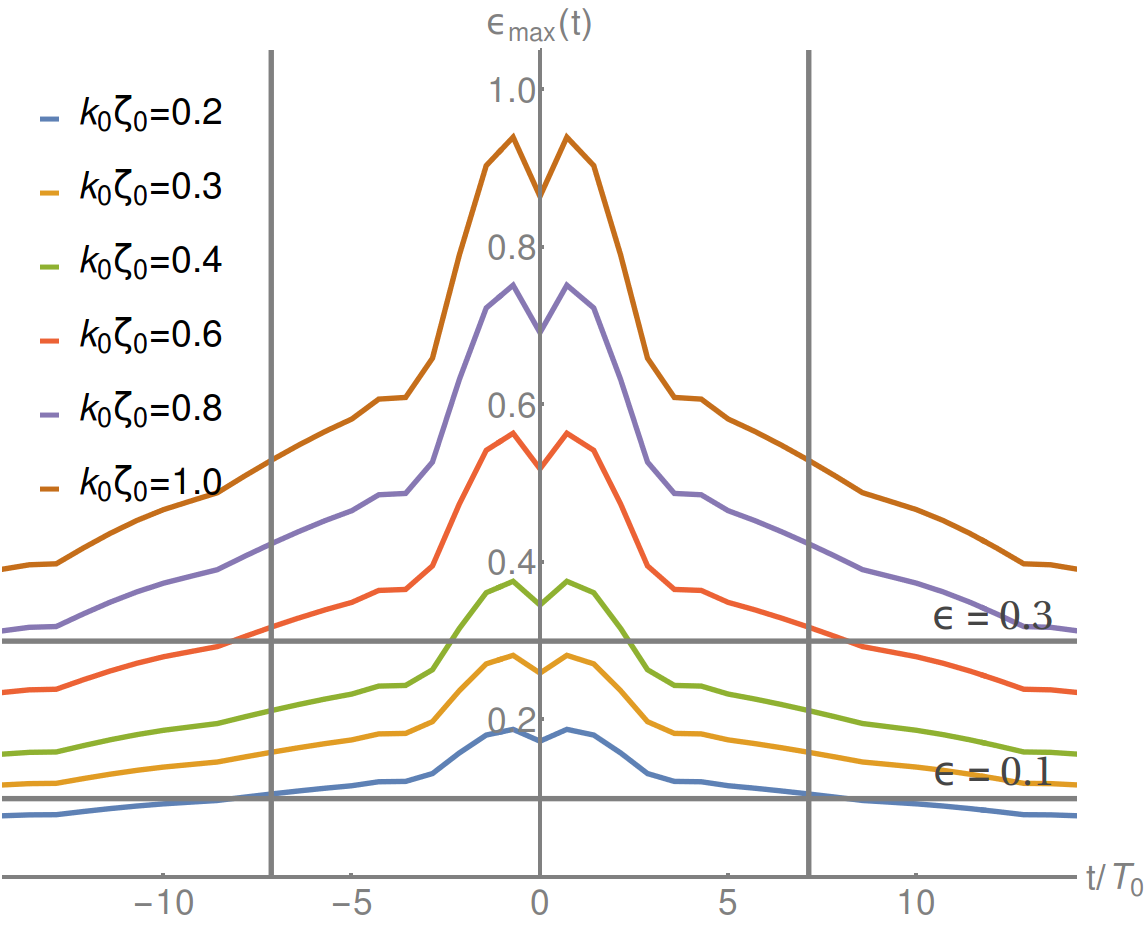}
    \caption{Temporal variation of the wave train steepness
             (\ref{eq:steepness}) for all considered cases.
             In this plot, the dispersive focusing appears at $t=0$.
             Two gray vertical lines depict the range
             $-7.14 \leq t/T_0 \leq +7.14$ particularly considered in
             the study. Within this time interval the full length of the
             wave train is present in the limits of the computational
             domain of both BEM and VOF models.}
    \label{fig:steepness}
\end{figure}

The wave train steepness (\ref{eq:steepness}) for all considered cases
is plotted in Figure~\ref{fig:steepness}. 
As expected for \replaced{a}{the} broad banded focusing wave train, the maximum value
of $\varepsilon_{max}$ is seen in the vicinity of the focal point, while 
it reduces farther away from this location.
Note that in the course of the study, particular attention \replaced{is}{will be} given
to the time interval $-7.14 \leq t/T_0 \leq +7.14$, when the full length
of the wave train is present within the limits of the computational domain
of both BEM and VOF models.
The steepness of all wave trains within the given time interval is
${\varepsilon_{max} > 0.1}$, thus showing the significance of nonlinearities.
The mild single breaking event may be expected when $k_0 \zeta_0 = 0.3$,
because the steepness satisfies the condition $\varepsilon_{max} > 0.3$ near the focal point for this case.
If $k_0 \zeta_0 \geq 0.6$, the steepness is
greater than $0.3$ within the entire time interval of interest as shown in the figure.
\replaced[id=ZM]{This}{which} means that waves may continuously break throughout the evolution
of the wave train.

Since the accuracy of \deleted[id=ZM]{the} wave generation is critical in the current
investigation, the ${2^{\text{nd}}\text{-order}}$ accurate method for
calculation of the wavemaker motion from the surface elevation 
was adopted \citep{Khait2019b}. The obtained wavemaker motion was then
used to control the displacement of the wave-generating boundary in
the BEM model, see Figure~\ref{fig:domain}.
The surface elevation variation with time at the wavemaker location
$\eta(x=0, t)$ was calculated according to (\ref{eq:spatial}).

\subsection{\label{sec:decompose} Spectrum decomposition}

It is known that the domains of the free and bound waves may overlap
each other in both frequency and wavenumber spectra \citep{Khait2019b}.
Despite limiting the analysis to a certain frequency range as discussed
above, the effect of bound waves on the surface elevation spectrum
may still be large leading to complication of the analysis. 
To facilitate the study, the bound waves should be separated from
the free waves \replaced[id=ZM]{by}{that can be done} using \deleted[id=ZM]{the} Zakharov's weakly-nonlinear
theory \replaced[id=ZM]{for}{of} surface water waves
\citep{Zakharov1968, Shemer1984, Shemer1987, Krasitskii1994}.
Within this theory, the surface elevation of nonlinear wave{s} may be
represented as a series of contributions appearing at different
orders of small parameter $\varepsilon$:
$\eta = \eta^{(1)} + \eta^{(2)} + O(\varepsilon^3)$; $\varepsilon$
is the characteristic wave steepness; $\eta^{(1)}$ and $\eta^{(2)}$
are contributions of the free and the $2^{\text{nd}}$-order bound waves,
respectively.

Application of the Discrete Fourier Transform to the spatial distribution
of the free waves surface elevation $ \eta^{(1)}$ gives the
complex wavenumber spectrum $A(k_m)$, where $k_m$ is the wavenumber
of the $m\text{-th}$ harmonic. The free wave surface elevation is now:
\begin{eqnarray}
   && \eta^{(1)}(x) = Re \left\{
      \sum_{m=0}^M A(k_m) e^{i k_m x}
                         \right\},
   \label{eq:freewave}
\end{eqnarray}
where $M$ is the number of the discrete harmonics.
Surface elevation of the $2^{\text{nd}}$-order bound waves is:
\begin{eqnarray}
   \begin{aligned}
   \eta^{(2)}(x) = Re \left\{ \sum_{m=0}^M \sum_{n=0}^M \left[
                         B(k_m, k_n) e^{i (k_m+k_n) x}
                       + C(k_m, k_n) e^{i (-k_m+k_n) x} 
                       \right. \right.\\
                       \left. \left.
                       + D(k_m, k_n) e^{i (-k_m-	k_n) x}
                       \right] \right\},
   \end{aligned}
   \label{eq:boundwave}
\end{eqnarray}
The complex amplitudes $B(k_m, k_n)$, $C(k_m, k_n)$ and $D(k_m, k_n)$
are expressed in terms of $A(k_m)$ as given in Appendix~\ref{app:B}.

At each instant $t$\deleted[id=ZM]{to be analysed in the study}, the results of
BEM and VOF simulations are processed to determine the distribution
of the surface elevation in space $\eta(x)$ as a series of discrete
values at 2048 points equidistantly distributed along the domains.
The range of the spatial coordinates considered in the analysis
is ${L_i \leq x \leq L_b}$, where $L_i$ corresponds to the inlet boundary
of the VOF domain and $L_b$ is the coordinate of the beginning of the
absorbing region, see Figure~\ref{fig:domain}.
To decompose the fully-nonlinear surface elevation $\eta(x)$ into free
and bound waves it is assumed that 
${\eta(x) \approx \eta^{(1)}(x) + \eta^{(2)}(x)}$.
From this expression it is possible to find the complex spectra
$A$, $B$, $C$ and $D$ iteratively \added[id=ZM]{by} following the algorithm presented in
\citep{Shemer2007, Khait2019a}.
\replaced{In}{At} the first iteration, the free waves spectrum can be taken as
$A(k) = \text{FFT}\{\eta(x)\}$, where $\text{FFT}$ stands for the
Fast Fourier Transform.
Usually, 10 to 20 iterations \replaced[id=ZM]{are}{is} sufficient to converge the spectrum
decomposition. Considering only the separated free waves spectrum
$A(k)$ and limiting the analysis to the wavenumbers range found in the
preceding section, see Figure~\ref{fig:math}, it is possible to \replaced[id=ZM]{minimise}{minimize}
the effect of bound waves.


\section{\label{sec:valid} Model validation and statement of the problem}

\replaced[id=ZM]{Breaking in wave trains}{Wave breaking in the wave train} subject to the modulational
instability (\ref{eq:mi}) was investigated by
\citet{Ducrozet2018}. Spatial evolution of waves was tracked
in experiments by measuring the surface elevation at several coordinates
along the wave flume. In particular, the \replaced[id=ZM]{emphasis}{emphasised} was given to the
following locations with respect to the wavemaker coordinate ($x=0$):
$x = 30.06,\; 34.26,\; 37.88 \;\text{and}\; 50.23 \; \text{m}$;
the total length of the wave flume \replaced{is}{was} $148 \;\text{m}$.
They implemented \added[id=ZM]{the} eddy viscosity approximation \replaced[id=ZM]{proposed}{of the wave breaking} by
\citet{Perlin2010, Perlin2012} in \replaced[id=ZM]{a}{the frame of FNP solver
based on the} High-Order Spectral (HOS) \replaced[id=ZM]{code}{method}.
To validate the BEM-VOF numerical model \replaced[id=ZM]{proposed}{suggested} in \replaced[id=ZM]{this}{the current} paper,
we compare \replaced[id=ZM]{our computations with the}{the results of our simulations with both} numerical
and experimental results of \citet{Ducrozet2018}.

\begin{figure}
\centering
\includegraphics[width=0.6\textwidth]{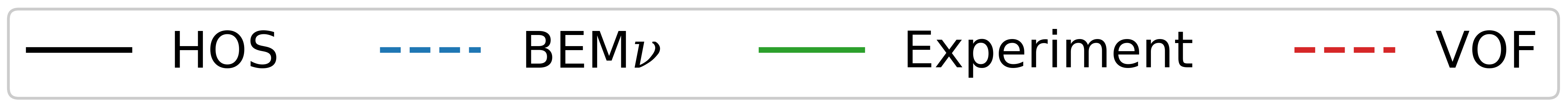} \\

\begin{overpic}[width=0.49\linewidth]{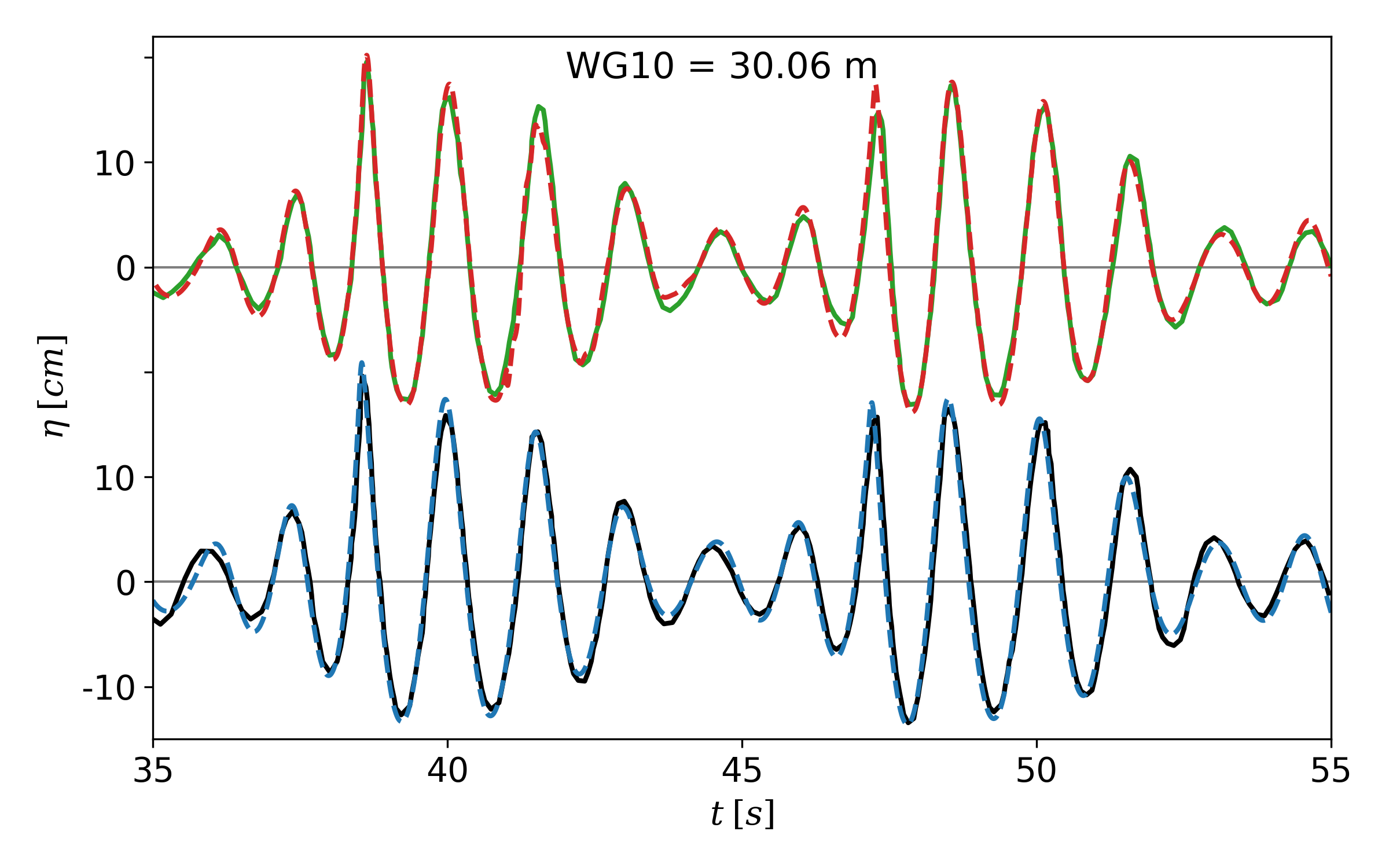}
    \put(2,55){\textbf{(a)}}
\end{overpic}
\begin{overpic}[width=0.49\linewidth]{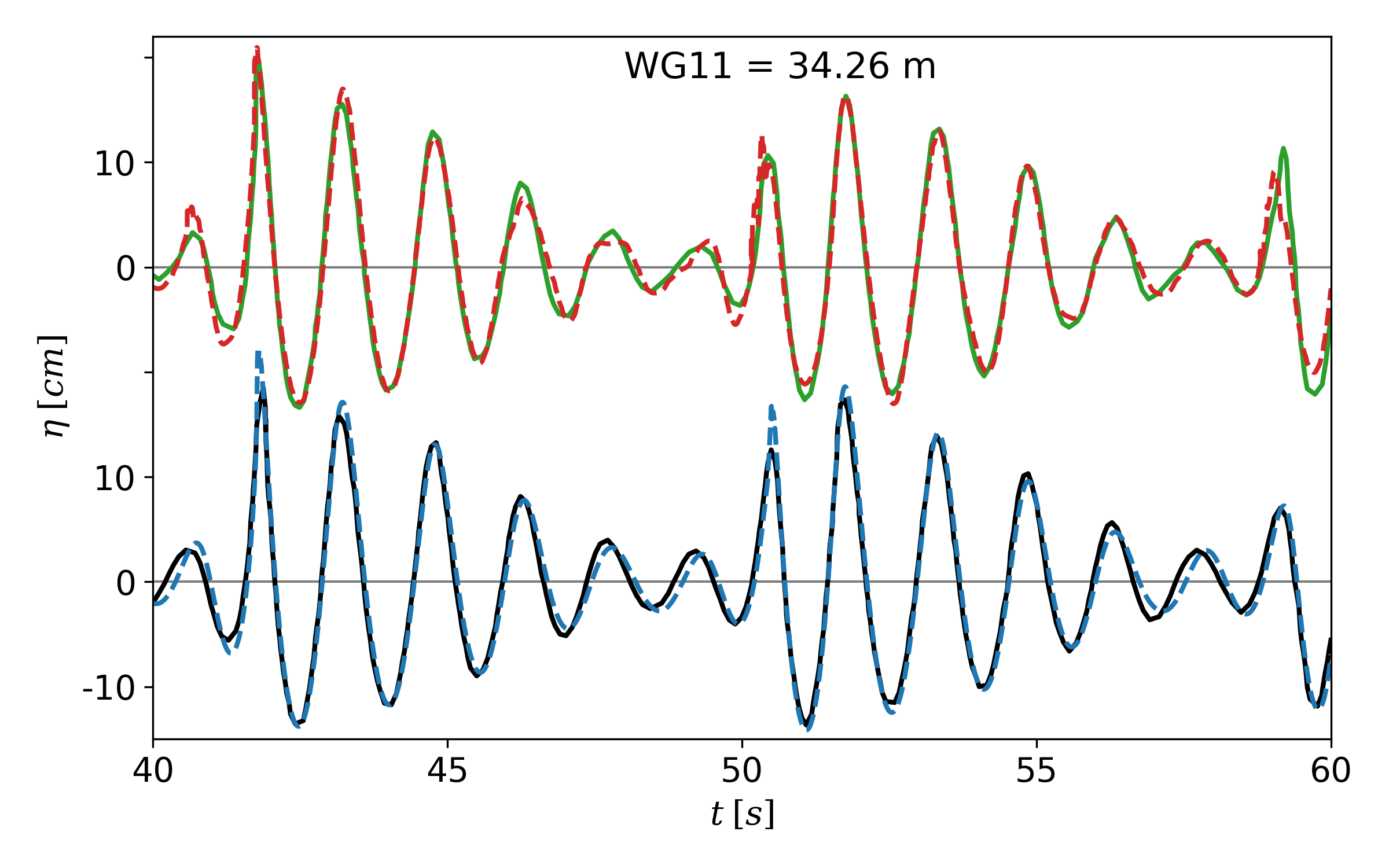}
    \put(-2,55){\textbf{(b)}}
\end{overpic}\\
\begin{overpic}[width=0.49\linewidth]{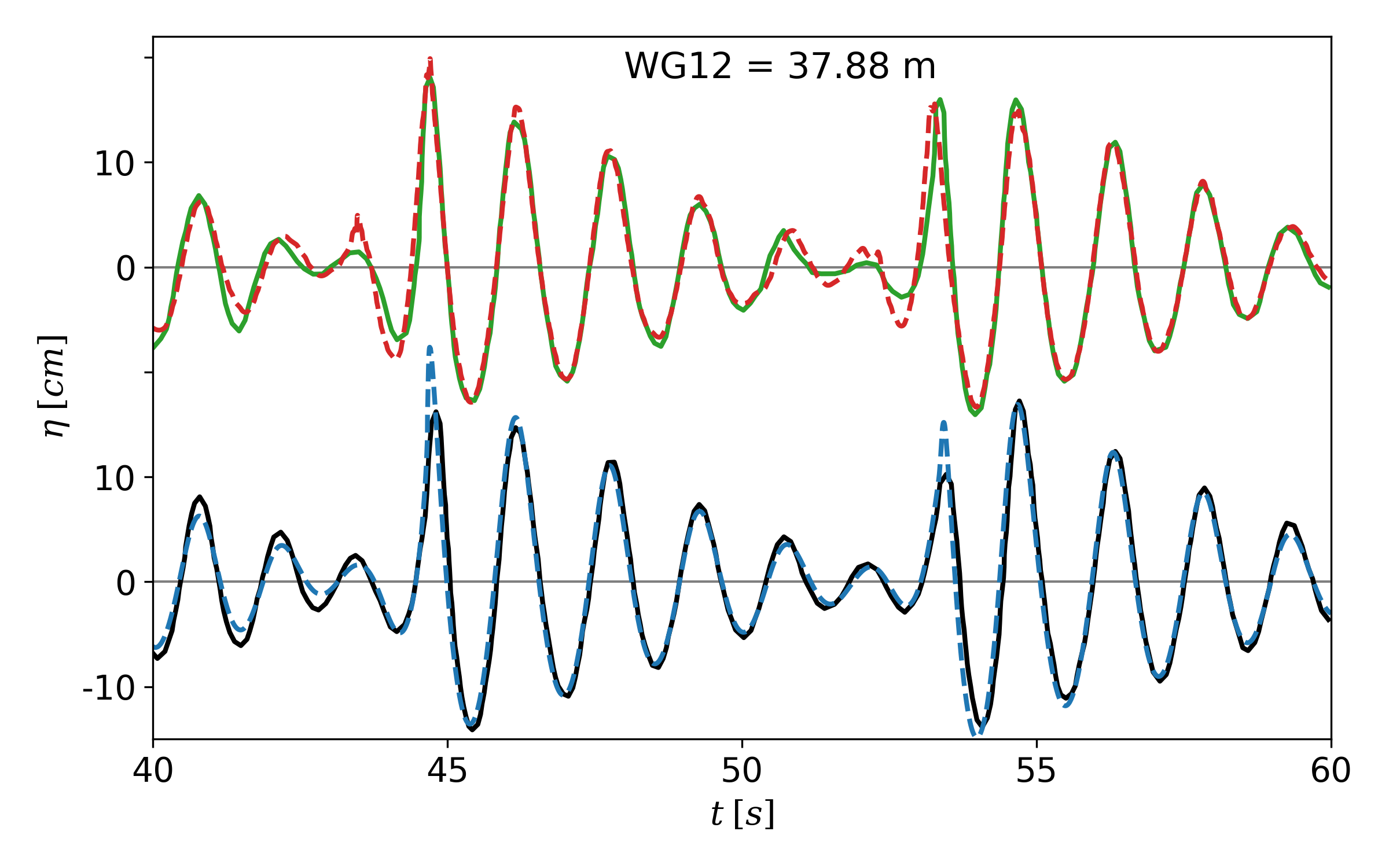}
    \put(2,55){\textbf{(c)}}
\end{overpic}
\begin{overpic}[width=0.49\linewidth]{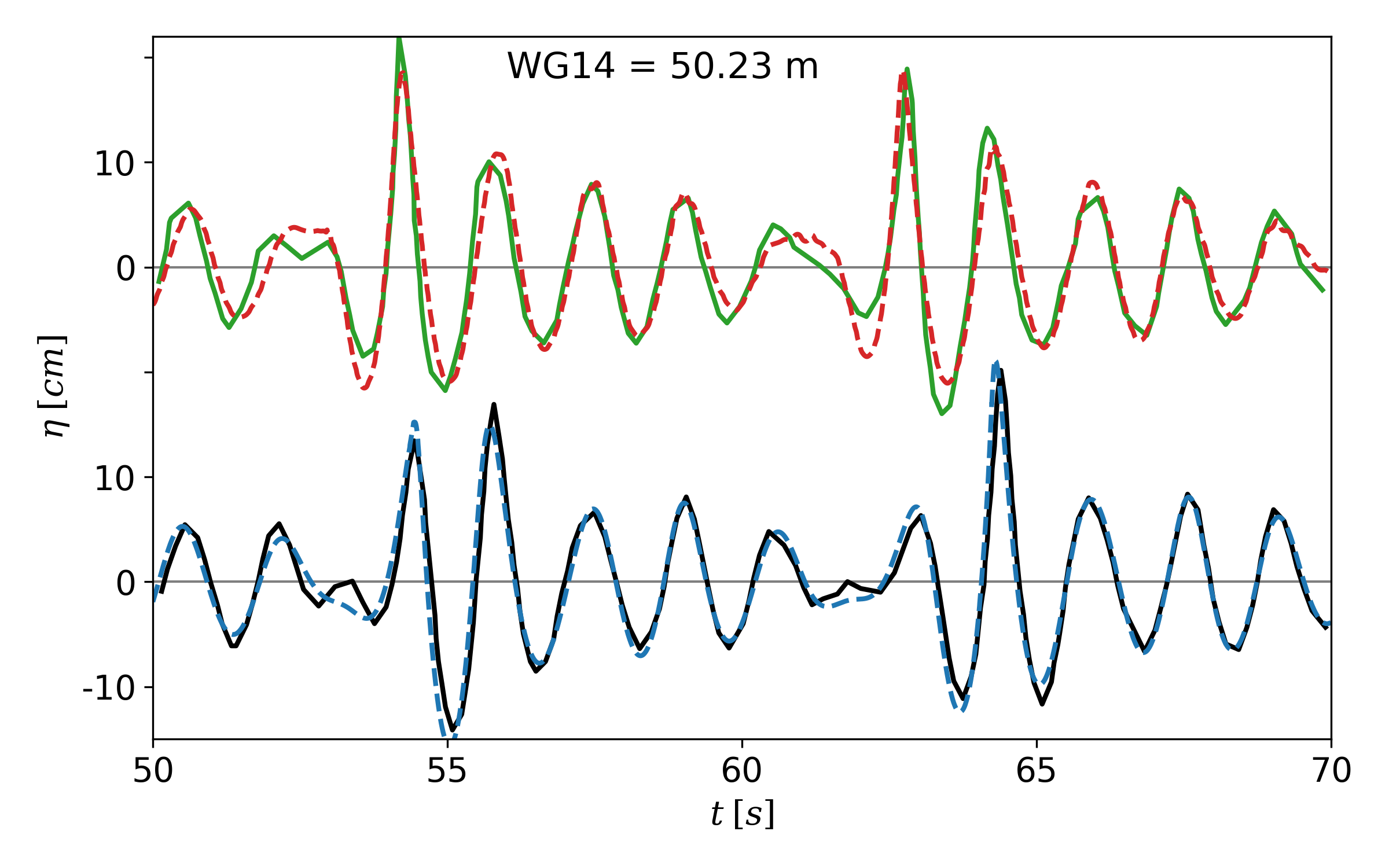}
    \put(-2,55){\textbf{(d)}}
\end{overpic}\\
\begin{overpic}[width=0.95\linewidth]{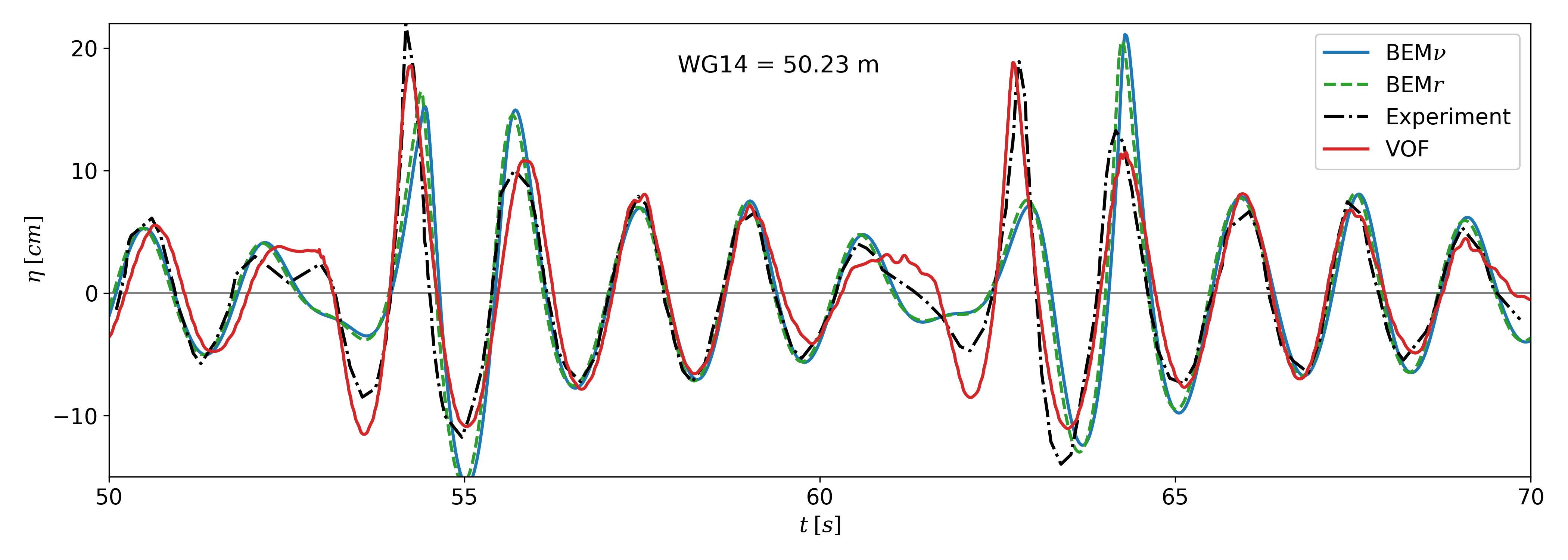}
    \put(0,32){\textbf{(e)}}
\end{overpic}
\caption{Measured and calculated surface elevations for a wave train subject to
modulational instability at four wave gauges: $\text{WG10} = 30.06 \; \text{m}$,
         $\text{WG11} = 34.26 \; \text{m}$,
         $\text{WG12} = 37.88 \; \text{m}$ and
         $\text{WG14} = 50.23 \; \text{m}$. Solid black line: high order spectral simulations of \citet{Ducrozet2018};
Solid green line: experiments of \citet{Ducrozet2018}; Blue dashed line: solution of the present
FNP model with the eddy viscosity enclosure; Red dashed line: result of the present NS model.
 }
\label{fig:SD}
\end{figure}

\replaced[id=ZM]{The computed and measured surface elevations are shown}
{Surface elevation variation with time recorded by different numerical
models and in the experiment are compared} in Figures~\ref{fig:SD} (a)-(d).
First, it can be noted that the results of the BEM$\nu$ model \deleted[id=ZM]{with the
eddy viscosity wave breaking closure} agree very well
with \deleted[id=ZM]{the results of} the HOS simulations \added[id=ZM]{of} \cite{Ducrozet2018}. \deleted{having
similar eddy viscosity correction.} \replaced[id=ZM]{This}{It} confirms the validity of
\replaced[id=ZM]{the}{implementation of the eddy viscosity closure in} BEM$\nu$ model.
\replaced[id=ZM]{At}{In} the same time, the results of the VOF simulations are very close
to the experimental measurements \replaced[id=ZM]{and this demonstrates}{confirming} the accuracy of the
\deleted[id=ZM]{used} two-phase high-fidelity model.

Now compare the results of BEM$\nu$, BEMr and VOF simulations,
see Figure~\ref{fig:SD}~(e). It is clearly seen that at a distant location
from the wavemaker, $WG1{4} = 50.23 \;\text{m}$, the plots of BEM$\nu$
and BEMr computations are close to each other. This suggests that \replaced[id=ZM]{the simple
remeshing technique}{a very
simple method of the free surface mesh regridding} implemented in the
BEMr model \citep{Grilli1994} \replaced[id=ZM]{can produce as good results 
as the complicated eddy viscosity approximation.}{produces a relatively good results
as compare to a more complicated eddy viscosity approximation.}
\replaced[id=ZM]{It can be clearly seen that there is a significant discrepancy between the low-fidelity calculations and 
the experiment \myhighlight{from about 62.5 to 64.5 seconds regarding the peak elevations, potentially critical to maritime safety}.  While the high-fidelity results are quite close to the measurements \myhighlight{in terms of peak surface elevation}.}
{However, both BEM$\nu$ and BEMr outcomes significantly disagree
with the VOF computations and, consequently, with the laboratory
measurements.} Similar \replaced[id=ZM]{observations were reported}{conclusions were made} by \citet{Perlin2012} and
\citet{Ducrozet2018}\deleted[id=ZM]{based on the measurements in the wave flume}.
\replaced[id=ZM]{\cite{Ducrozet2018} attempted to address this issue by modifying the eddy viscosity approximation without receiving much success.}{
The attempts to fix this disagreement by modifying the eddy viscosity
approximation for the wave breaking \citep{Ducrozet2018} did not show
any substantial improvement, suggesting more complicated reasons
for the observed disagreement.}

The eddy viscosity approximation for \deleted[id=ZM]{the} wave breaking is based on
the weakly-damped wave theory \citep{Ruvinsky1991, Dias2008},
\replaced[id=ZM]{which}{that} assumes the rotational components of the fluid velocity to be small.
\replaced[id=ZM]{However, strongly breaking waves may generate}{The strong wave breaking events may lead to generation
of} significant non-potential flows; such as sheared currents, vortices, etc. \citep{Iafrati2013, Melville2017, Melville2019}. This could
 be a reason for the eddy viscosity method \replaced{producing}{to produce} inaccurate prediction of surface elevation.
\deleted[id=ZM]{The current study is focused on the
investigation of the given phenomenon to shed more light on the role
of the non-potential flow components in the wave breaking events.}

\section{Results and Discussion}

\subsection{\label{sec:elevat} Surface elevation}

\begin{figure}
\centering
\begin{overpic}[width=0.49\linewidth]{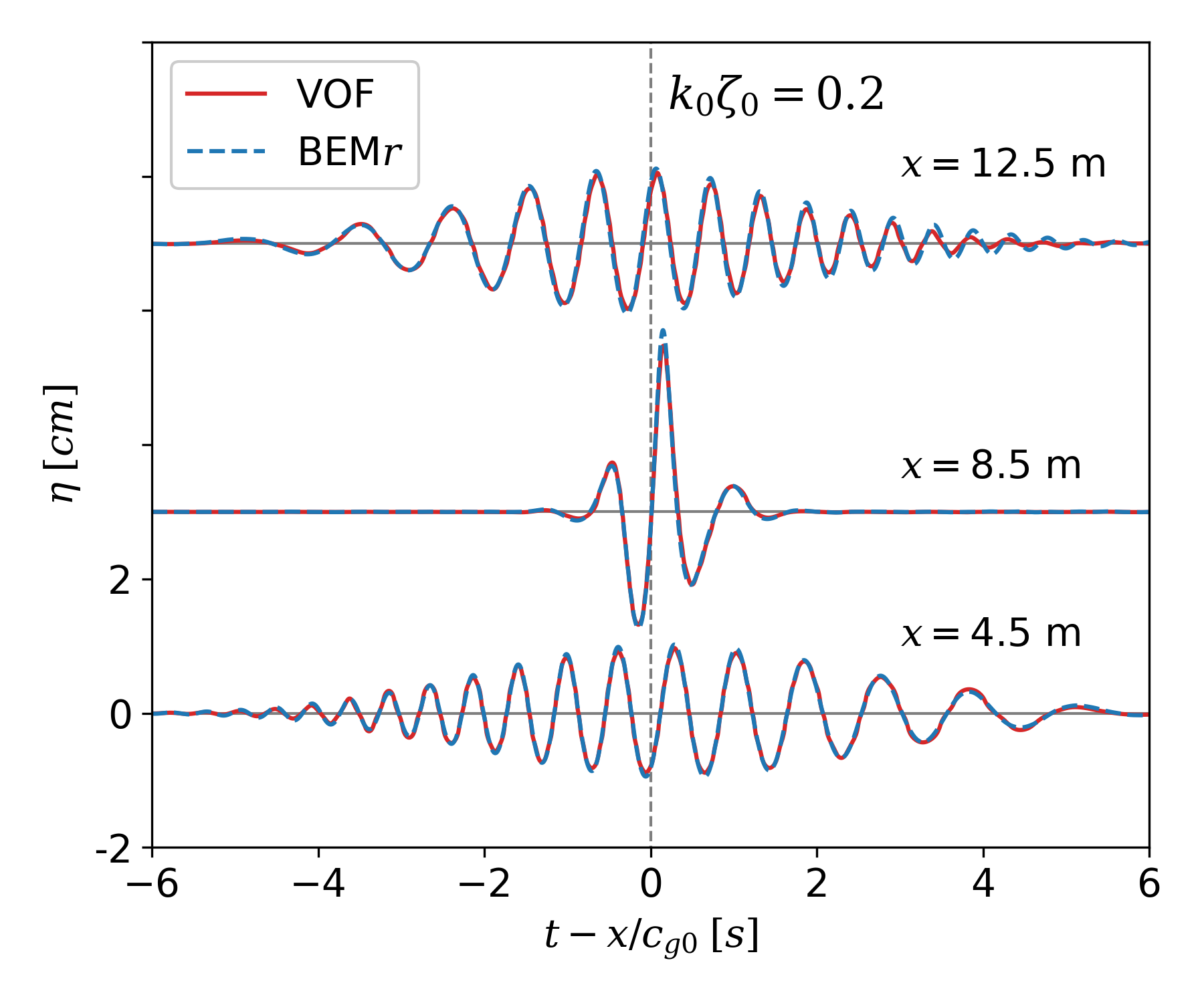}
    \put(2,74){\textbf{(a)}}
\end{overpic}
\begin{overpic}[width=0.49\linewidth]{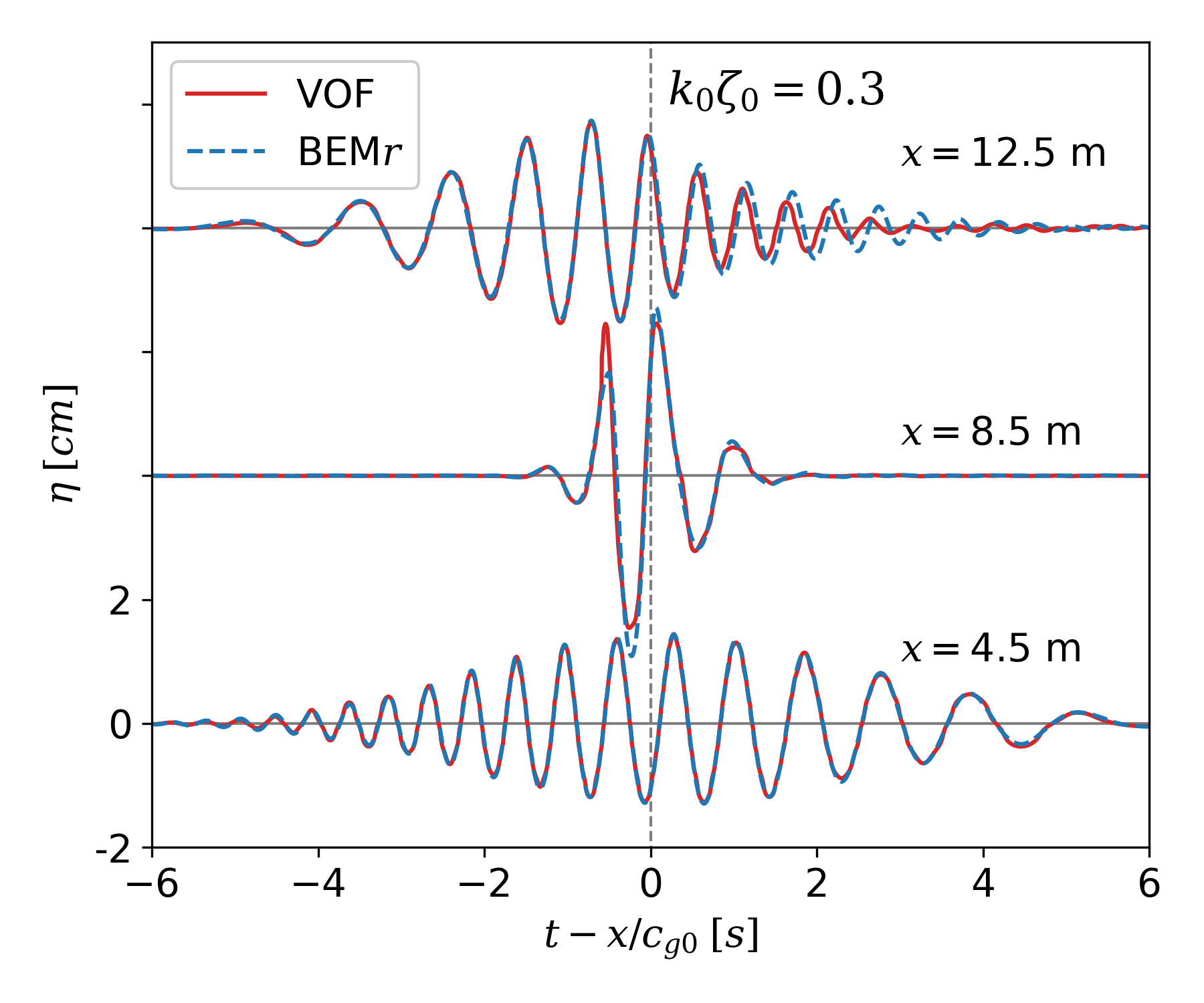}
    \put(2,74){\textbf{(b)}}
\end{overpic}\\
\begin{overpic}[width=0.49\linewidth]{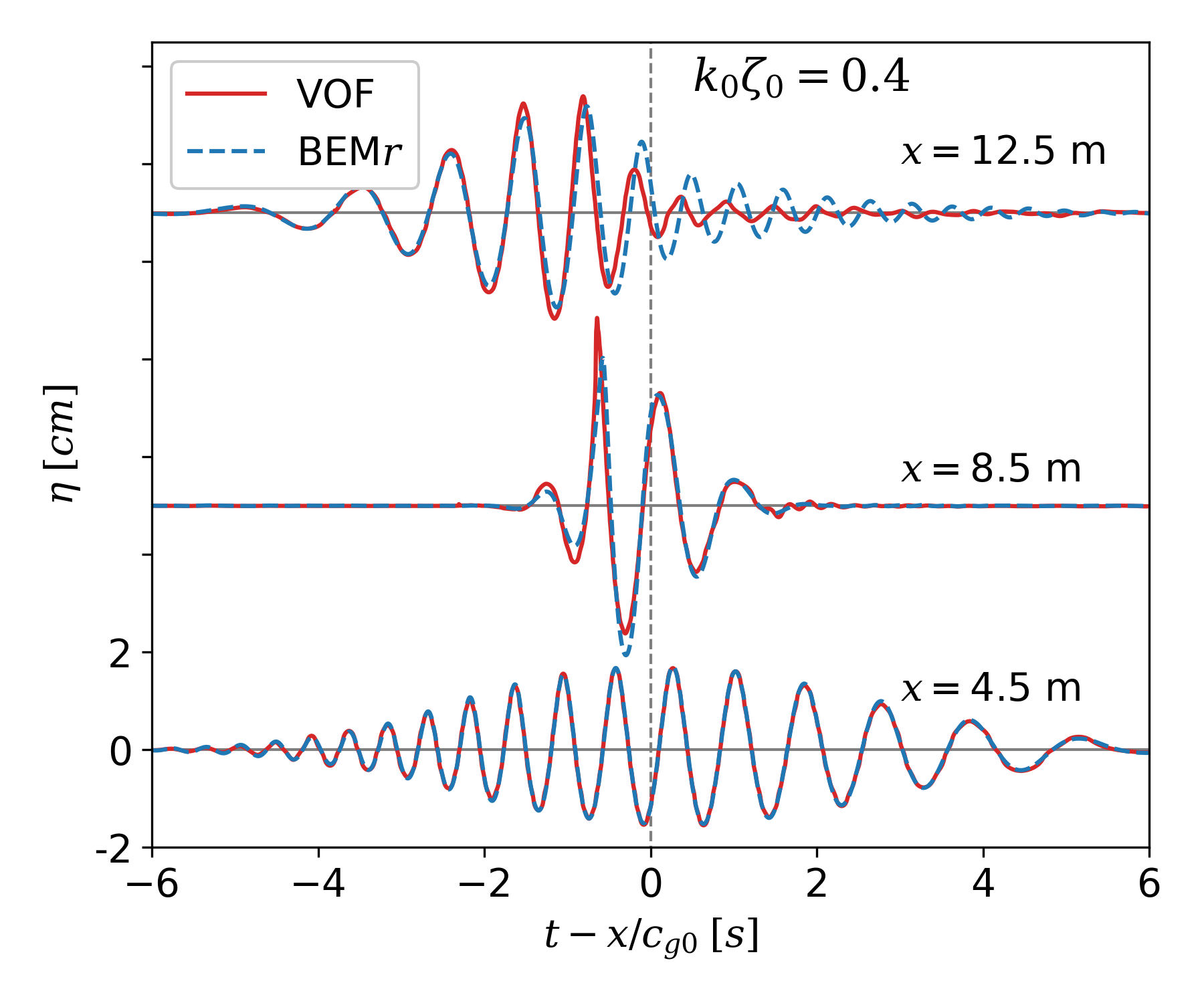}
    \put(2,74){\textbf{(c)}}
\end{overpic}
\begin{overpic}[width=0.49\linewidth]{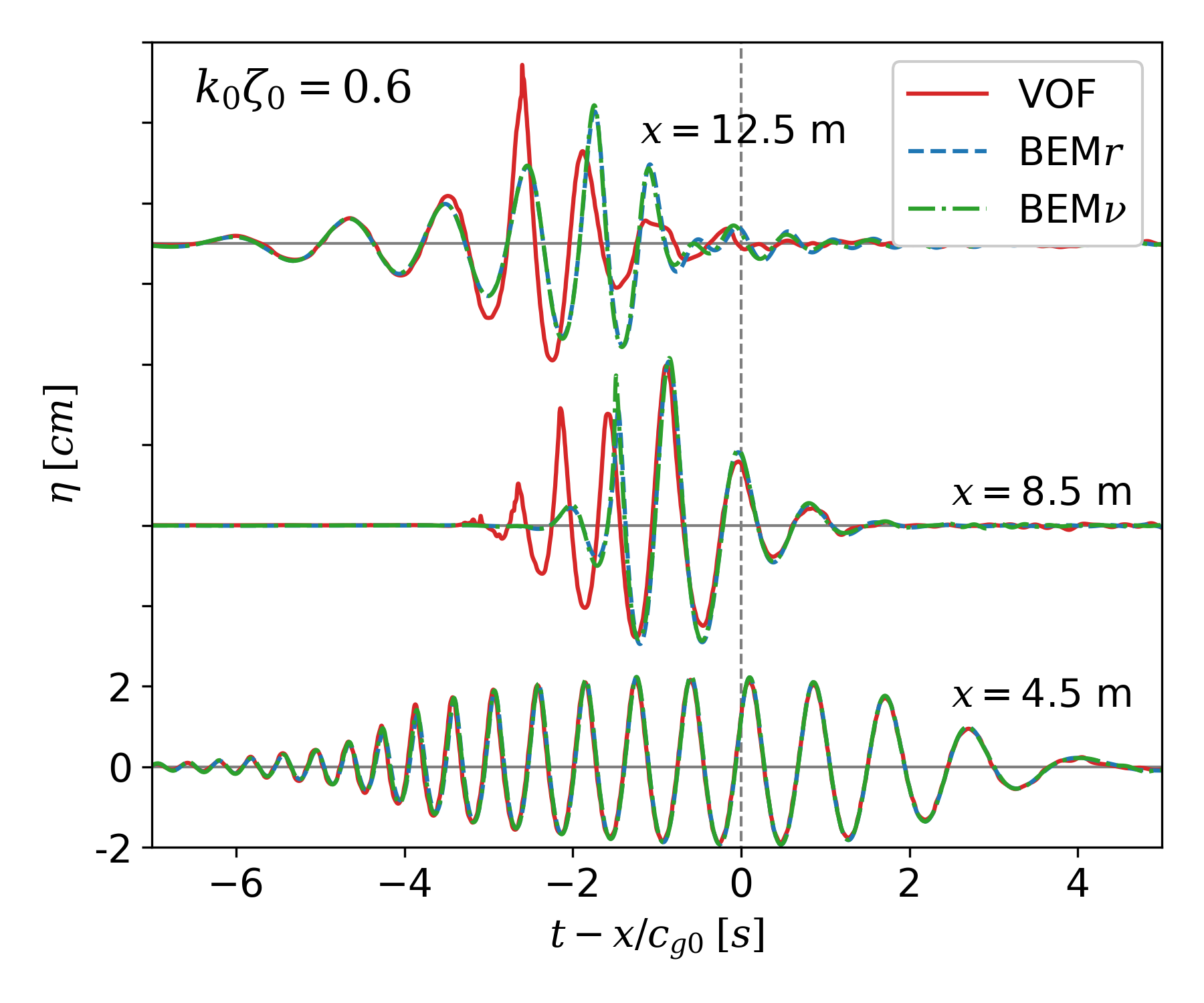}
    \put(2,74){\textbf{(d)}}
\end{overpic}\\
\begin{overpic}[width=0.49\linewidth]{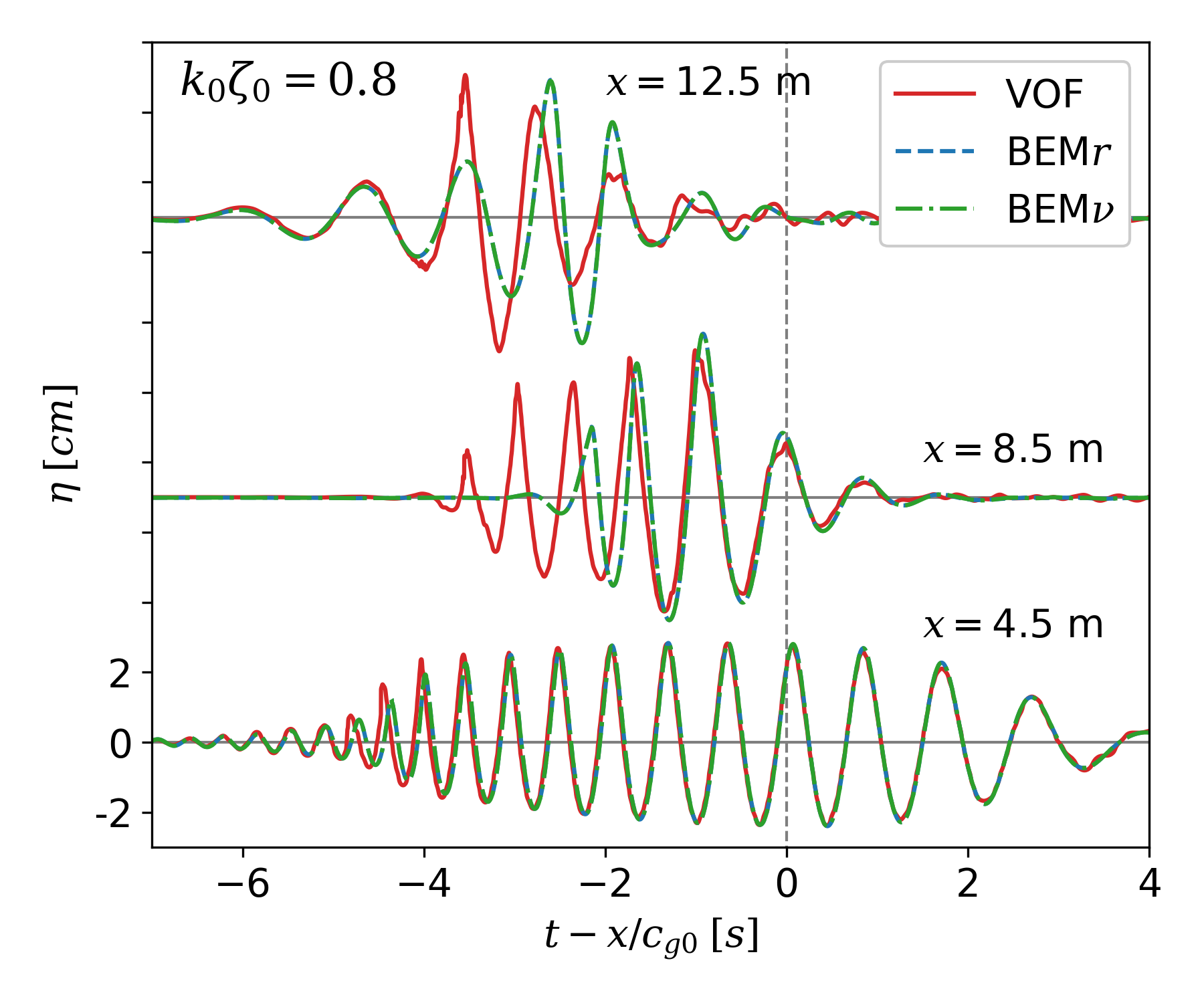}
    \put(2,74){\textbf{(e)}}
\end{overpic}
\begin{overpic}[width=0.49\linewidth]{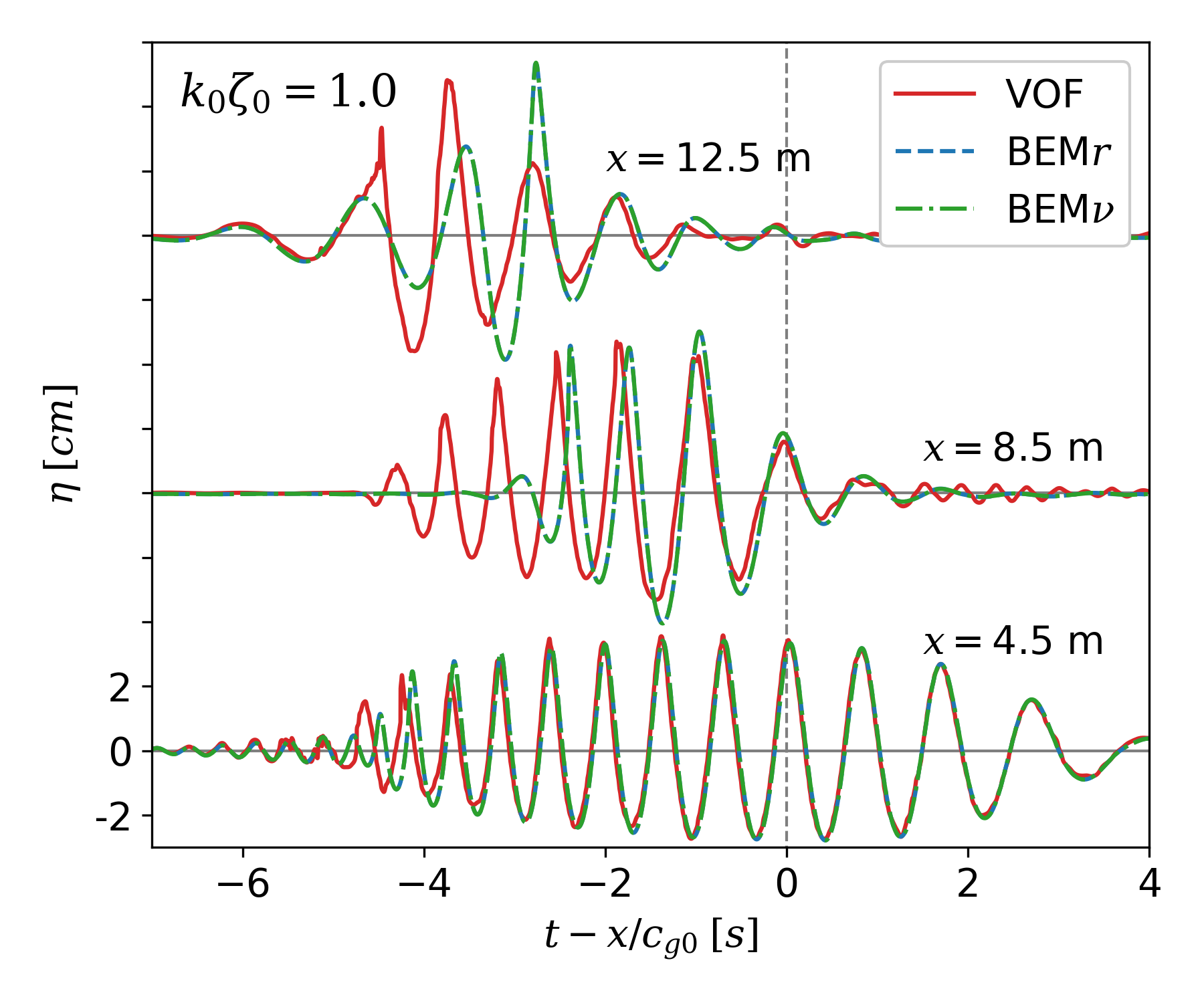}
    \put(2,74){\textbf{(f)}}
\end{overpic}
\caption{Surface elevation variation with time obtained by three numerical
         models: VOF, BEMr and BEM$\nu$. The value of time in the
         horizontal axes was displaced using the group velocity $c_{g0}$
         calculated for the carrier (peak) frequency of the wave train
         spectrum. The linear dispersive focusing is expected at
         $t-x/c_{g0} = 0$.} 
\label{fig:elev}
\end{figure}

Here we investigate the evolution of Gaussian-shaped broad-banded wave trains
(\ref{eq:gauss}) by using the standalone FNP model and the hybrid FNP-NS 
model.  We selected a series of representative wave trains with steepness of
$k_0 \zeta_0=0.2$, $0.3$, $0.4$, $0.6$, $0.8$ and $1$. Computed surface
elevations recorded at $x = 4.5 \; \text{m}$, $x = x_f =8.5 \; \text{m}$ ($x_f$
is the expected focal point), and $x = 12.5 \; \text{m}$ are plotted in
Figure~\ref{fig:elev}.  For wave trains with low steepness $k_0 \zeta_0 \leq
0.4$, there is only one mild breaking event or even no breaking at all. Thus the
eddy viscosity closure was not used for the cases illustrated in the panels (a)-(c).

Figure~\ref{fig:elev}(a) shows perfect coincidence of the results
obtained by BEMr and VOF simulations when no breaking is present.
It confirms the effectiveness of the BEM-VOF coupling used
in the current study. The shape of the
wave train at the focal point $x_f = 8.5 \; \text{m}$ is very close
to the linear prediction shown in Figure~\ref{fig:math}(a).
However, since the wave train is substantially nonlinear, a certain amount of
 asymmetry of the surface elevation before and after the focal point can be noticed.
\replaced{An increase in steepness}{Increase of the wave train steepness}, i.e.
Figures~\ref{fig:elev}(b) and (c), leads to a significant deviation of
either the BEMr solution or the VOF result from the linear estimation at the focal
point, as expected for the steeper nonlinear waves.
For the cases with higher steepness parameter $k_0 \zeta_0 \geq 0.6$ shown
in Figure~\ref{fig:elev} (d)-(f), both the BEMr and BEM$\nu$ models were
used. In compliance with the previous results,
see Section~\ref{sec:valid}, no significant difference was noticed
between the BEMr and BEM$\nu$ results.
\replaced{Therefore}{Nevertheless}, further analysis in the paper will be based on the
BEM$\nu$ model.

For wave trains with strong breaking shown in Figure~\ref{fig:elev} (d)-(f), the
BEM$\nu$ and VOF models produced quite similar results at $x = 4.5 \; \text{m}$.
On the contrary, in the vicinity and beyond the focal point, a considerable
deviation between BEM$\nu$ and VOF results is found and it increases with the
steepness $k_0 \zeta_0$.  This indicates that certain physical processes
associated with the breaking events are not accurately reflected by the
quasi-potential eddy viscosity approximation in BEM$\nu$.  It
\replaced{seems}{is also suspected} that \replaced{this}{the given} phenomenon
is similar to the discrepancy we observed in Section~\ref{sec:valid} for the 
experiments of \citet{Perlin2012} and \citet{Ducrozet2018}.

\begin{figure}
\centering
\begin{overpic}[width=0.8\linewidth]{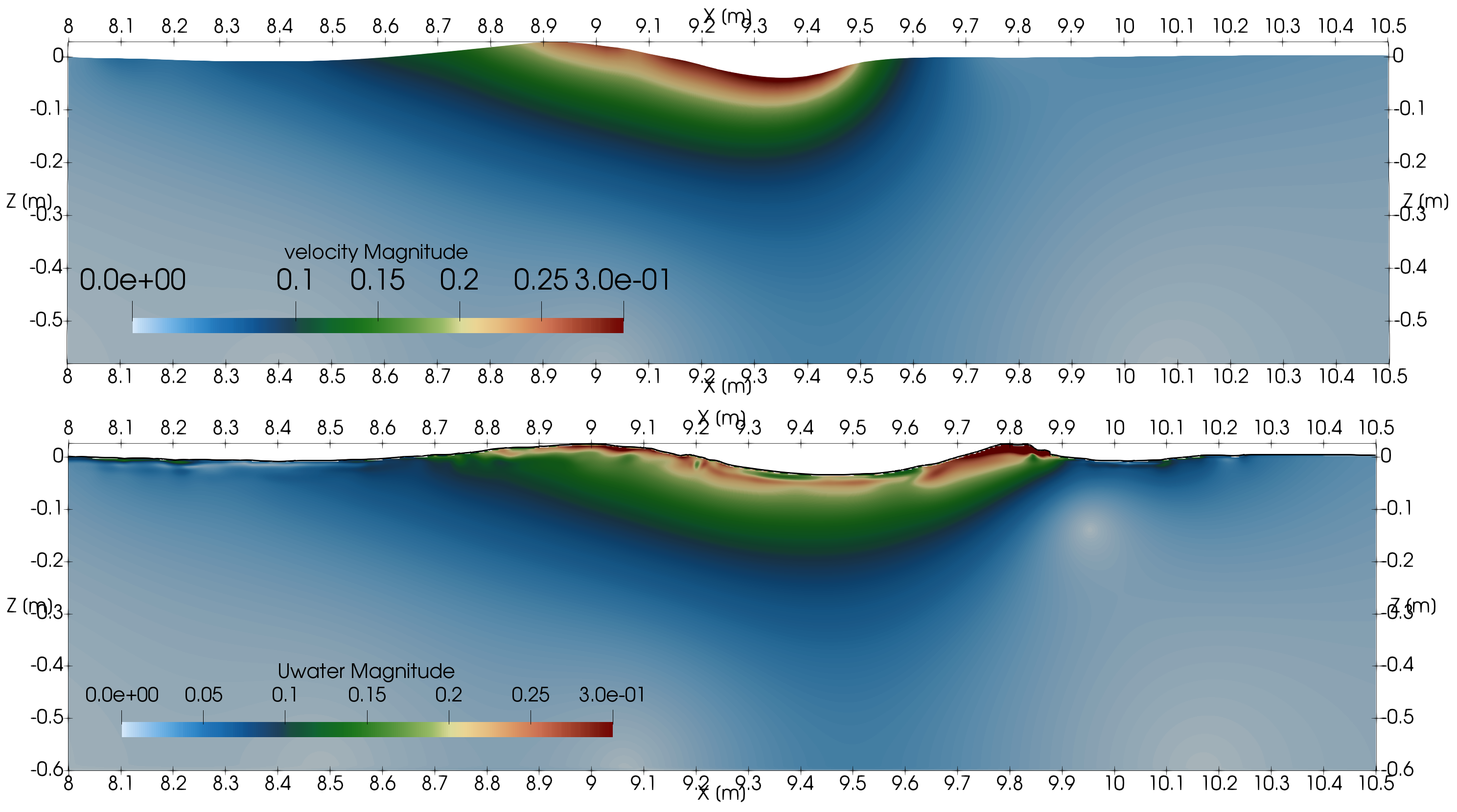}
    \put(-5,50){\color{black}\textbf{(a)}}
    \put(80,35){\color{black}\textbf{BEM$\nu$}}
    \put(78,10){\color{black}\textbf{VOF}}
    \put(75,5){\color{black}$k_0 \zeta_0 = 0.6$}
\end{overpic}\\
\begin{overpic}[width=0.8\linewidth]{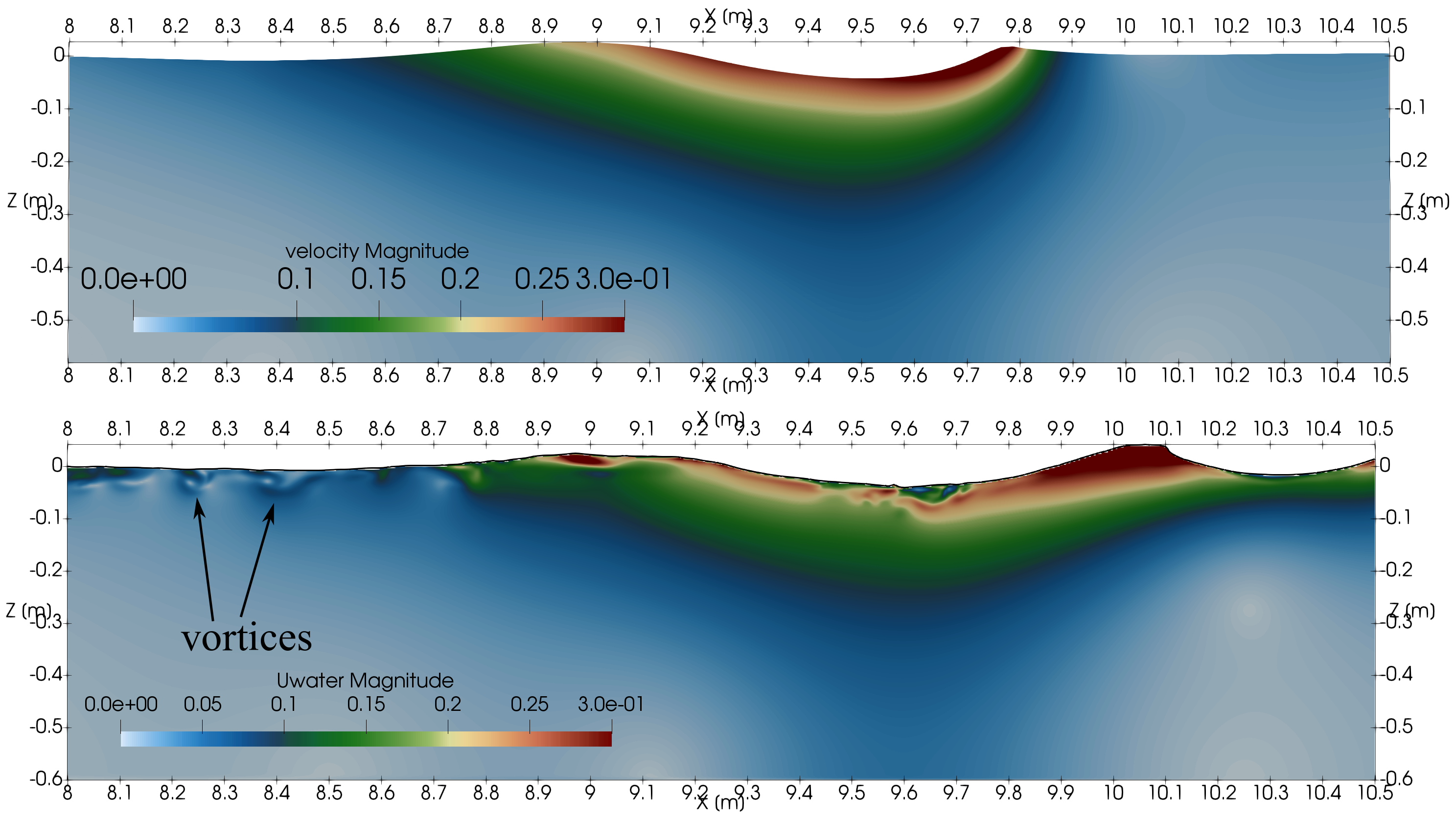}
    \put(-5,50){\color{black}\textbf{(b)}}
    \put(80,35){\color{black}\textbf{BEM$\nu$}}
    \put(78,10){\color{black}\textbf{VOF}}
    \put(75,5){\color{black}$k_0 \zeta_0 = 0.8$}
\end{overpic}\\
\begin{overpic}[width=0.8\linewidth]{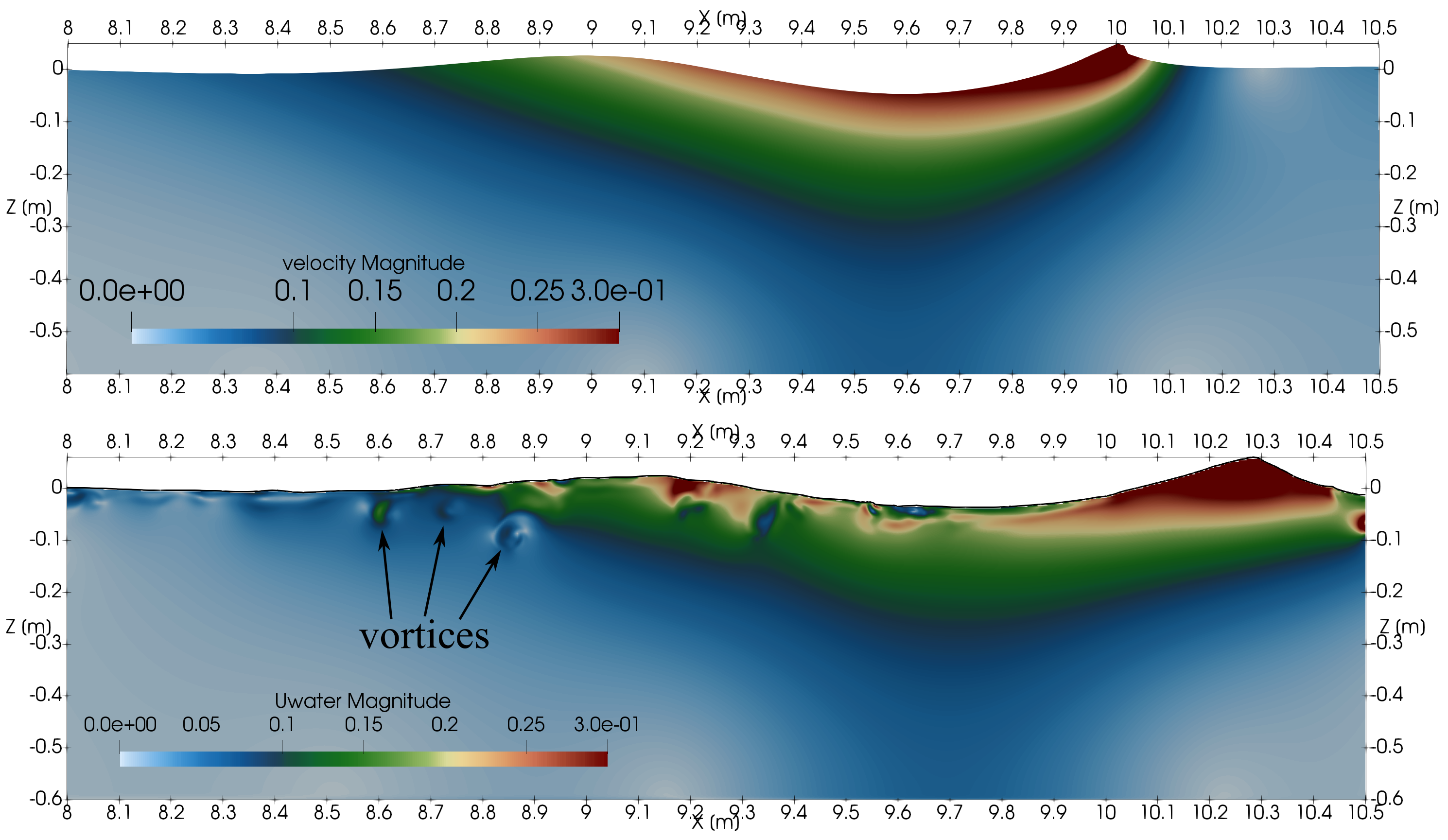}
   \put(-5,50){\color{black}\textbf{(c)}}
    \put(80,35){\color{black}\textbf{BEM$\nu$}}
    \put(78,10){\color{black}\textbf{VOF}}
    \put(75,5){\color{black}$k_0 \zeta_0 = 1.0$}
\end{overpic}\\
\caption{Distribution of the velocity magnitude $|\mathbf{U}|$ beneath
         the free surface obtained in the BEM$\nu$ and VOF models for
         three strongly breaking cases:
         (a) $k_0 \zeta_0 = 0.6$, (b) $k_0 \zeta_0 = 0.8$,
         and (c) $k_0 \zeta_0 = 1.0$. The plots are obtained at
         the instant of focusing $t_f = 35 \; \text{s}$ and in the
         vicinity of the focal point $x_f = 8.5 \; \text{m}$: 
         the horizontal scale is $8 \leq x \leq 10.5 \; \text{m}$.}
\label{fig:veloc}
\end{figure}

Flooding contours of the velocity magnitude underneath the free surface are plotted in
Figure~\ref{fig:veloc} so that we can have a close look at the flow field to
study the difference between the two-phase VOF
solutions and the fully-nonlinear potential BEM$\nu$ results.
The fields are derived at $t = 35 \; \text{s}$ near
$x_f = 8.5 \; \text{m}$ corresponding to the temporal
and spatial location of the expected focal point according to the
linear wave dispersion.
The BEM$\nu$ model shows a quite smooth distribution of the velocity
\replaced{in}{along} the domain, contrary to the VOF solution, which embraces
a certain perturbation component due to the vortical part of the flow.
The plots \replaced{show}{confirm} that the \added{amplitude of the} vortical velocity
$-\mathbf{\nabla} \times \mathbf{\Psi}$ is no longer small as assumed
in the weakly-damped theory \replaced{used by}{of} \citet{Perlin2010, Perlin2012}. \deleted{used in
BEM$\nu$, equations (\ref{eq:freebc1}) and (\ref{eq:freebc2}),}
\deleted{which is considered to be the reason for the inaccuracy of the eddy
viscosity wave breaking approximation.}
\added{Consequently, the applicability of the eddy viscosity model for these problems is in question.}

It is expected that the vortical flow consists of the sheared currents
and other local and distributed non-potential fluid motions
\citep{Iafrati2013, Melville2017, Melville2019}.
For instance, several vortical structures are clearly \replaced{shown}{distinguished}
in Figures~\ref{fig:veloc}~(b)~and~(c).
A more detailed investigation of the non-potential flows
supposes the need of quantitative comparison of the BEM$\nu$ and VOF
velocity fields. But a meaningful comparison is practically impossible for the
considered cases because the shapes of free surface obtained by these models
are very different.

\subsection{\label{sec:dissip} Energy dissipation due to wave breaking}

\begin{figure}
\centering
\begin{overpic}[width=0.32\linewidth]{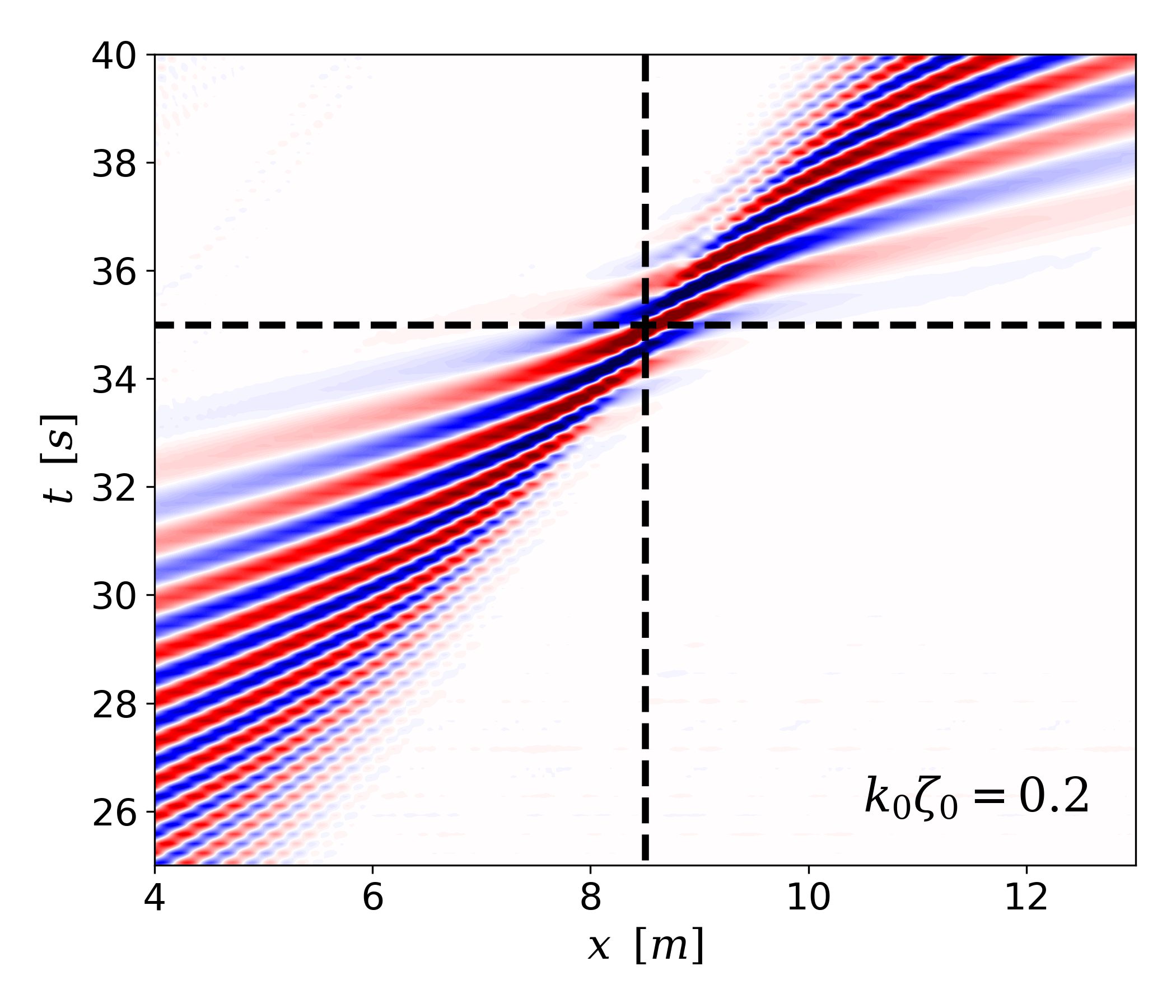}
    \put(15,72){\textbf{(a)}}
\end{overpic}
\begin{overpic}[width=0.32\linewidth]{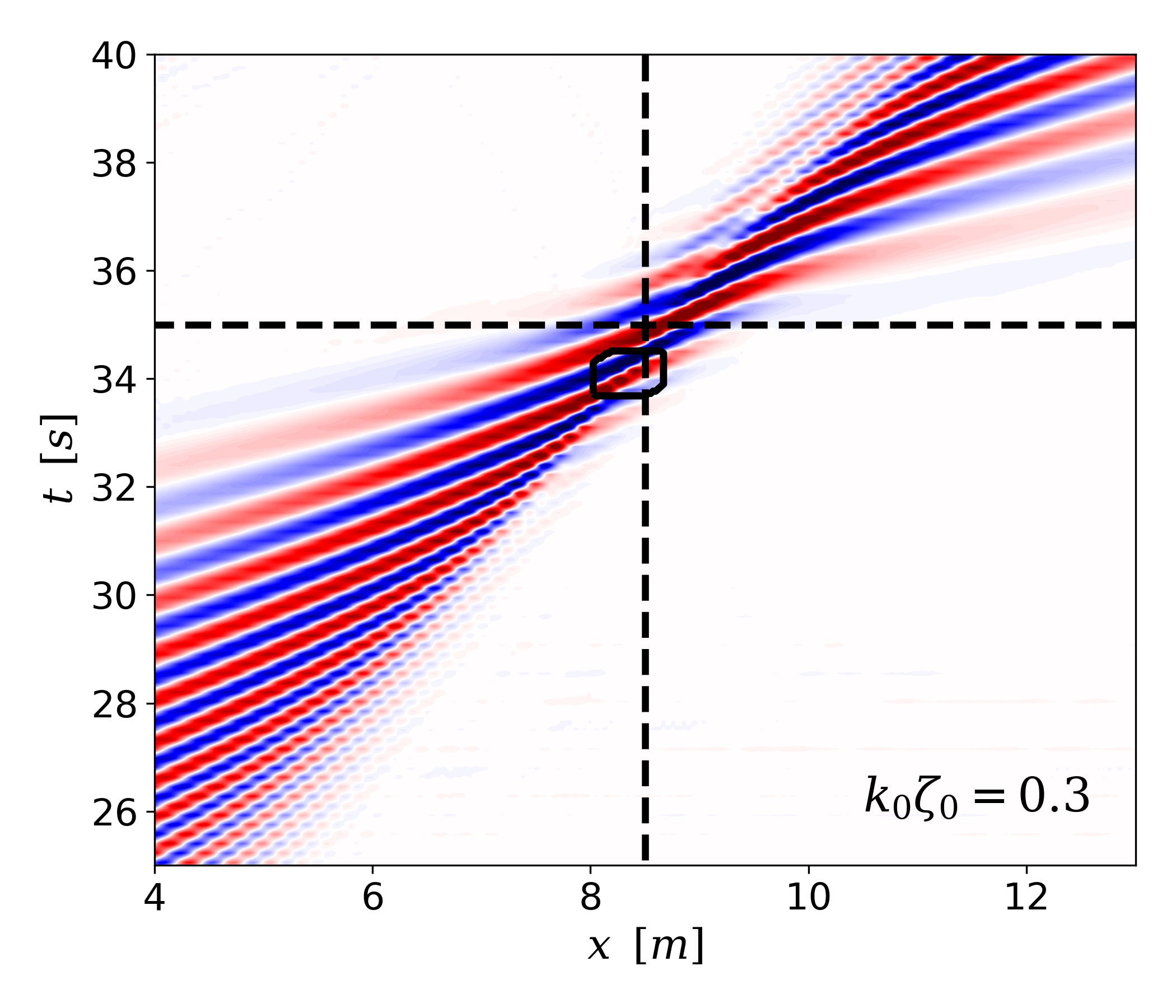}
    \put(15,72){\textbf{(b)}}
\end{overpic}
\begin{overpic}[width=0.32\linewidth]{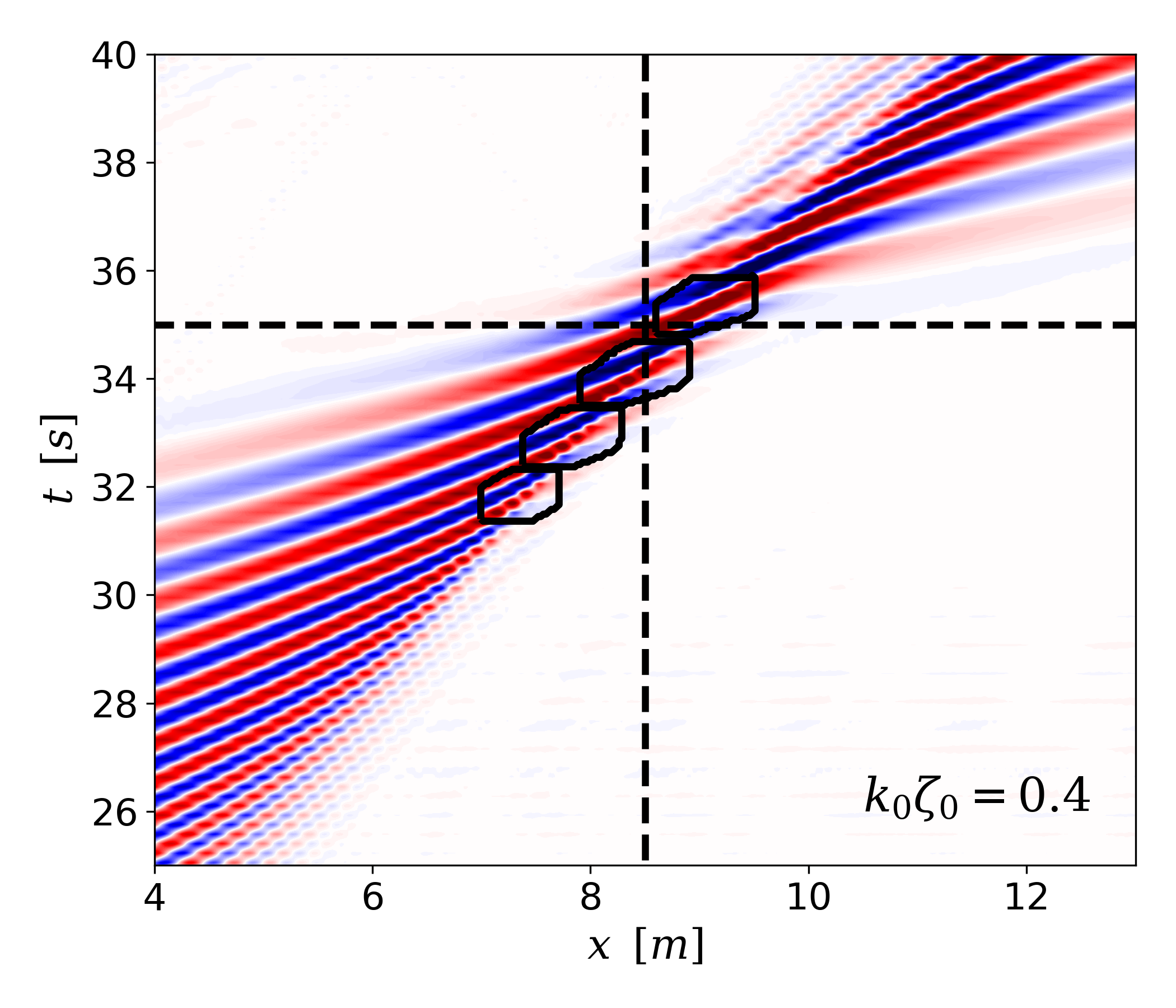}
    \put(15,72){\textbf{(c)}}
\end{overpic}\\
\begin{overpic}[width=0.32\linewidth]{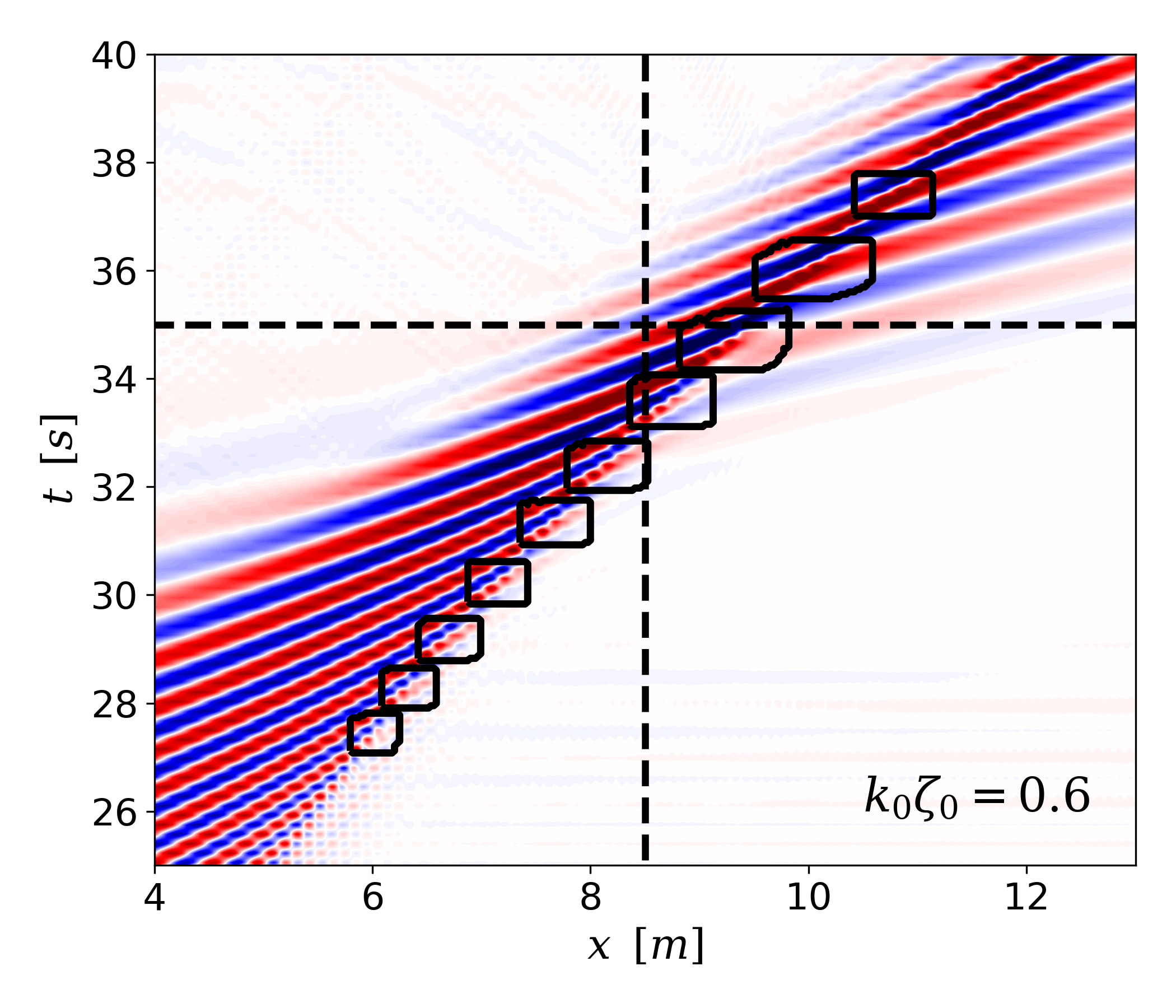}
    \put(15,72){\textbf{(d)}}
\end{overpic}
\begin{overpic}[width=0.32\linewidth]{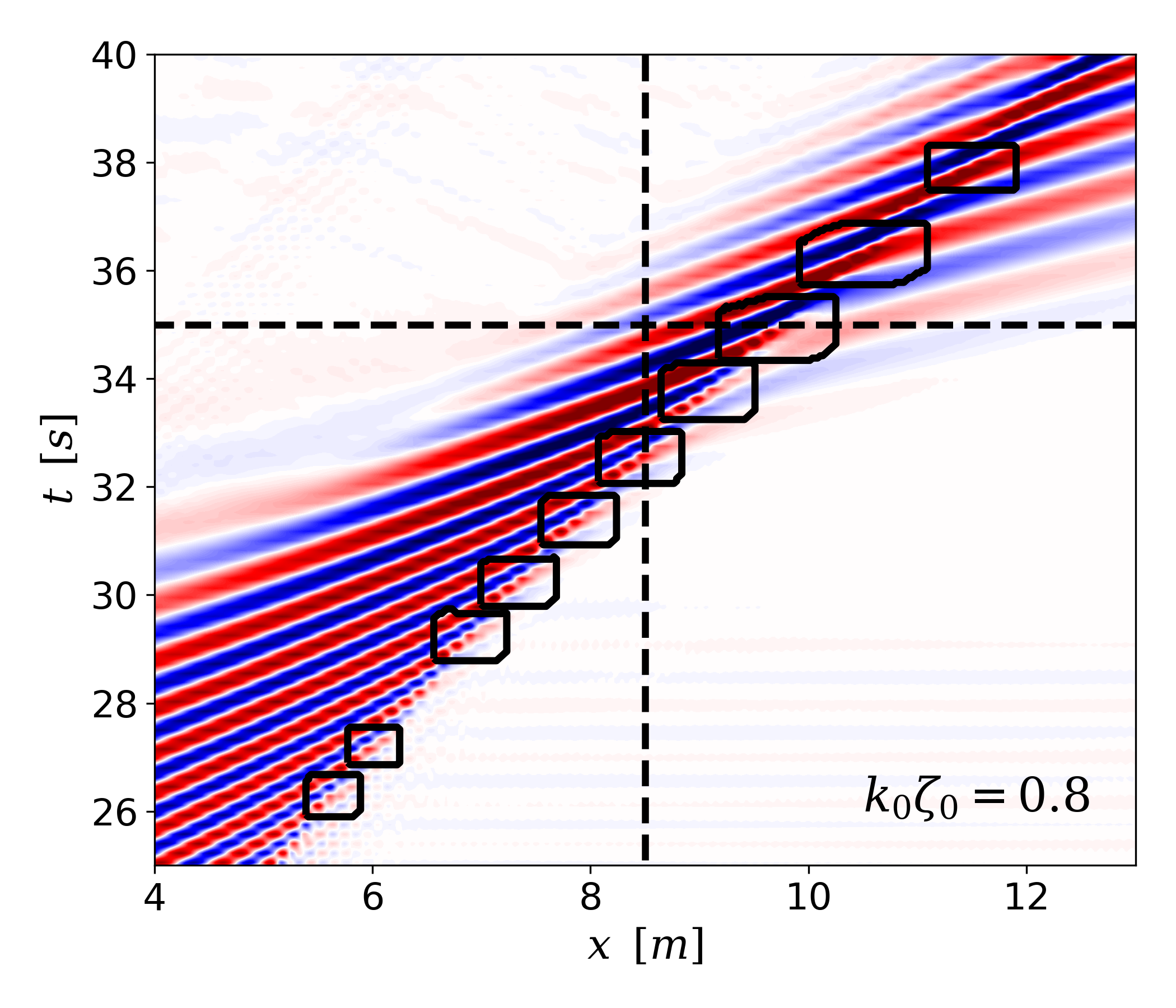}
    \put(15,72){\textbf{(e)}}
\end{overpic}
\begin{overpic}[width=0.32\linewidth]{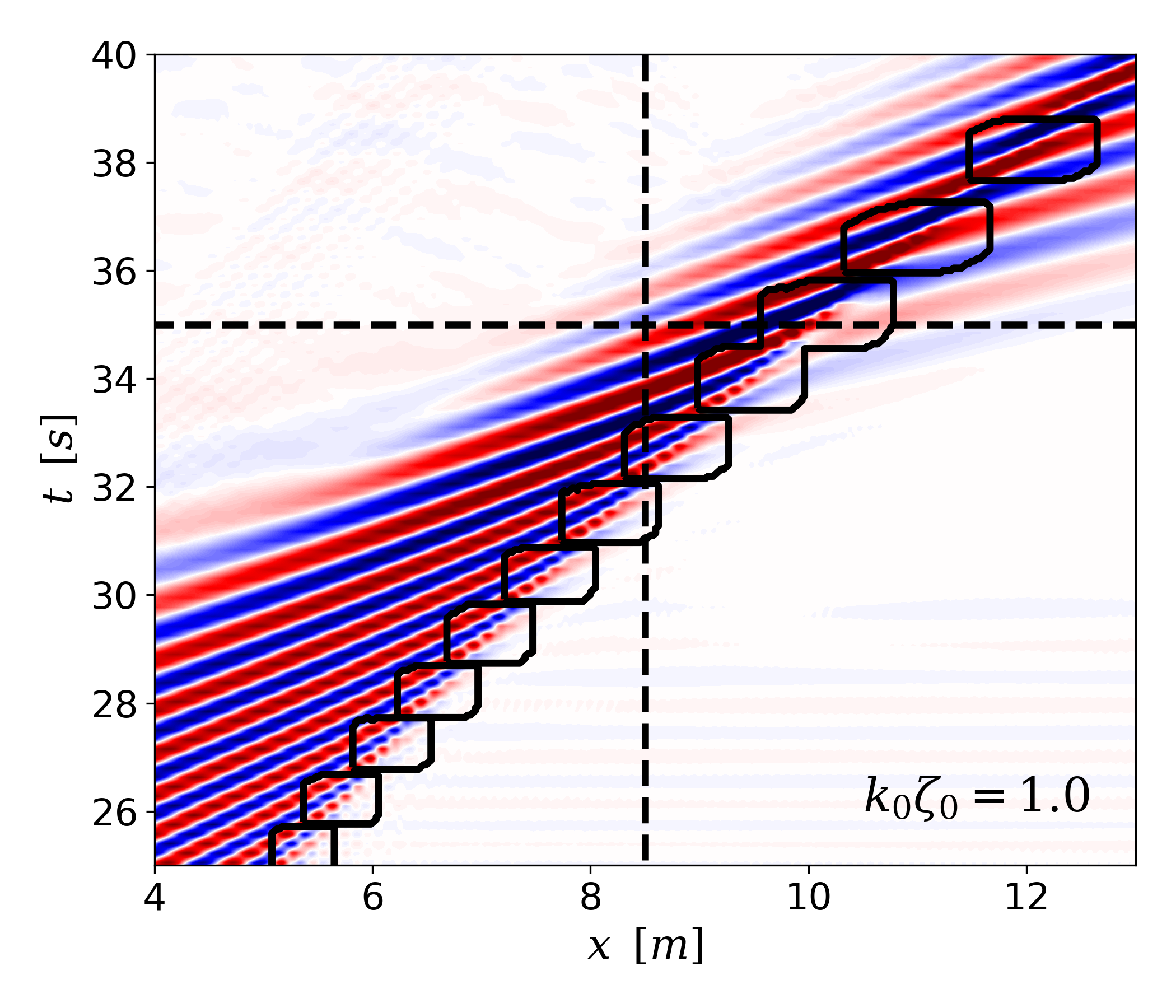}
    \put(15,72){\textbf{(f)}}
\end{overpic}
\caption{Energy dissipation regions predicted by the eddy viscosity
         approximation. Panels (a)-(f) present the wave trains having
         steepness parameter $k_0 \zeta_0 = 0.2, \; 0.3, \; 0.4, \; 0.6,
         \; 0.8, \;\text{and}\; 1.0$, respectively. The breaking regions
         are depicted by black solid lines. Color shows the
         spatio-temporal variation of the surface elevation $\eta(x,t)$
         calculated by the BEM$\nu$ model. Dashed lines show the
         location of the focal point expected according to the linear
         dispersion.}
\label{fig:bemnu}
\end{figure}

The spatio-temporal evolution of the wave train \replaced{calculated}{determined} by the
BEM$\nu$ model is \replaced{illustrated}{studied} in Figure~\ref{fig:bemnu} in terms of 
surface elevation\deleted{variation}. \replaced{Wave breaking regions are highlighted in the figure as rectangular areas enclosed by black solid lines. The dimensions of these regions $L_b$ in space and $T_b$ in time were determined by the eddy viscosity closure \eqref{eq:HLT}.}{The regions of the wave breaking having
size $L_b$ and $T_b$ provided by the eddy viscosity closure
(\ref{eq:HLT}) have the rectangular shape as shown by solid lines.}
For each region the constant value of the eddy viscosity
$\nu_{\text{eddy}}$ was calculated by using the formula~(\ref{eq:eddy}).
As expected, the non-breaking case, i.e. Figure~\ref{fig:bemnu}(a),
does not have any predicted breaking locations.
We also note that \replaced{increasing}{the increase of} the steepness \deleted{parameter} $k_0 \zeta_0$
\replaced{can cause}{leads  to appearance the} multiple breaking events instead of a
single very strong one. This is because the wave train loses its
stability much ahead of \deleted{reaching} the focal point.
\replaced{At}{In} the same time, \deleted{it is generally expected that} \replaced{larger}{higher} values
of $\nu_{\text{eddy}}$ correspond to \replaced{more}{the stronger} energy dissipation.
\deleted{during the given breaking event}

\replaced{The energy}{Energy} dissipation locations shown in Figure~\ref{fig:bemnu} do not
overlap \added{with} each other. \replaced{This}{It} means that in the studied wave trains, waves
always break at different locations. 
Before the focal point, breaking always appears at the leading
edge of the wave train because of the presence of short steep waves.
On the contrary, after the focal point breaking locations move
closer to the centre of wave train.
Taking into account the fact that the leading edge of the wave train after the
focal point consists of long waves, it can be assumed that those
waves are more stable.
This observation is usually involved in the spectral models of water
waves, where the energy dissipation mostly at
high frequencies is incorporated
\citep{Babanin2011, Shrira2018}.

Within the eddy viscosity quasi-potential approach,
the energy dissipation process can be seen as a
transformation of the nonlinear surface wave energy into the
rotational fluid flow energy.
Since the  considered broad-banded wave train occupies a restricted
space, it is convenient to estimate the strength of the wave breaking
by tracing the amount of energy transferred by the wave train through
different cross sections, i.e. integral energy flux
\citep{Banner2007, Melville2008, Perlin2008, Kirby2016}.
Taking into account that \deleted{the} wave breaking is \added{a} strongly \replaced[id=ZM]{localised}{localized}
phenomenon, the energy loss is associated with a particular location
and appears as a reduction \replaced{of}{in} the integral energy flux \replaced{across}{before and after}
the breaking location.

The total integral energy flux \replaced{passing}{passed} through the cross section \deleted{having
coordinate} $x$ in \replaced{the VOF}{FOAM} model is \added{given by}:
\begin{eqnarray}
   && F_{VOF}^{NL}(x) = \int_{-\infty}^{+\infty} dt
                         \int_{-h}^{+h} \left\lbrace
               \frac{1}{2} \rho_w \left| \mathbf{U} \right|^2
             + \rho_w g z + p
   \right\rbrace u \beta   dz,
   \label{eq:fluxFOAM}
\end{eqnarray}
where $\mathbf{U} = (u, w)$. This expression does not involve any
physical simplification and thus determines the fully-nonlinear (NL)
value of the energy flux.  \replaced{In the VOF model, the volume fraction $\beta$ is used to determine the percentage of water
contained in a mesh cell.}{Utilization of the VoF method in the FOAM
model determines the fluid mixture contained in each cell of the domain
by the volumetric fraction of water $\alpha$.}
Thus the height of the water layer in each cell is \replaced{$\beta dz$}{$\alpha dz$}.
The so-called dry and wet cells are distinguished by \replaced{$\beta=0$}{$\alpha = 0$}
and \replaced{$\beta = 1$}{$\alpha = 1$}, respectively.
Determination of \added{the} nonlinear energy flux \replaced{in the}{from the results of} BEM$\nu$
\replaced{model}{computations} is \added{more} complicated \replaced{because the needed}{by unavailability of the value of} pressure
$p$ \added{is not readily available} in the solution.
Simplification of (\ref{eq:fluxFOAM}) \replaced{can be achieved}{is possible} \replaced{by using}{involving} the 
Bernoulli equation (\ref{eq:freebc2}) and taking
$\mathbf{U} = -\mathbf{\nabla} \varphi$:
\begin{eqnarray}
   && F_{BEM}^{NL}(x) = - \rho_w \int_{-\infty}^{+\infty} dt
                                 \int_{-h}^{\eta(x,t)} 
                    \frac{\partial \varphi}{\partial t} 
                    \frac{\partial \varphi}{\partial x} dz
   \label{eq:fluxBEM}
\end{eqnarray}

\replaced{In laboratory it is not feasible yet to measure the nonlinear energy flux.}
{In practice, experimental measurement of the nonlinear energy flux
even in the laboratory conditions is not feasible.} \replaced{Instead}{Thus}, a linearisation
of (\ref{eq:fluxBEM}) is usually applied 
\citep{Banner2007, Melville2008, Perlin2008, Kirby2016, Perlin2010,
Perlin2012, Ducrozet2018}.
\replaced{Researchers usually assume}{Assume} the equipartition of total energy between the
kinetic and potential parts, which is admissible for linear wave system.
The linear approximation for the total energy density is
$E = \rho_w g \overline{\eta^2}$, where \added{the} overbar represents averaging
over the local wavelength. The linear approximation for the energy
flux is then \added{given by} \citep{D&D1991, Melville2008, Kirby2016}:
\begin{eqnarray}
   && F^{L}(x) = \int_{-\infty}^{+\infty} E c_{gs} dt
               = \int_{-\infty}^{+\infty} \rho_w g c_{gs} \eta^2 dt
   \label{eq:fluxLIN}
\end{eqnarray}
Here $c_{gs}$ is the spectral-weighted group speed approximating
the velocity of energy \replaced{transfered}{transfer} by the given wave train:
\begin{eqnarray}
   && c_{gs} = \frac{\sum_j c_{g,j} a_j^2}{\sum_j a_j^2},
   \label{eq:cgs}
\end{eqnarray}
where $a_j$ and $c_{g,j}$ are the amplitude and the group velocity
of the $j^{\text{th}}$ component of the wave packet spectrum.

\begin{figure}
    \centering
    \includegraphics[width=0.6\textwidth]{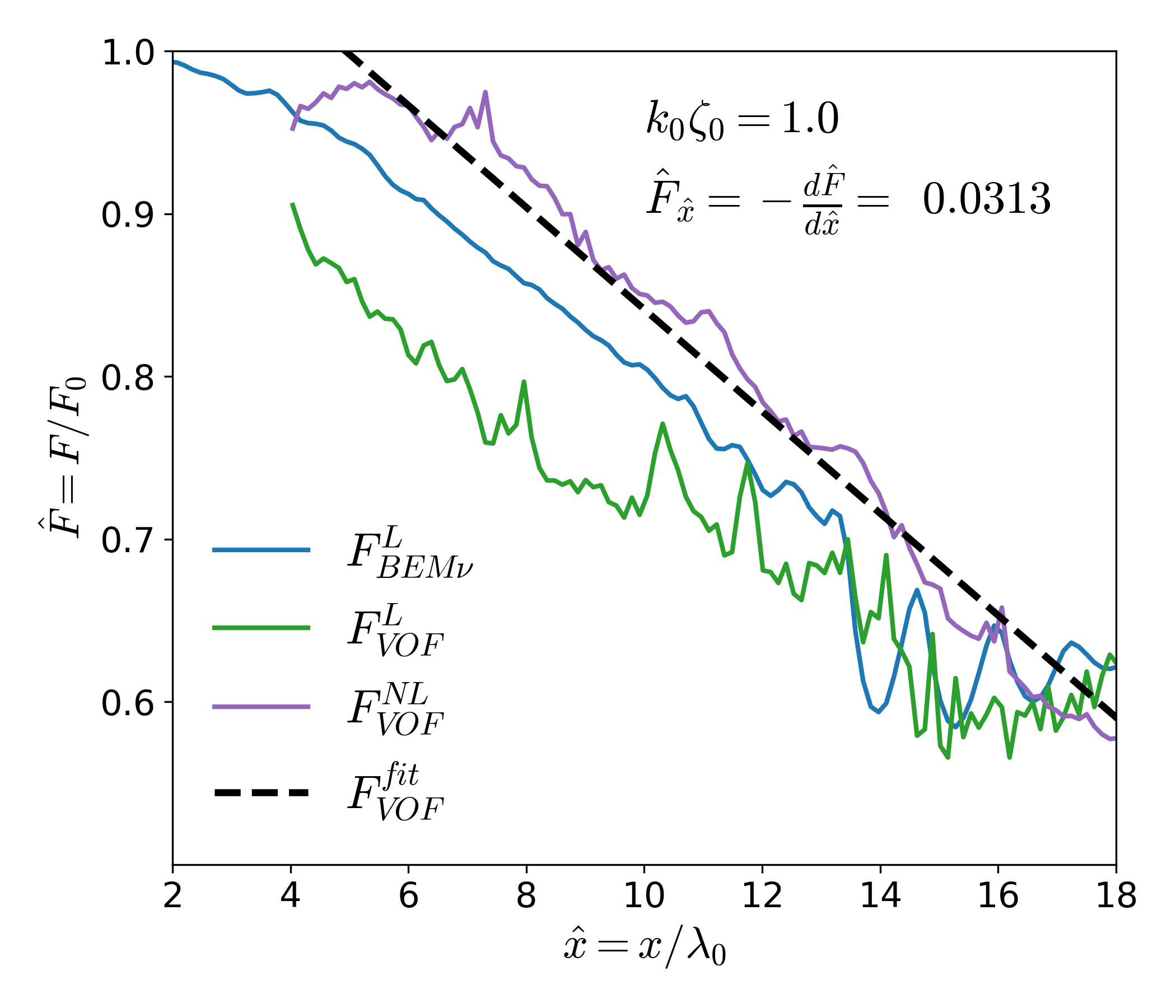}
    \caption{Distribution of the dimensionless integral energy flux 
             $\hat{F} = F / F_0$ along the computational domain
             $\hat{x} = x / \lambda_0$ for the wave train with the
             steepness parameter $k_0 \zeta_0 = 1.0$.}
    \label{fig:flux1}
\end{figure}

The outcomes of (\ref{eq:fluxFOAM}) and (\ref{eq:fluxLIN})
are compared in Figure~\ref{fig:flux1} for the steepest
wave train with $k_0 \zeta_0 = 1.0$.
The dimensionless variables are introduced: energy flux
$\hat{F} = F / F_0$ and spatial coordinate  $\hat{x} = x / \lambda_0$;
$F_0$ is the initial energy flux computed at $x \approx 0$.
Figure~\ref{fig:flux1} shows that the distributions of linear energy
flux computed \replaced{by}{based on} the BEM$\nu$ and VOF \replaced{models}{results}, i.e.
$F_{\text{BEM}\nu}^L$ and $F_{\text{VOF}}^L$, somewhat deviate from
each other. The nonlinear energy flux $F_{\text{VOF}}^{NL}$ \replaced{obtained by}{computed
based on} the VOF model provides the highest values of $\hat{F}$.
In spite of slightly different trajectories,
the values of $F_{\text{BEM}\nu}^L$, $F_{\text{VOF}}^L$ and
$F_{\text{VOF}}^{NL}$ at the outlet \replaced{of}{from} the domain, i.e. behind
the active breaking region, are \replaced{quite close to each other}{practically identical}.
This suggests that the eddy viscosity approximation \deleted{of the wave breaking}
used in the BEM$\nu$ model predicts quite accurately the energy dissipation
process, in agreement with earlier investigations
\citep{Perlin2010, Perlin2012, Ducrozet2018, Sriram2019}. 
\replaced{It is interesting to note that the discrepancy in surface elevation calculation (see Figure~\ref{fig:elev}) doesn't significantly affect the amount of wave energy dissipated by breaking.}{Inaccuracies
in the surface elevation plots in figure~\ref{fig:elev} are not linked
to the amount of energy dissipated by the breaking.}

\begin{figure}
    \centering
    \includegraphics[width=0.6\textwidth]{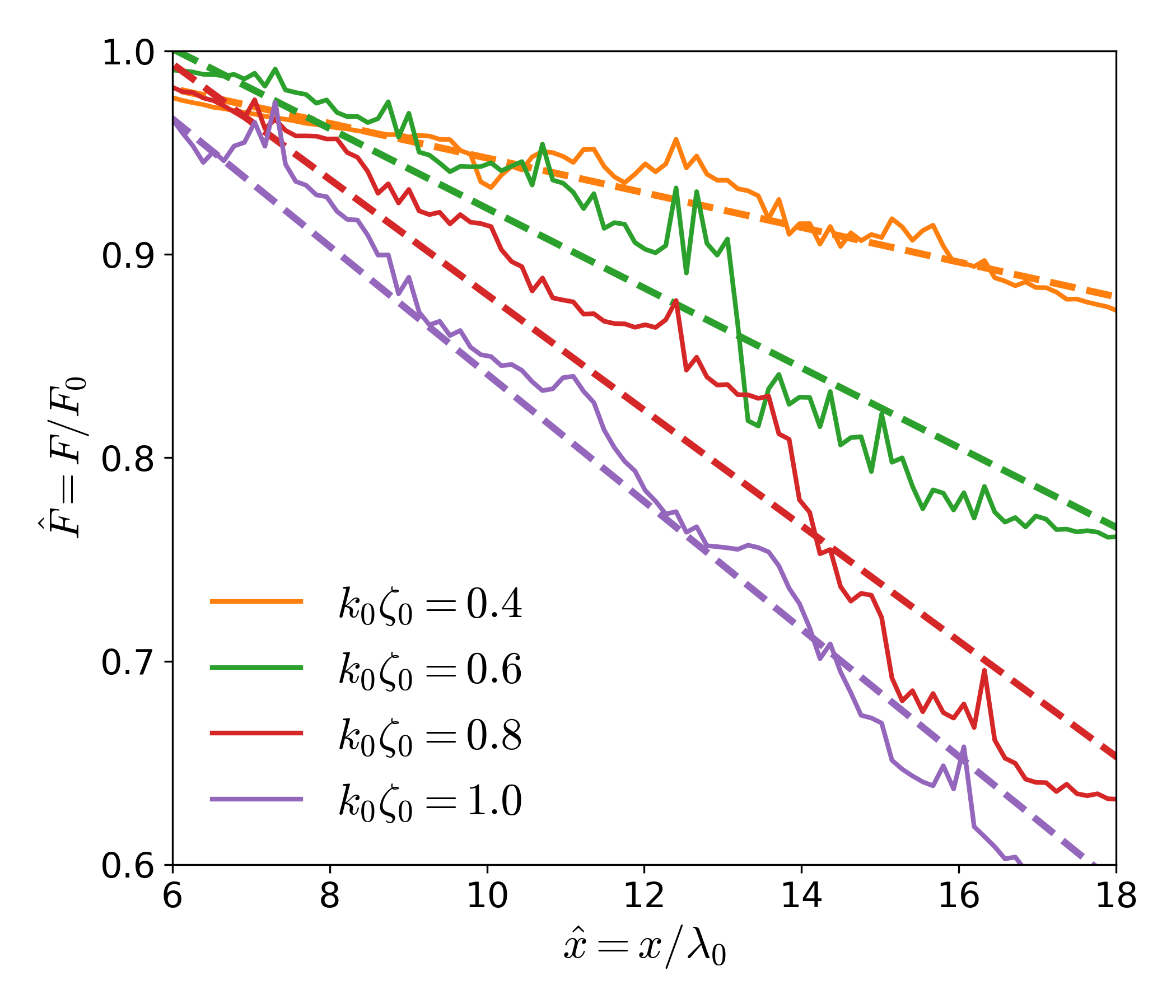}
    \caption{Distribution of the nonlinear energy flux
             $F_{\text{VOF}}^{NL}$ obtained by the
             VOF model. The solid curves correspond to energy
             fluxes produced by the wave trains of different steepness
             $k_0 \zeta_0$. The dashed lines show the best fit of the
             data needed to compute the energy dissipation rate
             $\hat{F}_{\hat{x}}$.}
    \label{fig:flux2}
\end{figure}

\replaced{In the present work, we use the mean gradient of nonlinear energy flux $\hat{F}_{\hat{x}} = - d \hat{F} / d \hat{x}$ to characterise the strength of breaking. Such a quantify can be interpreted
	as the wave energy dissipation rate in space. For the VOF model, we produced the best fit of the energy flux $F_{\text{VOF}}^{fit}$ and illustrated it in Figure \ref{fig:flux1} as the black dashed line, which has a gradient of 0.0313 in magnitude.}{
We assume that the strength of the wave breaking is characterized
by the mean gradient of the nonlinear energy flux that may be interpreted
as the energy dissipation rate and calculated as
$\hat{F}_{\hat{x}} = - d \hat{F} / d \hat{x}$ using the linear fit
$F_{\text{VOF}}^{fit}$ shown in figure~\ref{fig:flux1}.}
\replaced{The distributions}{Distributions} of $F_{\text{VOF}}^{NL}$ and $F_{\text{VOF}}^{fit}$
for wave trains \replaced{with}{having} different
steepness $k_0 \zeta_0$ are plotted in Figure~\ref{fig:flux2}. The calculated values of $\hat{F}_{\hat{x}}$
are \replaced[id=ZM]{summarised}{summarized} in Table~\ref{tab:flux}.
It \replaced{can be}{is} seen from \deleted{both} Figure~\ref{fig:flux2} and
Table~\ref{tab:flux} that \replaced{the energy dissipation rate $\hat{F}_{\hat{x}}$ increases with the wave train steepness $k_0 \zeta_0$}{the increase of $k_0 \zeta_0$ leads to
corresponding increase in the energy dissipation rate $\hat{F}_{\hat{x}}$
confirming the applicability of the introduced indicator}.	

\begin{table}
\centering
\caption{\label{tab:flux} The values of the energy dissipation rate
         computed for the wave trains of different steepness.}
\begin{tabular}{ l c }
 \hline
 $k_0 \zeta_0$ \;\;\;\;\; & $\hat{F}_{\hat{x}} = - d \hat{F} / d \hat{x}$ \\
 \hline 
 0.3 & 0.00695 \\
 0.4 & 0.00852 \\
 0.6 & 0.0196 \\
 0.8 & 0.0284 \\
 1.0 & 0.0313 \\
 \hline
\end{tabular}
\end{table}

\subsection{\label{sec:shift} Shift of phase due to wave breaking}

\begin{figure}
\centering
\begin{overpic}[width=0.48\linewidth]{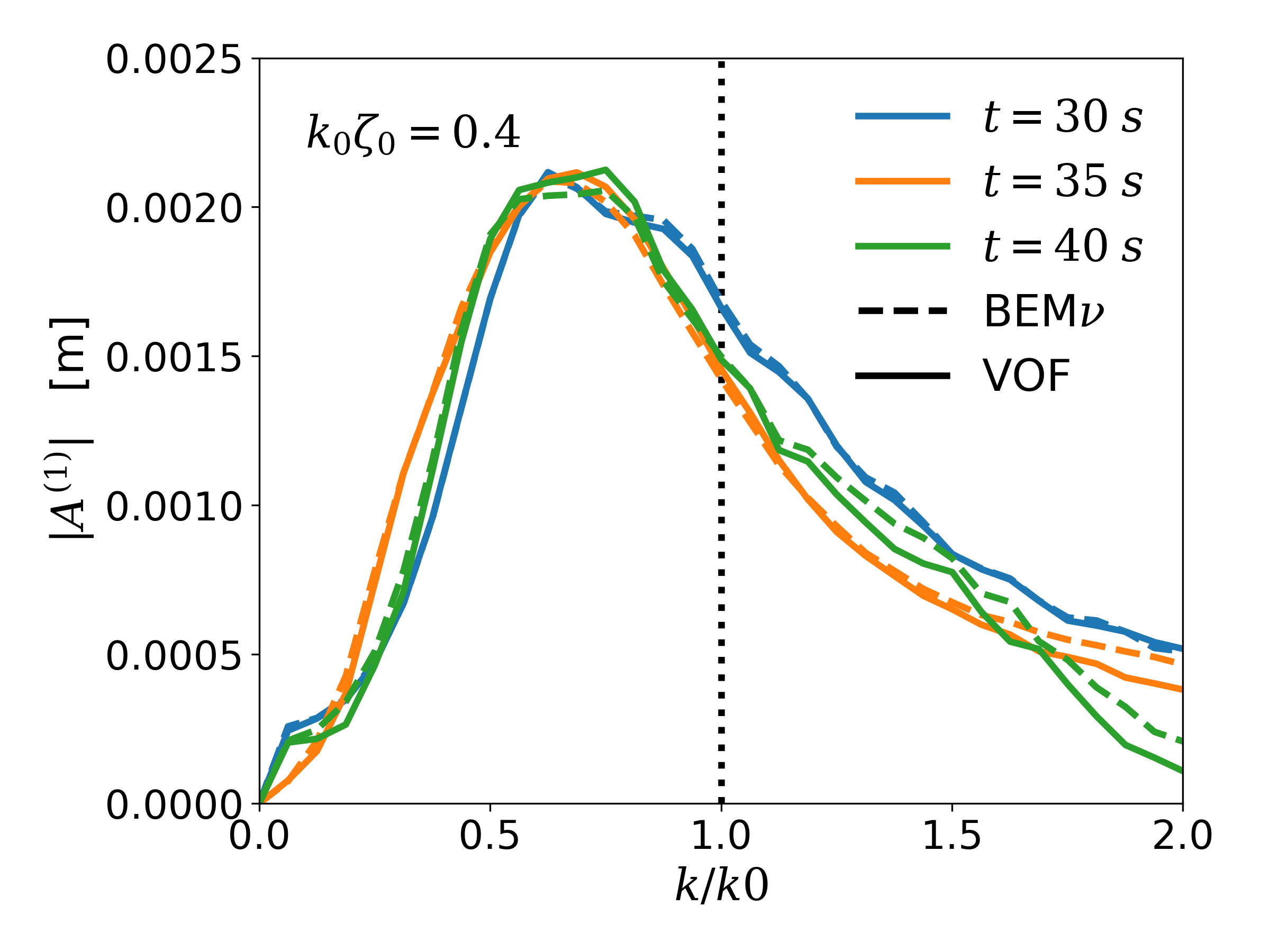}
    \put(0,65){\textbf{(a)}}
\end{overpic}
\begin{overpic}[width=0.48\linewidth]{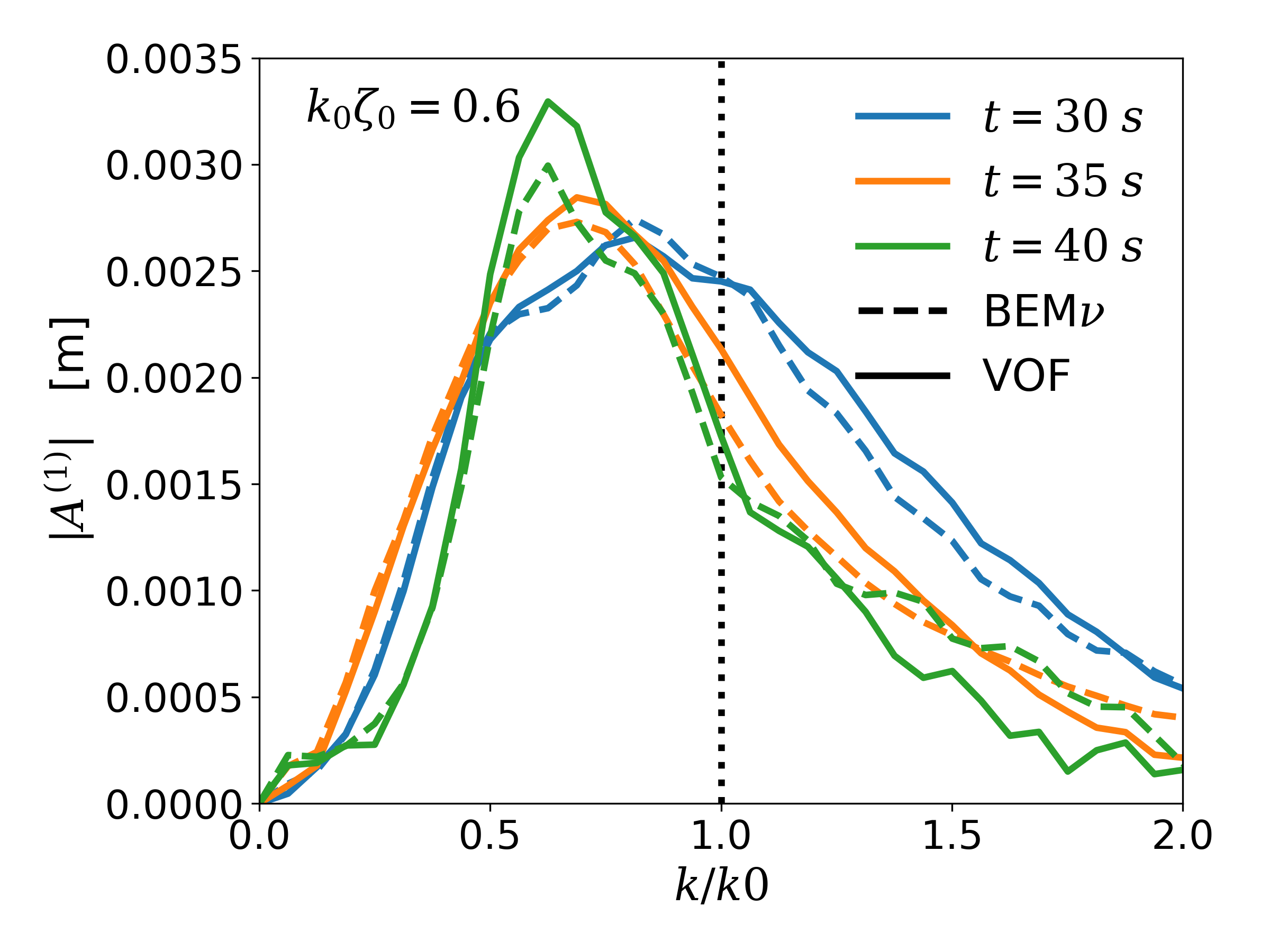}
    \put(0,65){\textbf{(b)}}
\end{overpic}\\
\begin{overpic}[width=0.48\linewidth]{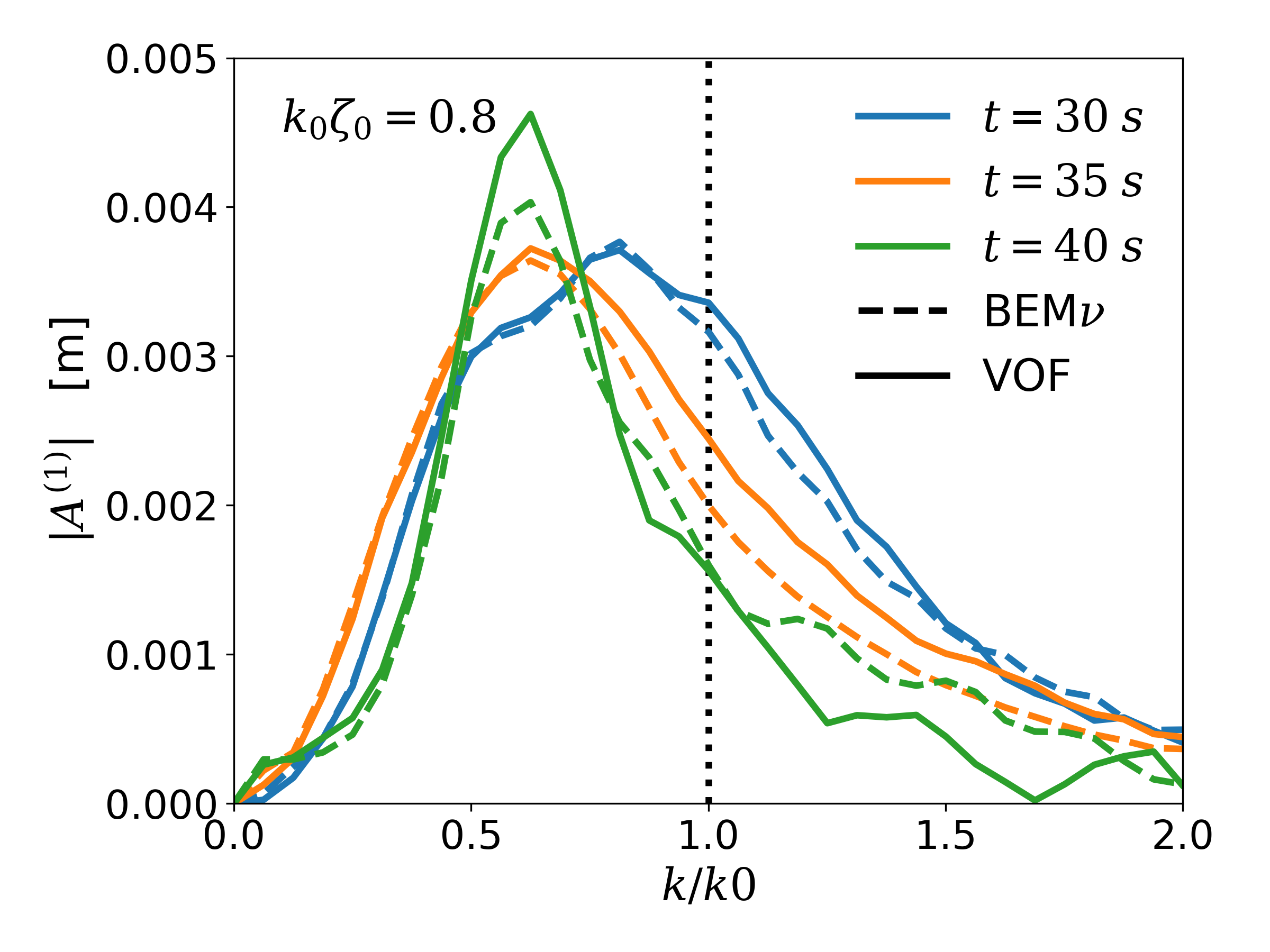}
    \put(0,65){\textbf{(c)}}
\end{overpic}
\begin{overpic}[width=0.48\linewidth]{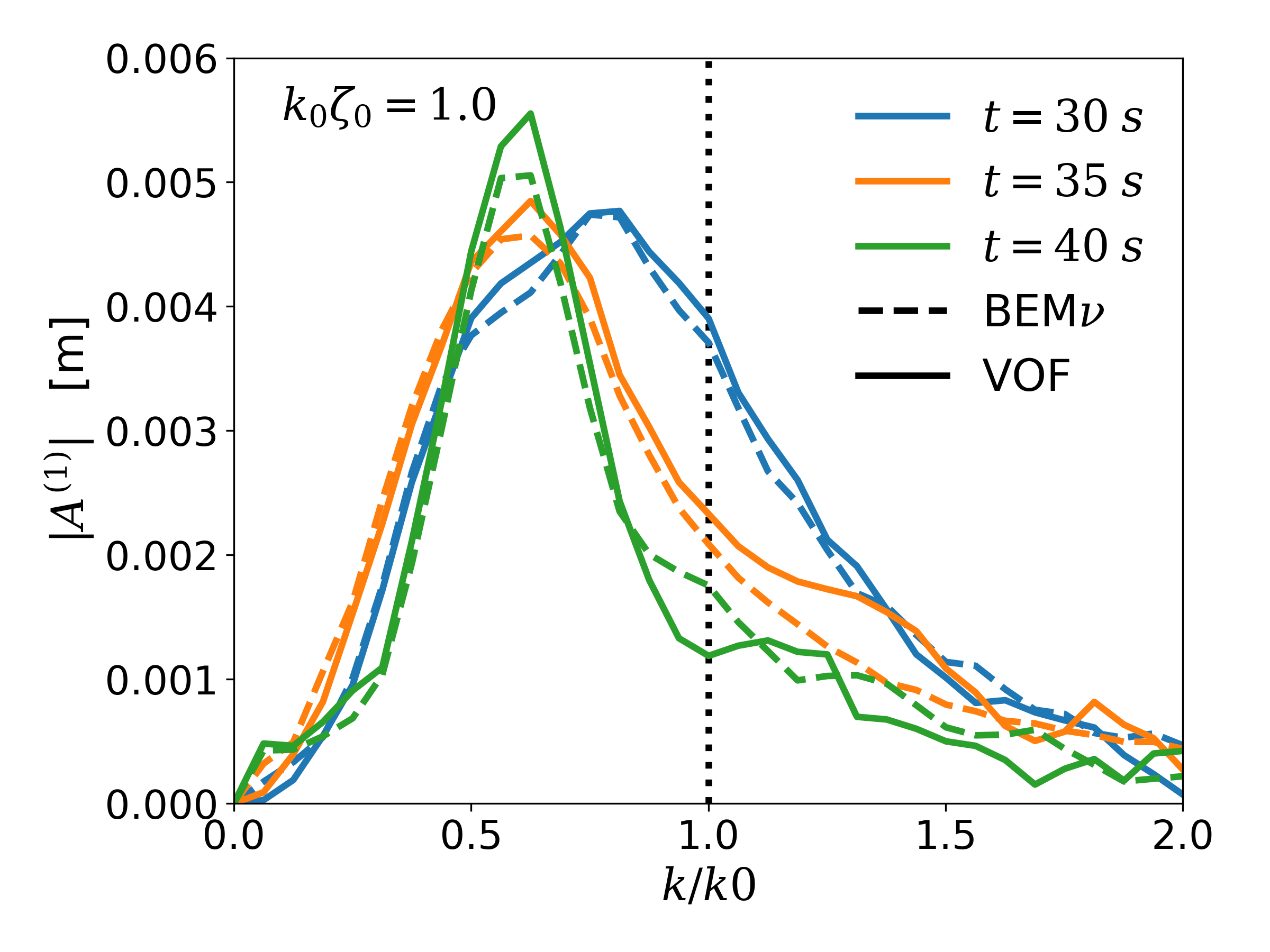}
    \put(0,65){\textbf{(d)}}
\end{overpic}
\caption{Free wave spectra obtained in VOF (solid lines) and BEM$\nu$
         (dashed lines) models.
         Panels (a)-(d) present the wave trains of different steepness parameter
         $k_0 \zeta_0$. Plots of different colour are obtained at various times
         of simulations.}
\label{fig:amplmagn}
\end{figure}

\begin{figure}
\centering
\begin{overpic}[height=0.4\linewidth]{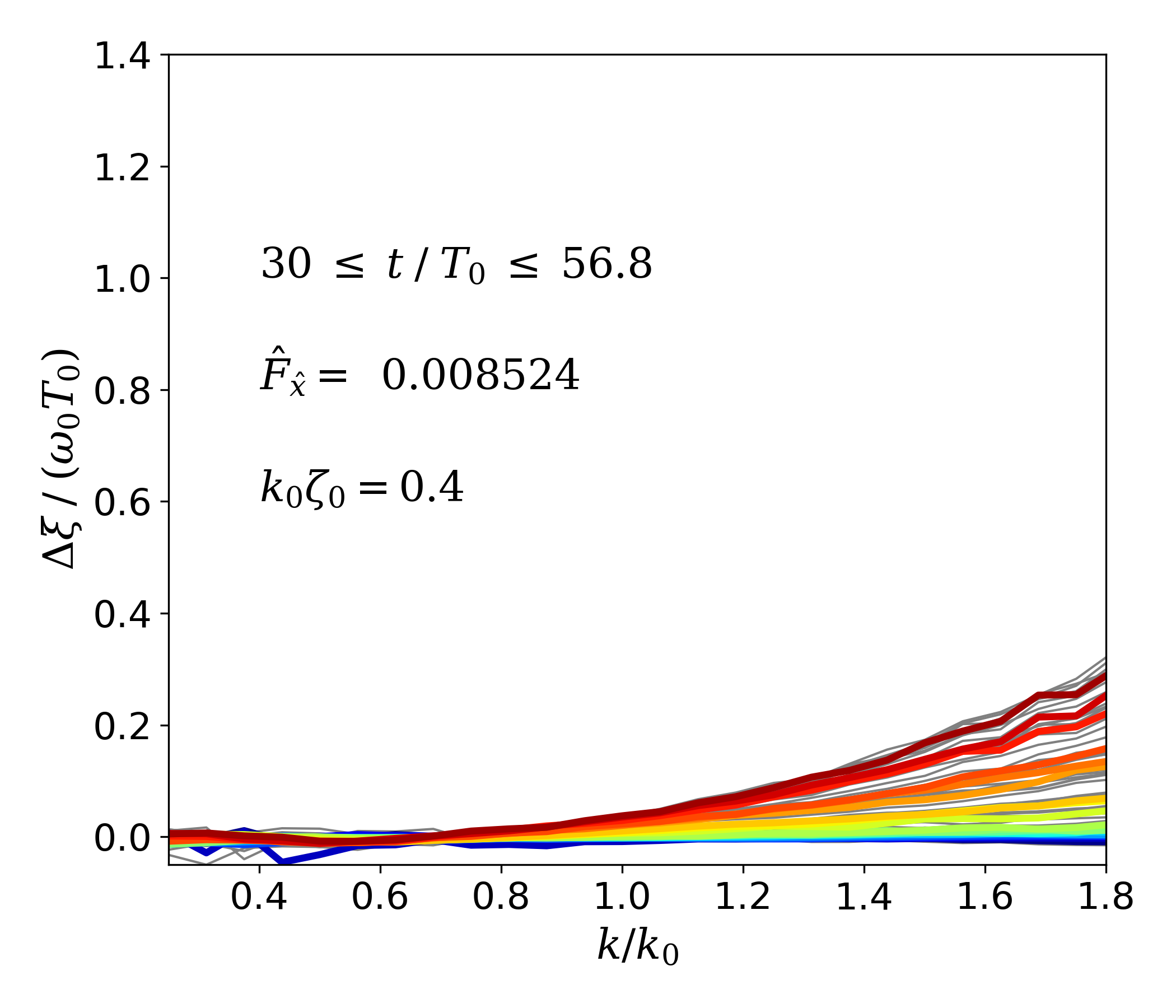}
    \put(18,73){\textbf{(a)}}
\end{overpic}
\begin{overpic}[height=0.4\linewidth]{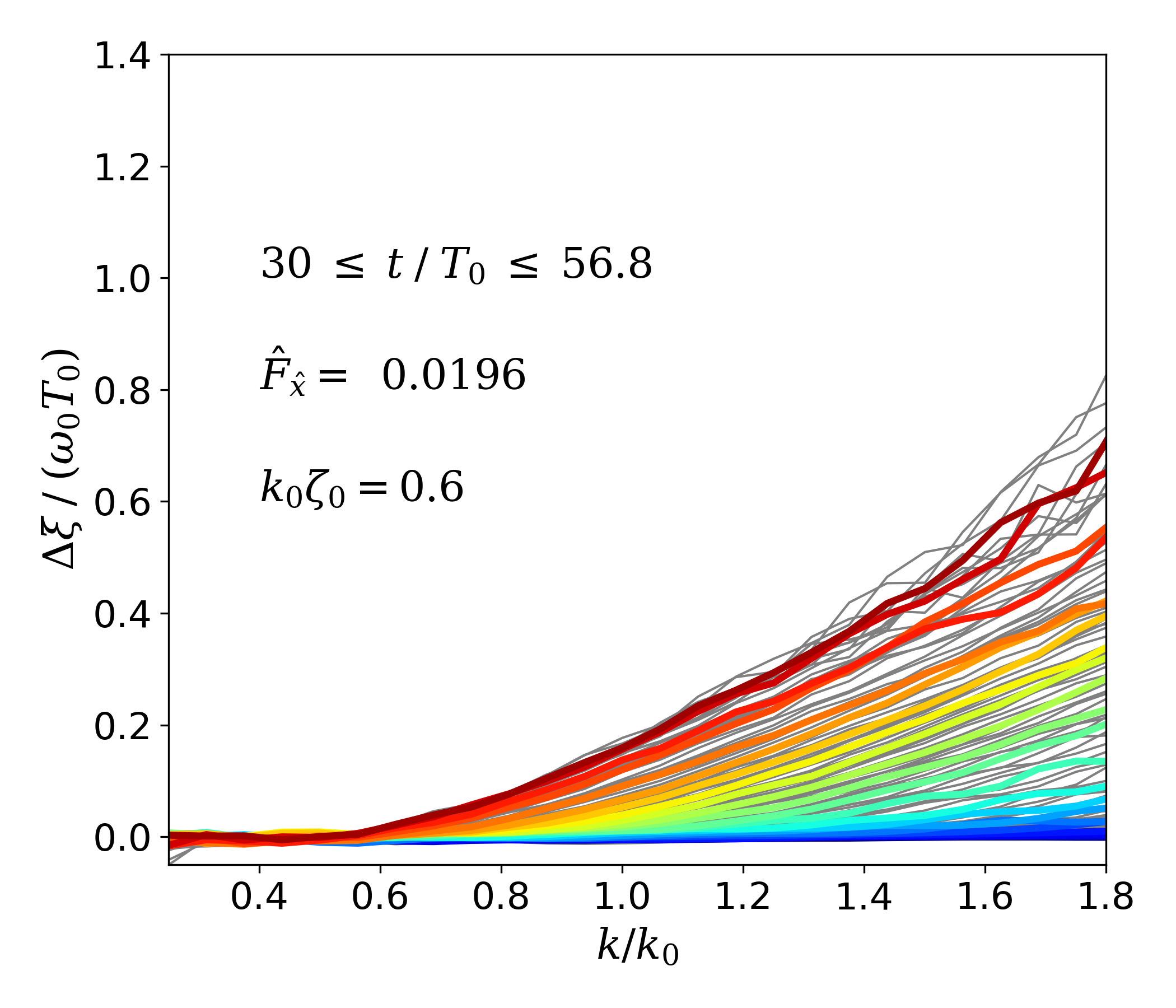}
    \put(18,73){\textbf{(b)}}
\end{overpic}
\includegraphics[height=0.4\linewidth]{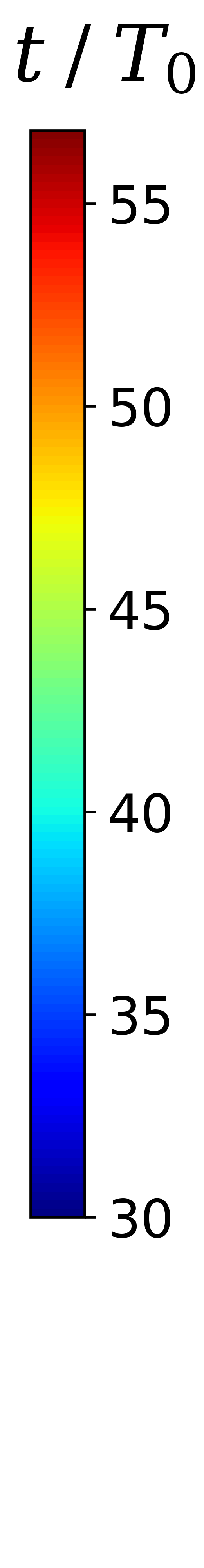}\\
\begin{overpic}[height=0.4\linewidth]{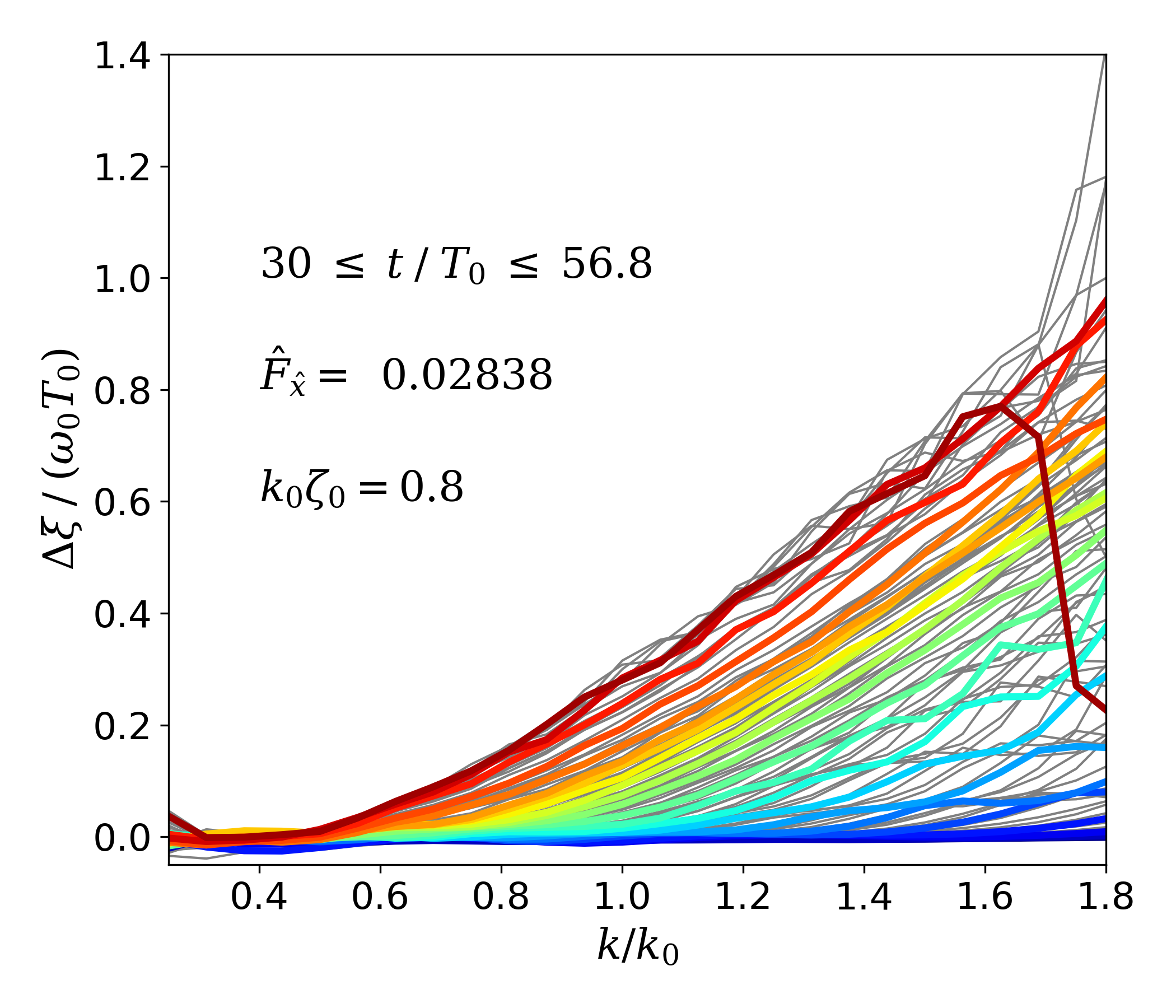}
    \put(18,73){\textbf{(c)}}
\end{overpic}
\begin{overpic}[height=0.4\linewidth]{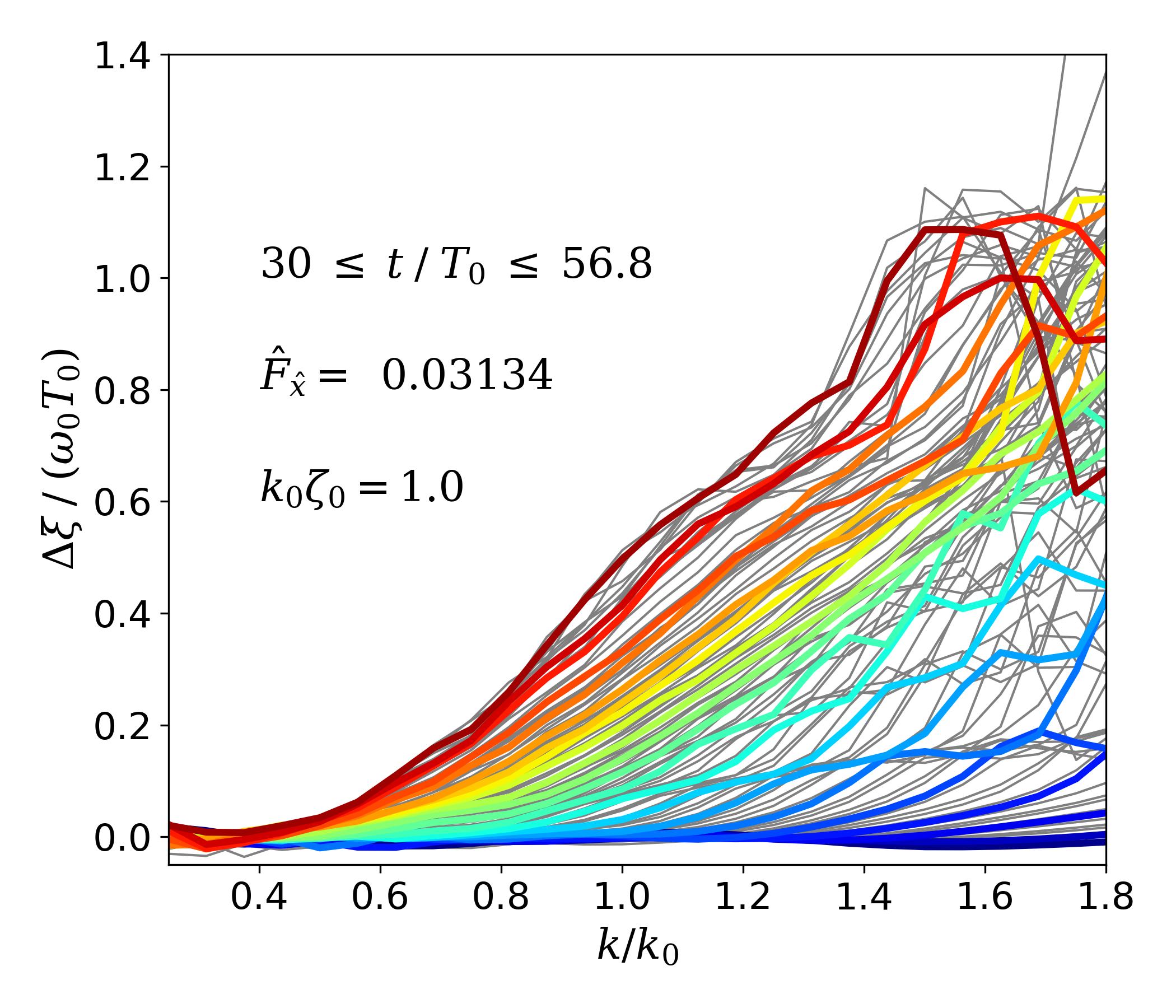}
    \put(18,73){\textbf{(d)}}
\end{overpic}
\includegraphics[height=0.4\linewidth]{figs/phase/bar}
\caption{Evolution of difference in the free waves phases $\Delta \xi_m$
         between VOF and BEM$\nu$ simulations. The color scheme
         from blue to red displays the plots obtained at different
         instances within the range $30 \leq t / T_0 \leq 56.8$.
         Gray plots show other not colored data within the same time
         interval. Panels (a)-(d) correspond to the wave trains of
         various steepness $k_0 \zeta_0$.}
\label{fig:phase}
\end{figure}

In the beginning and \replaced{at}{in} the end of computations only a part of the wave
train is present within the domain. Therefore, in the analysis
we \replaced{focus on}{consider only} the interval ${30 \leq t / T_0 \leq 56.8}$, when the
full length of the wave train is present within the domain.
The surface elevation of free waves (\ref{eq:freewave}) obtained
\replaced{by}{after} the spectral decomposition (see Section~\ref{sec:decompose})
of \deleted{the} simulation results \replaced{can}{may} be written \replaced{as}{in the form}:
\begin{eqnarray}
   && \eta^{(1)}(x) = \sum_{m=0}^M
          \left| A(k_m) \right| \cos(k_m x + \xi_m),
   \label{eq:freewave2}
\end{eqnarray}
where \added{the} phase is \added{given by}:
\begin{eqnarray}
   && \xi_m = \tan^{-1} \frac{ Im \{ A(k_m) \} }
              { Re \{ A(k_m) \} }.
\end{eqnarray}
\myhighlight{It was found that the amplitude spectra obtained from the BEM$\nu$ and VOF
simulations are practically identical in terms of the absolute value
of the amplitude $|A(k_m,\omega_m)|$ regardless of wavenumber and
angular frequency, cf. Figure~\ref{fig:amplmagn} (a)-(d).
This observation confirms the capability of the eddy
viscosity approximation in predicting wave energy dissipation,
which is in line with previous studies
\citep{Perlin2010, Perlin2012, Ducrozet2018, Sriram2019},
see also Figure~\ref{fig:flux1}.} Note that the energy
contained in the spectrum is $\sum_{m=0}^M |A(k_m,\omega_m)|^2$.
\replaced{Therefore the visible divergence}{Divergence} in surface elevation \replaced{shown}{visible} in Figure~\ref{fig:elev}
is \replaced{very likely}{in turn} caused by \deleted{the influence of} the phase $\xi_m$.
The phase difference between the VOF and BEM$\nu$ results
\replaced{relative}{related} to the carrier wave characteristics $\omega_0 T_0$ is \added{given by}:
\begin{eqnarray}
   && \frac{\Delta \xi_m}{\omega_0 T_0} = 
      \frac{1}{2 \pi} \left( 
      \xi_{m,\text{VOF}} - \xi_{m,\text{BEM}\nu} \right)
   \label{eq:phasediff}
\end{eqnarray}
The evolution of $\Delta \xi_m$ in time for different 
wavenumber $k_m$ is plotted in Figure~\ref{fig:phase} for the wave
trains with steepness $k_0 \zeta_0 \geq 0.4$.

Figure~\ref{fig:phase} reveals \added{a} quite smooth and deterministic rather than
stochastic evolution of the phase shift $\Delta \xi$ in time.
Moreover, there is a strong dependence of $\Delta \xi$ on the wavenumber.
The phase shift is \replaced{relatively small}{almost unaffected} at low wavenumbers, while
at a high wavenumber \replaced{it has a much}{there is its} pronounced growth with time.
For instance, at the end of the wave breaking
region the phase shift \replaced{for a high wavenumber}{at} $k/k_0 \approx 1.5$ \replaced{can increase by more than}{may be higher than} one full revolution, see Figure~\ref{fig:phase}(d).
If the wave train steepness $k_0 \zeta_0$ and subsequently the energy
dissipation rate $\hat{F}_{\hat{x}}$ are relatively low, \replaced{the corresponding phase shift is much weaker}{a weaker
phase shift is found}, see Figure~\ref{fig:phase}(a).
\replaced{The phase difference between the BEM$\nu$ and VOF calculations could be due to the fact that
the highly nonlinear rotational flows generated by breaking cannot be properly handled
by the weakly-potential eddy viscosity approximation.}{It allows assumption that the revealed phase shift
between VOF and BEM$\nu$ computations appears because of nonlinearities
in the coexisting rotational and irrotational flows not taken into account
in the weakly-potential eddy viscosity approximation.}

\begin{figure}
    \centering
    \includegraphics[width=0.6\textwidth]{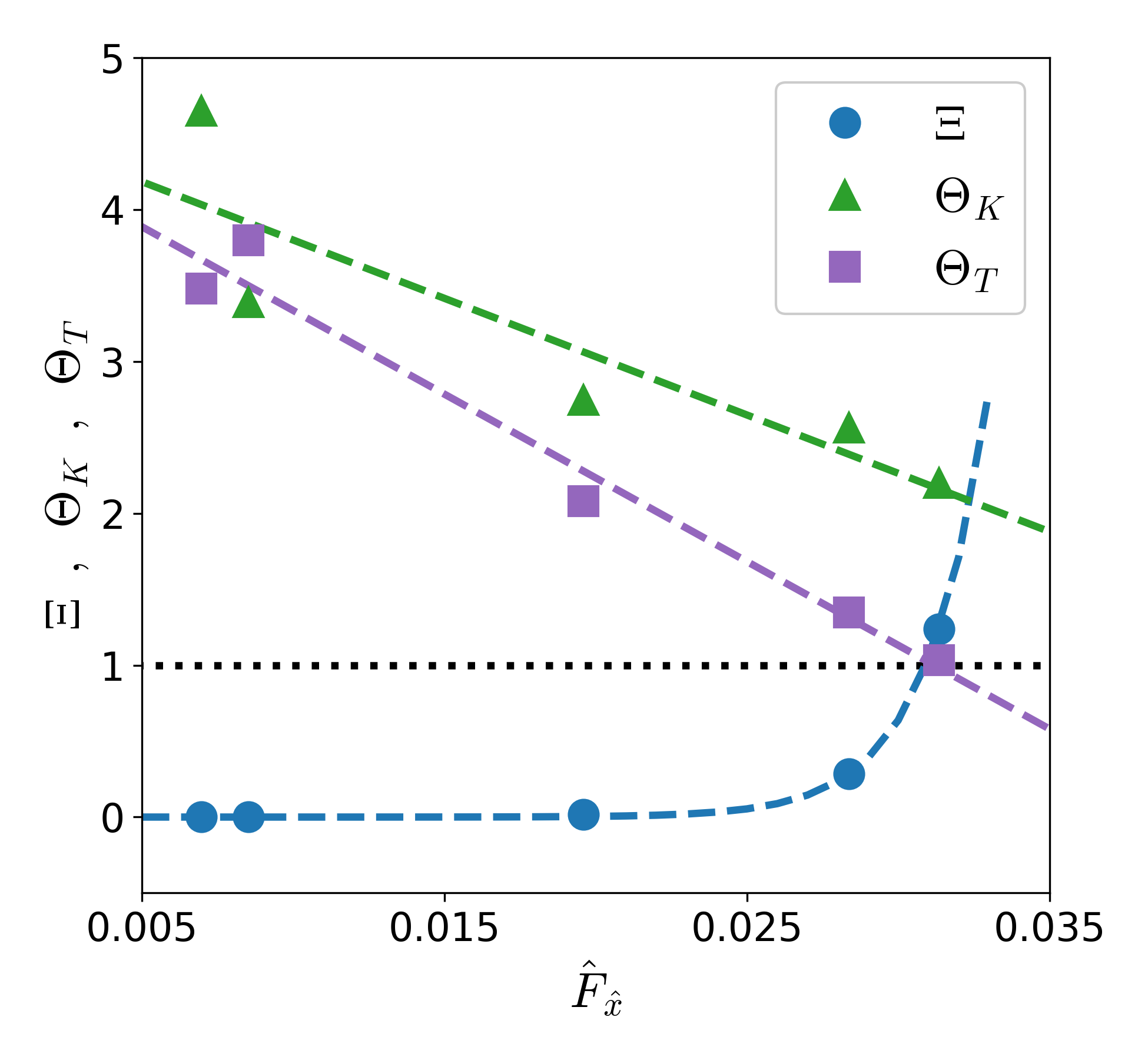}
    \caption{Dependence of the fitting coefficients $\Xi$, $\Theta_K$
             and $\Theta_T$ (\ref{eq:phaseapprox1}) on the wave
             breaking strength $\hat{F}_{\hat{x}}$.}
    \label{fig:coeff}
\end{figure}

\replaced{Considering the}{Consider} dependence of \deleted{the} phase shift on the wavenumber $k$ and time $t$
for \replaced{the}{the fixed wave} breaking strength $\hat{F}_{\hat{x}}$, 
\replaced{we approximate it}{Approximate this dependence} with the following expression:
\begin{eqnarray}
   && \left. \frac{\Delta \xi}{\omega_0 T_0}
      \right|_{\hat{F}_{\hat{x}}} =
      \Xi \left[ \frac{k}{k_0} \right]^{\Theta_K}
          \left[ \frac{t}{T_0} \right]^{\Theta_T}
   \label{eq:phaseapprox1}
\end{eqnarray}
The values of three coefficients $\Xi$, $\Theta_K$ and $\Theta_T$ \replaced{change}{are varied}
from one wave train to another due to different breaking strength.
\replaced{We applied the least squares method to obtain the dependencies $\Xi (\hat{F}_{\hat{x}})$, $\Theta_K (\hat{F}_{\hat{x}})$,
	and $\Theta_T (\hat{F}_{\hat{x}})$ from the numerical simulations, and present them in Figure~\ref{fig:coeff}. }
{The dependencies $\Xi (\hat{F}_{\hat{x}})$, $\Theta_K (\hat{F}_{\hat{x}})$,
and $\Theta_T (\hat{F}_{\hat{x}})$ found using the least squares
interpolation of outcomes of the numerical simulations are present
in figure~\ref{fig:coeff}.}
It is \replaced{shown}{seen from the figure} that the rate of phase shift has a
nonlinear dependence on time and wavenumber for relatively weak \deleted{wave}
breaking; i.e. the values of the power coefficients
$\Theta_K \approx \Theta_T \approx 4$ when $\hat{F}_{\hat{x}} < 0.01$.
If \deleted{the wave} breaking is strong, the dependence of the phase shift on time
tends to be linear with $\Theta_T \approx 1$, while the dependence
on \deleted{the} wavenumber is \deleted{still} close to quadratic with $\Theta_K \approx 2$.
Both $\Theta_K$ and $\Theta_T$ are reduced almost linearly with the
increase  \replaced{of}{in the} breaking strength $\hat{F}_{\hat{x}}$.
In turn, the proportionality coefficient
$\Xi (\hat{F}_{\hat{x}})$ demonstrates \added{an} exponential dependence on the
energy dissipation rate $\hat{F}_{\hat{x}}$. \replaced{Therefore it seems reasonble
to infer from the analysis that the phase shift can become quite significant for strong breaking events.}{This leads to an assumption
that the phase shift may be significantly larger if the breaking event
is stronger.}

\replaced{We applied the}{Assume that} similar interpolation function \replaced{to all the considered wave trains and obtained the corresponding phase shift in a dimensionless form}{can be applied to approximate
the phase shift inherent to all considered wave trains}:
\begin{eqnarray}
   && \frac{\Delta \xi}{\omega_0 T_0} =
      \Xi \left[ \hat{F}_{\hat{x}} \right]^{\Theta_F}
          \left[ \frac{k}{k_0} \right]^{\Theta_K}
          \left[ \frac{t}{T_0} \right]^{\Theta_T}
   \label{eq:phaseapprox2}
\end{eqnarray}
\replaced{Appling the least squares fitting of the numerical results, we obtained the coefficents and listed them in Table \ref{tab:phaseapprox}.}
{The least squares fitting of the data obtained in all executed simulations
gives the values of the coefficients \replaced[id=ZM]{summarised}{summarized} in
table~\ref{tab:phaseapprox}.} \replaced{In addition}{Additionally} to the previous findings,
the value of $\Theta_F \approx 2$ implies \replaced{a}{the} quadratic dependence of the
phase shift on the energy dissipation rate.

\begin{table}
\centering
\caption{\label{tab:phaseapprox} Values of the dimensionless coefficients
         in the expression (\ref{eq:phaseapprox2})}
\begin{tabular}{ c c c c }
 \hline
 $\Xi$ & $\Theta_F$ & $\Theta_K$ & $\Theta_T$ \\
 \hline
 $\;\;5.02\;\;$ & $\;\;1.926\;\;$ & $\;\;2.362\;\;$ & $\;\;1.2105\;\;$ \\
 \hline
\end{tabular}
\end{table}

\begin{figure}
\centering
\begin{overpic}[width=0.49\linewidth]{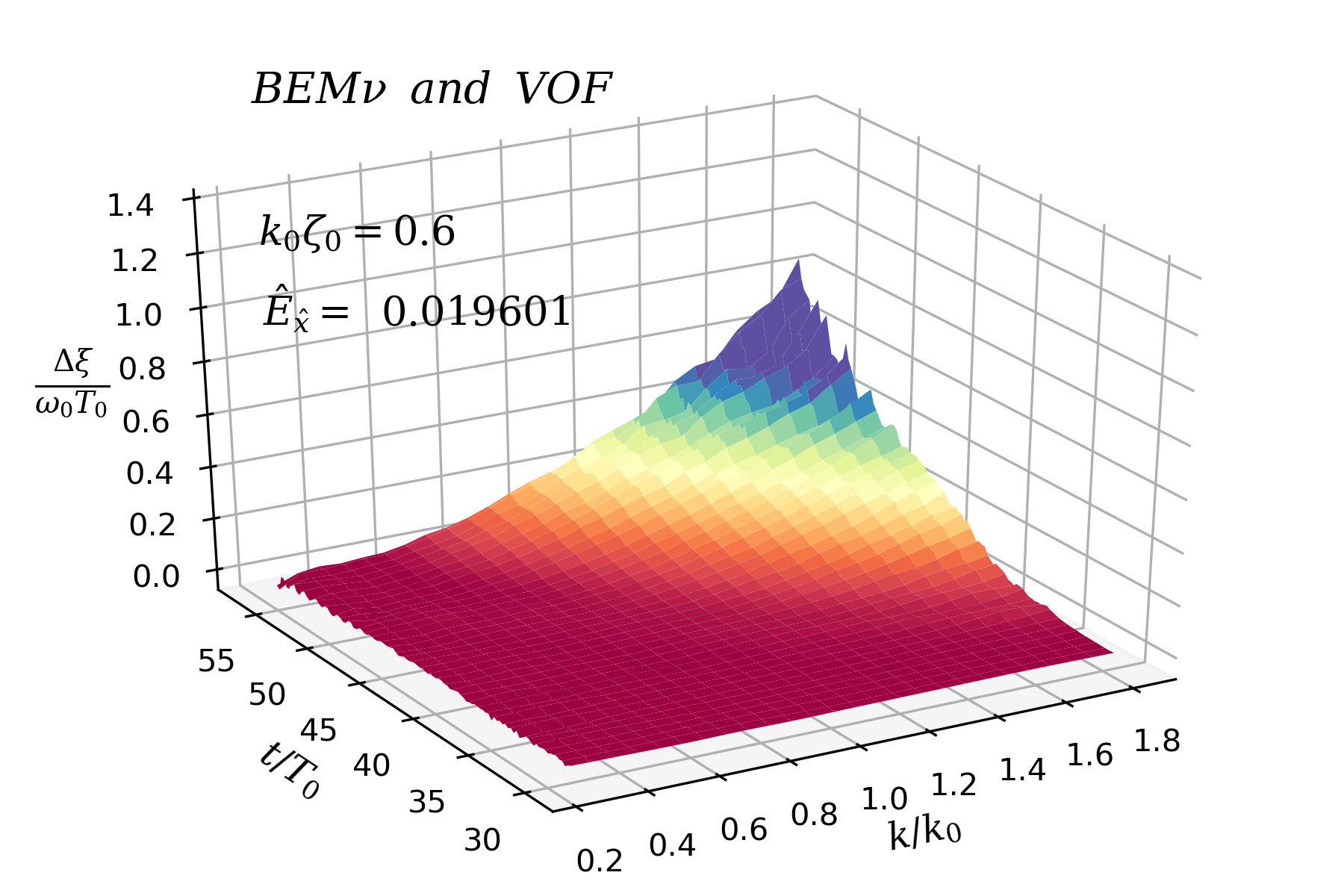}
    \put(10,59){\textbf{(a)}}
\end{overpic}
\begin{overpic}[width=0.49\linewidth]{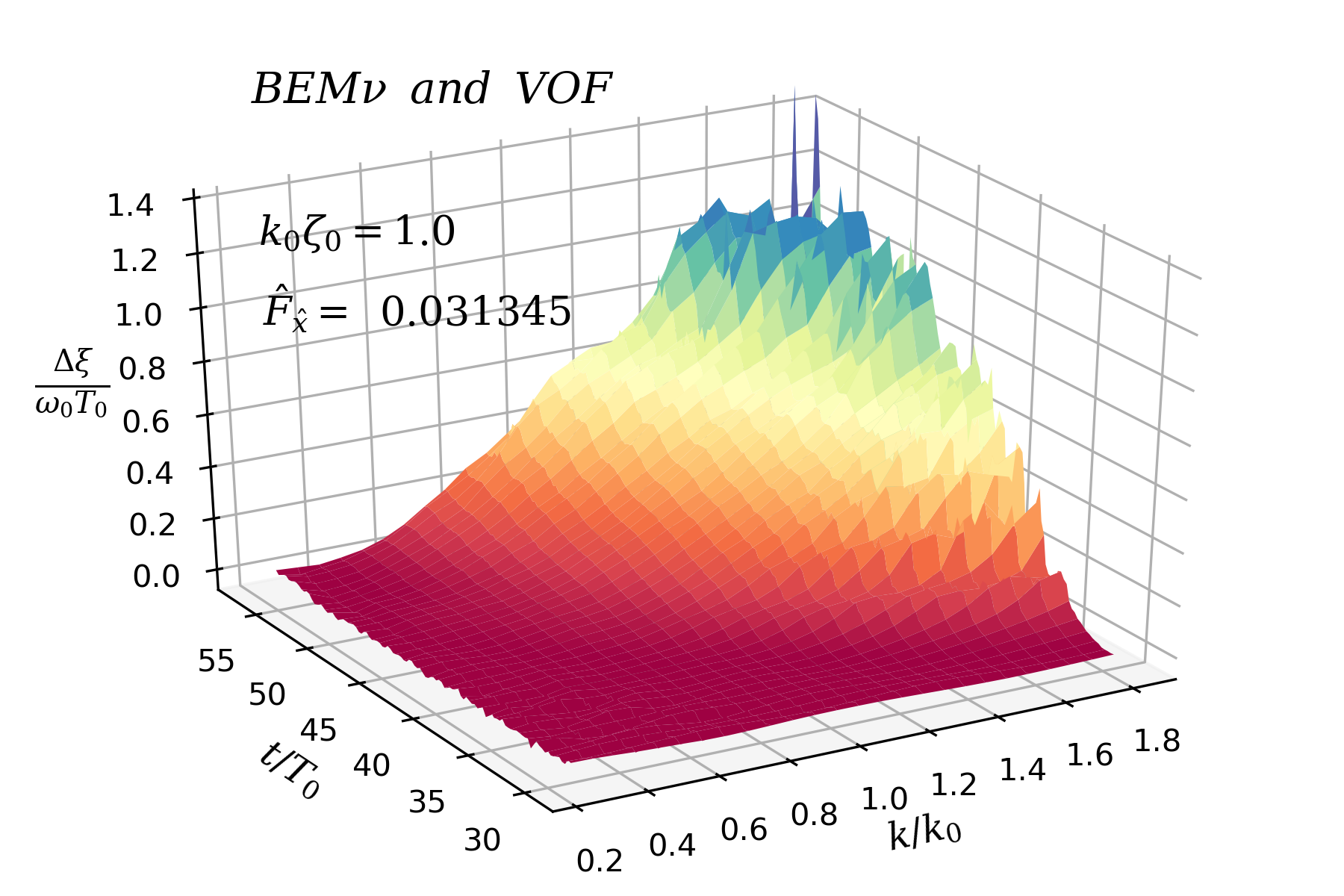}
    \put(10,59){\textbf{(c)}}
\end{overpic}\\
\begin{overpic}[width=0.49\linewidth]{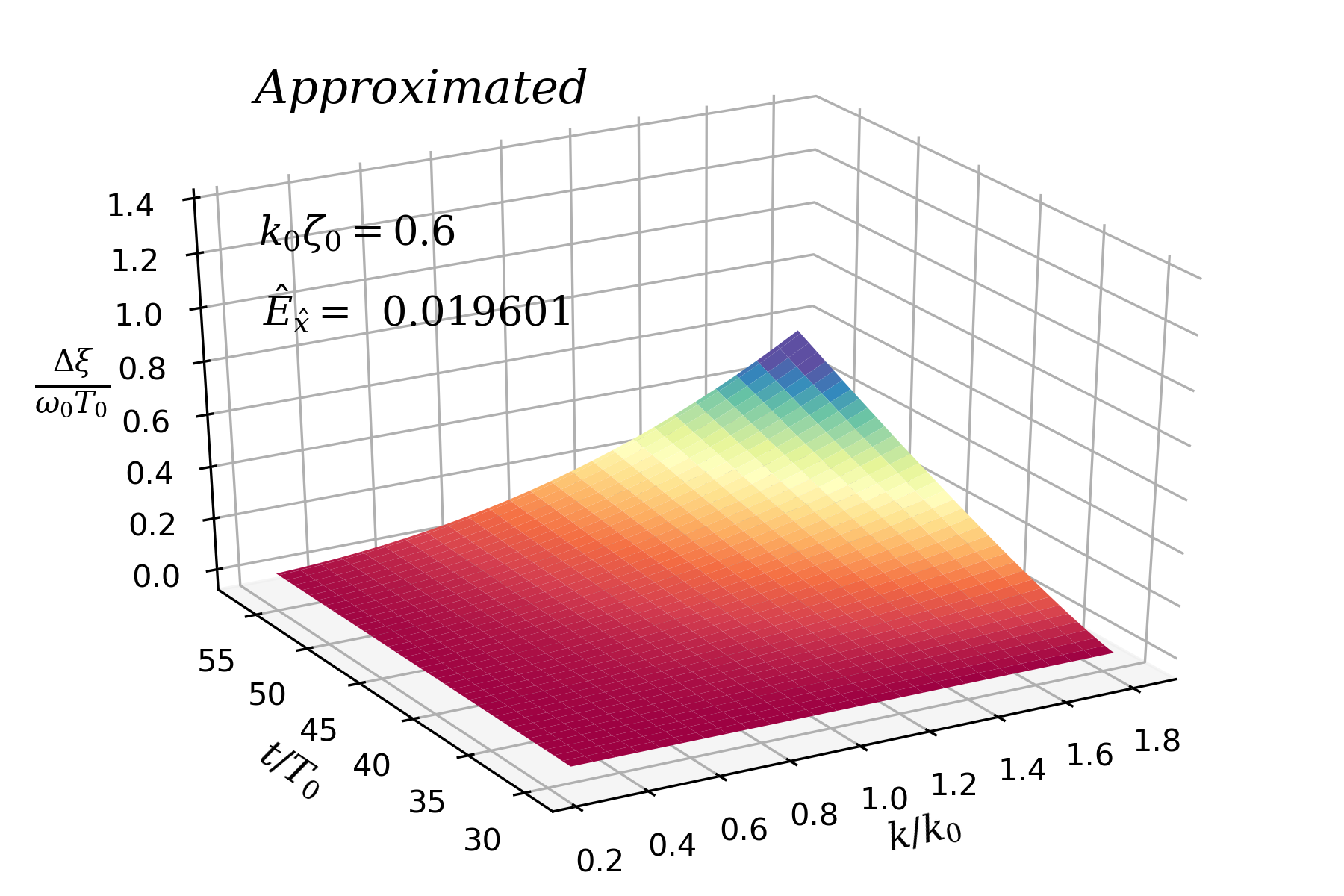}
    \put(10,59){\textbf{(b)}}
\end{overpic}
\begin{overpic}[width=0.49\linewidth]{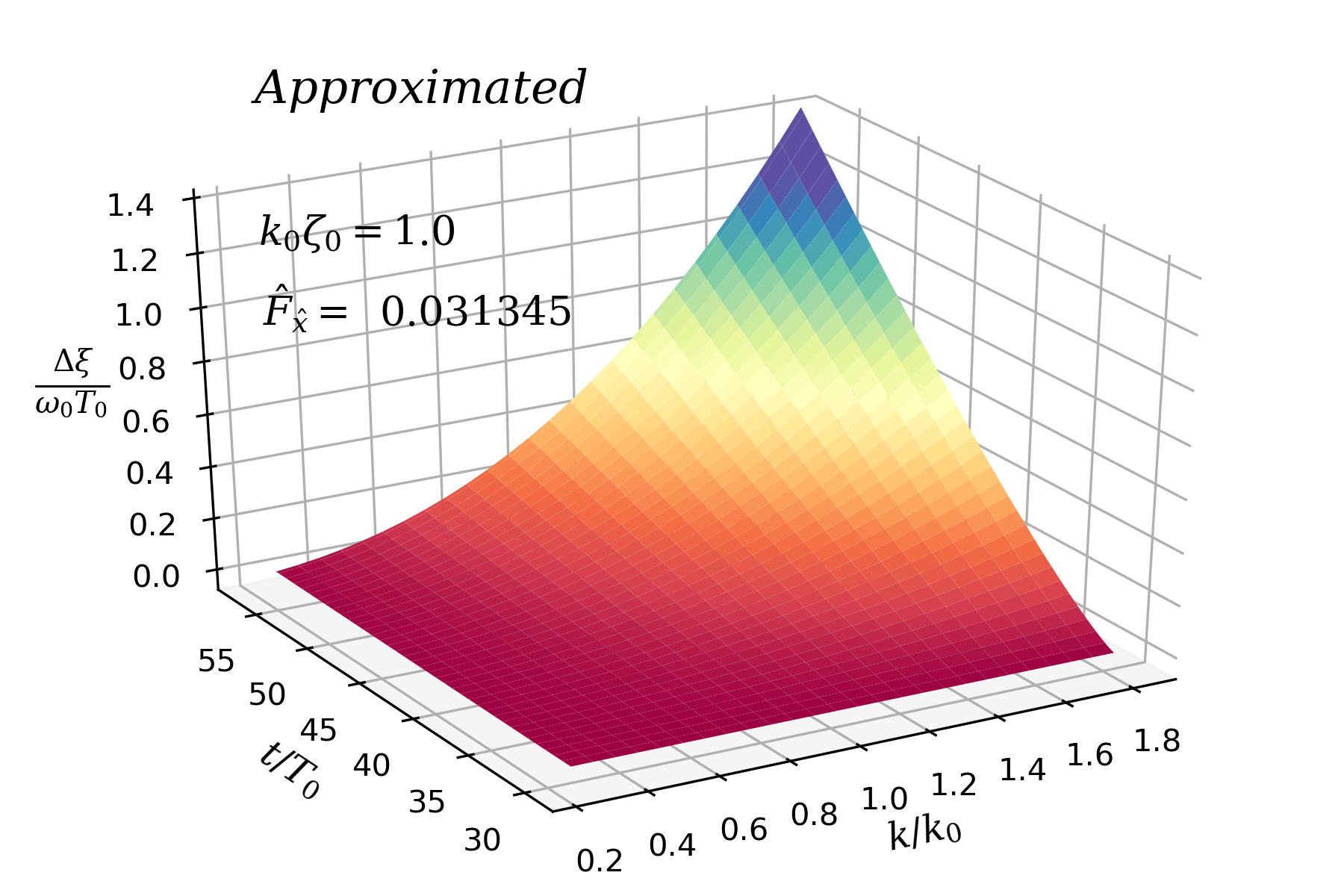}
    \put(10,59){\textbf{(d)}}
\end{overpic}
\caption{Phase shift $\Delta \xi / \omega_0 T_0$ as a function of wave
         number and time obtained for the wave trains of different
         breaking strength: (a) and (b) $\hat{F}_{\hat{x}} = 0.0196$;
         (c) and (d) $\hat{F}_{\hat{x}} = 0.0313$.
         Panels (a) and (c) present the raw phase shift calculated using
         (\ref{eq:phasediff}), while panels (b) and (d) are plotted
         using the interpolation (\ref{eq:phaseapprox2}).}
\label{fig:phase3d}
\end{figure}

In Figure~\ref{fig:phase3d} the raw plots of $\Delta \xi (k, t)$
obtained from the numerical simulations are compared with those
reconstructed from (\ref{eq:phaseapprox2}) in order to study the accuracy
of the suggested \replaced{approximation}{approximate expression}.
It can be noticed that the plots obtained directly from the simulations
using (\ref{eq:phasediff}), see panels (a) and (c), exhibit fluctuations
at high wavenumbers because the bound waves are not perfectly separated
from the free waves by the decomposition method
(section~\ref{sec:decompose}).
Even minor presence of unseparated bound waves significantly influences
phases of the harmonics.
Function (\ref{eq:phaseapprox2}) filters the fluctuations from the
plots as presented in panels (b) and (d), while keeping
a good qualitative and quantitative correspondence with the raw data
extracted from the simulations.

The observed shift in \replaced{phase is possibly}{the phases can be} related to the so-called
phase-locking phenomenon \replaced{reported}{found} by \citet{Kirby2016}.
The phase-locking processes is considered as \added{a} nonlinear linkage between
\replaced{high- and low-frequency wave components}{the evolution of the high-frequency components with those of the lower
frequency}. This linkage was demonstrated by analysing the
wavelet spectra of the surface elevation in the vicinity of the
breaking event \citep{Kirby2016}. The propagation velocity of the
high frequency components of the breaking wave was found to be
different from that of the pre-breaking but stable wave.

\subsection{\label{sec:disper} Dispersion variation}

It \replaced{has been}{was} shown in various studies that the relationship between
wavenumber and frequency may deviate from the linear dispersion relation
(\ref{eq:disprel}) when the frequency spectrum is narrow.
\citet{Trulsen2010} studied the dynamic nonlinear
evolution of unidirectional Gaussian wave packets using the nonlinear
Schr\"{o}dinger equation and its \replaced{generalisations}{generalizations}. It was observed that
in the $k$--$\omega$ space the spectrum does not maintain a thin well-defined
dispersion surface but  develops into continuous distribution.
The spectral components above and below the spectral peak were found
to have the phase and group velocities close to that of the spectral peak.
It was concluded that in some cases it is inappropriate to use the
linear dispersion relation for the post-processing of experimental data. 

\citet{Waseda2018b, Waseda2018a} have demonstrated that 
\replaced{the dispersion characteristics of the Akhmediev breather solution to the nonlinear
Schr\"{o}dinger equation in deep water deviates significantly from the linear relationship \eqref{eq:disprel}.}{in deep water the Akhmediev breather solution of the nonlinear
Schr\"{o}dinger equation exhibits the linear dependence of frequency on
wavenumber significantly deviating from (\ref{eq:disprel}).}
It was also shown that these findings extend beyond the applicability
of nonlinear Schr\"{o}dinger equation.
Accordingly, the highly nonlinear non-breaking modulated wave trains
also maintain the straight line relationship between the wavenumber and
the instantaneous frequency; this line is tangent to the dispersion
relation curve (\ref{eq:disprel}) at the carrier wavenumber.
\citet{Swan2006} \deleted[id=ZM]{have} investigated the waves dispersion
using the numerical simulations based on the ${3^{\text{rd}}\text{-order}}$
Zakharov equation. Inconsistency of the simulation results with the
linear wave dispersion (\ref{eq:disprel}) was explained by analysing
the third-order resonant interactions. The nonlinear energy transfer
between harmonics in the free wave spectrum alters the values of the
complex amplitudes. If interacting wave components are out of phase,
this energy transfer will change the instantaneous phases of waves,
which is \replaced{reflected}{seen} in the $k$--$\omega$ space.
Adopting a similar Zakharov equation based theoretical model, 
the nonlinear correction to the dispersion relation for gravity waves
\replaced{in a}{on water of} constant depth was derived analytically by
\citet{Stuhlmeier2019}.

The phase of each free wave component is \citep{Waseda2018a, Waseda2018b}:
\begin{eqnarray}
   && \xi (k, \omega, t) = k x - \omega t - \delta^{NL} (k, t),
   \label{eq:phasenl}
\end{eqnarray}
where $\delta^{NL} (k, t)$ is the slowly varying nonlinear phase
\replaced{induced by}{appearing due to} the ${3^{\text{rd}}\text{-order}}$ resonant
interaction between the waves \citep{Swan2006}.
The angular frequency can be found from (\ref{eq:phasenl}) by involving
(\ref{eq:disprel}):
\begin{eqnarray}
   && \omega (k) = - \frac{\partial \xi(k, \omega, t)}{\partial t}
      = \sqrt{g k \tanh (k h)}
      + \left< \frac{\partial \delta^{NL}(k, t)}{\partial t} \right>
   \label{eq:freqnl}
\end{eqnarray}
When analysing a long time evolution of the wave group, the influence of
$\left< \partial \delta^{NL} / \partial t \right> \neq 0$
seen in $k$--$\omega$ spectrum may be interpreted as \added{a} deviation of the
relationship $\omega(k)$ from (\ref{eq:disprel}).

\begin{figure}
\centering
\begin{overpic}[width=0.45\linewidth]{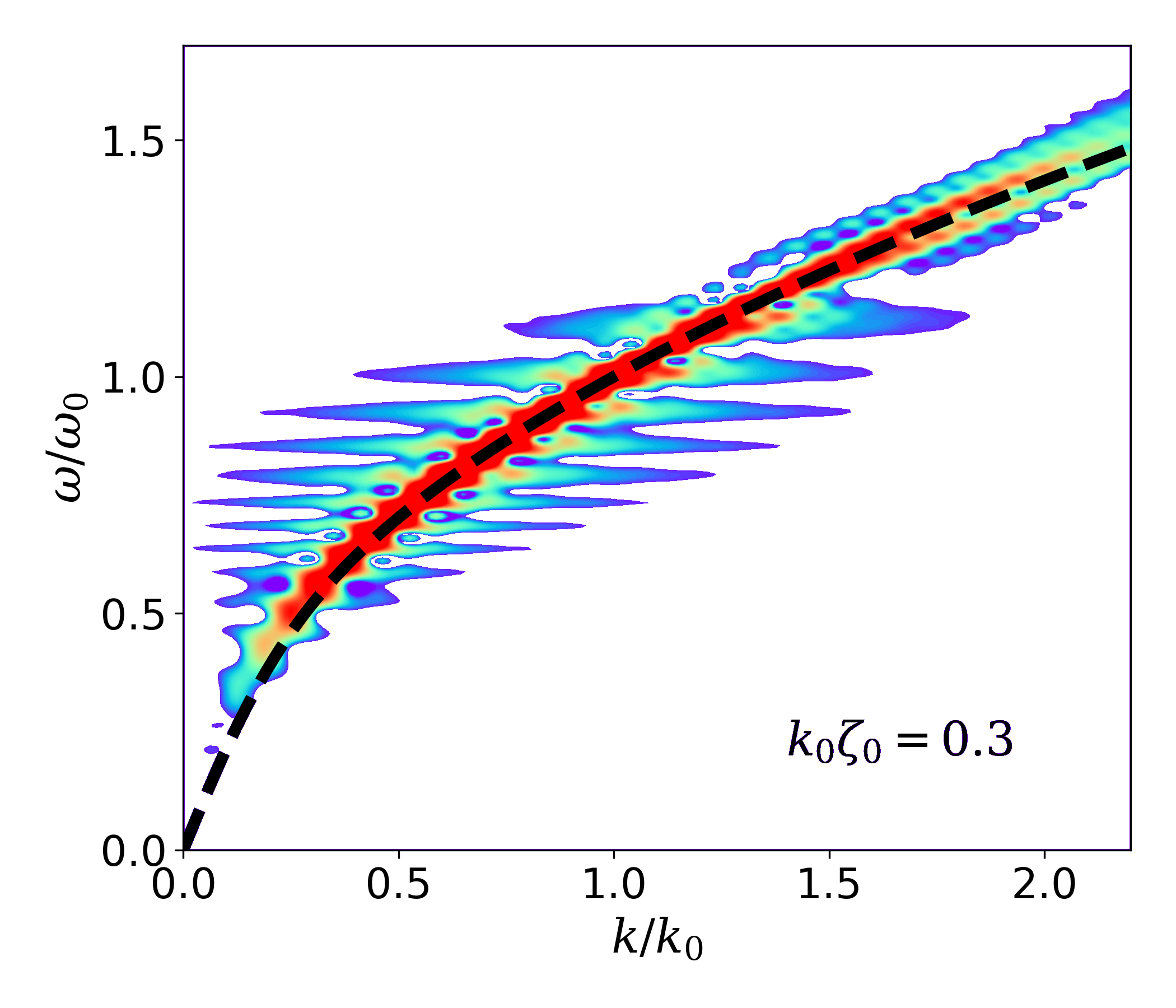}
    \put(20,75){\textbf{\color{black}(a)}}
\end{overpic}
\begin{overpic}[width=0.45\linewidth]{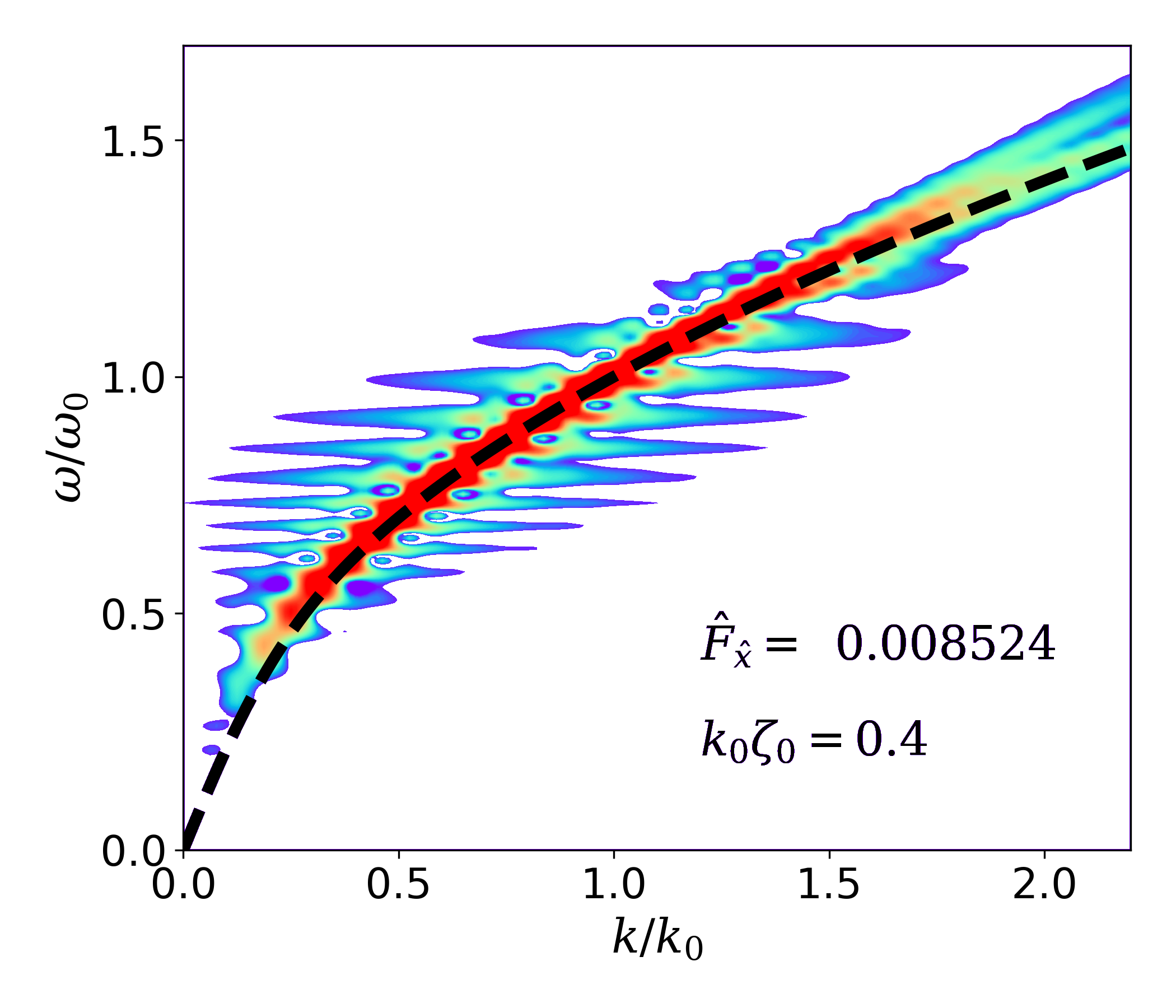}
    \put(20,75){\textbf{\color{black}(b)}}
\end{overpic}\\
\begin{overpic}[width=0.45\linewidth]{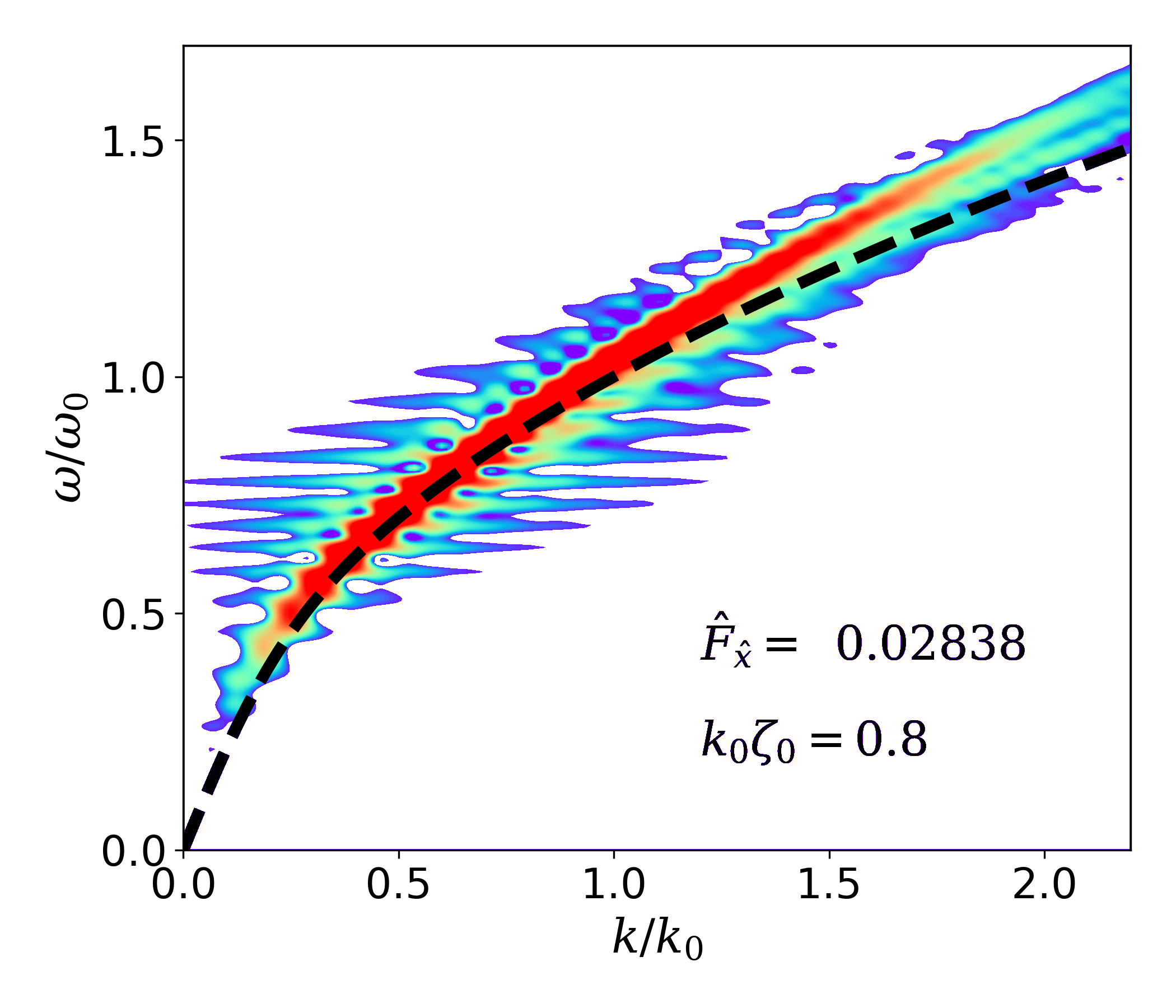}
    \put(20,75){\textbf{\color{black}(c)}}
\end{overpic}
\begin{overpic}[width=0.45\linewidth]{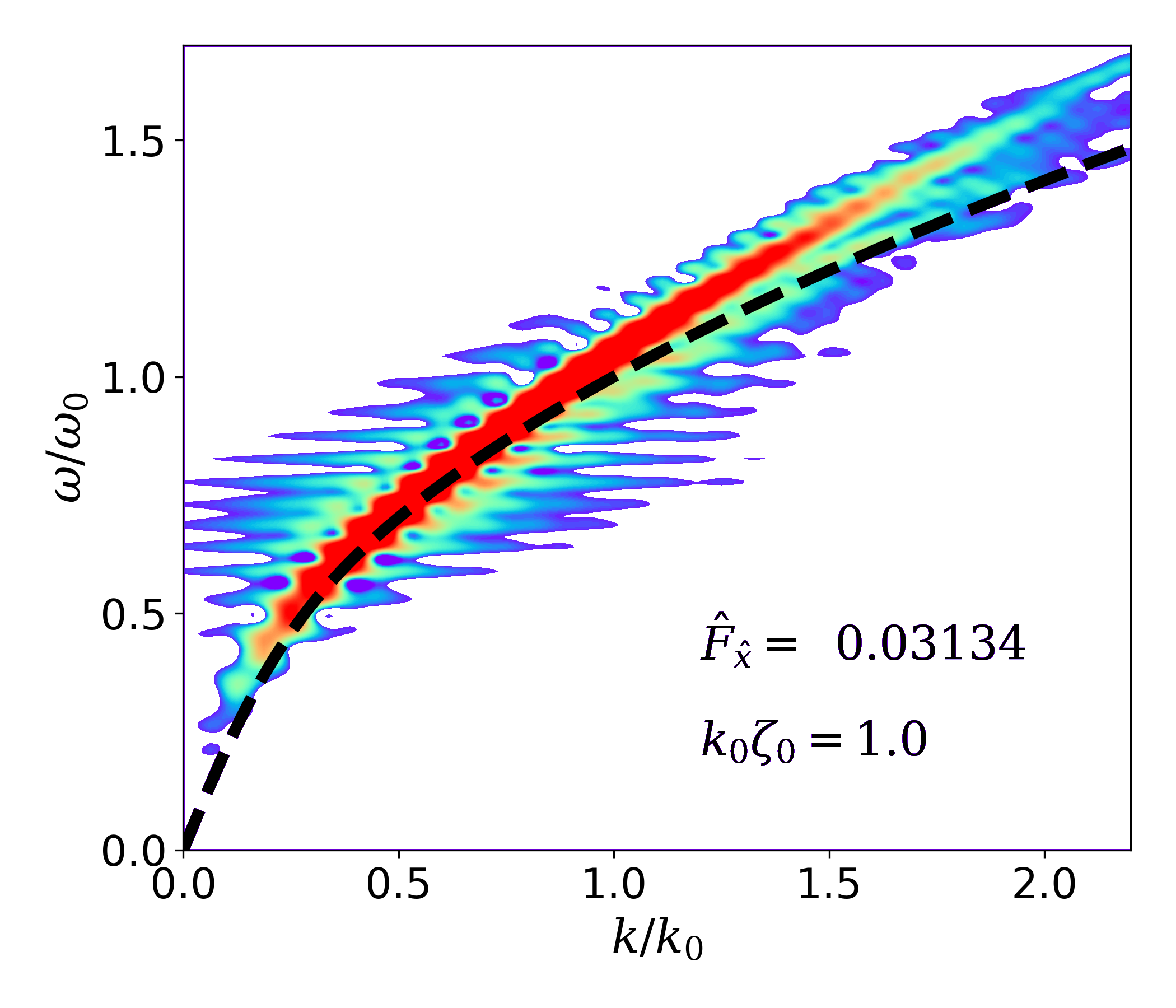}
    \put(20,75){\textbf{\color{black}(d)}}
\end{overpic}
\caption{Dimensionless free wave $k$--$\omega$ spectrum
         $|\hat{\eta} (k, \omega)| / \max(|\hat{\eta} (k, \omega)|)$
		 obtained from the VOF
         simulations for wave packets with different steepness.
         The logarithmic color scale has the range
         $|\hat{\eta}| / \max(|\hat{\eta}|) = \left[ 3\times10^{-3},
         3\times10^{-1} \right]$. Dashed line represents the linear
         dispersion curve (\ref{eq:disprel}) with
         accounted capillary effects.}
\label{fig:disp}
\end{figure}

In the current study it was found that the nonlinear contribution
$\delta^{NL}(k, t)$ can be caused \deleted{not only} by the resonant interactions
\replaced{as well as}{but also by contribution of} the breaking-induced rotational flows.
To construct the $k$--$\omega$ spectrum from BEM$\nu$ and VOF
computations the discrete distribution of the surface elevation in space
is extracted from the numerical results.
After applying spectrum decomposition (see Section~\ref{sec:decompose}) 
the elevations of free waves at different instants are expressed as a function
$\eta(x,t)$. The discrete 2D Fourier transform of this function
in the $k$--$\omega$ space is given by:
\begin{eqnarray}
   && \hat{\eta} (k, \omega) = \frac{1}{N M} \sum_{x} \sum_{t}
      \eta (x, t) e^{i (k x - \omega t)},
   \label{eq:2dfft}
\end{eqnarray}
where $N$ and $M$ are the number of discrete values of $\eta$ in $x$ and $t$
directions, respectively.
Distributions of the absolute values $|\hat{\eta} (k, \omega)|$
obtained from the VOF simulations are shown in
Figure~\ref{fig:disp}.

The highest values of $|\hat{\eta} (k, \omega)|$ are located \replaced{in}{along} a
relatively narrow region approximately defining the dispersion relation $\omega(k)$.
For the considered broad-banded
Gaussian wave trains the distributions of $|\hat{\eta} (k, \omega)|$
have no visible deviation from the linear dispersion relation for both
non-breaking and weakly-breaking cases $k_0 \zeta_0 = 0.3 - 0.4$,
cf. Figure~\ref{fig:disp} (a) and (b). According to
 Figure~\ref{fig:steepness}, the maximum steepness of these wave trains
\replaced{is}{varies} within the interval ${0.2 < \varepsilon_{max} < 0.4}$, indicating
strong non-linearity. Such dispersive properties are
different from those discussed in \citep{Swan2006, Trulsen2010,
Waseda2018a, Waseda2018b}, where the linear dispersion relation was
found to be inaccurate.
The difference in the wave packet evolution may be related to the
width of the spectra that is significantly higher for the wave
trains considered in the current study.

Subsequent increase of the wave train steepness, i.e. $k_0 \zeta_0 > 0.4$,
accompanied by the intensification of wave breaking leads to the deviation
of the distribution of $|\hat{\eta} (k, \omega)|$ from the linear
dispersion curve (\ref{eq:disprel}), cf. Figure~\ref{fig:disp} (c) and (d).
The \replaced{magnitude}{value} of this deviation is dependent on the wave train steepness
parameter $k_0 \zeta_0$. Note, the wave train with ${k_0 \zeta_0 = 1.0}$
is subject to breaking within the entire range of $t$ and $x$ used for
computing the ${k-\omega}$ spectrum.
Moreover, the deviation is observed only for the wavenumbers above the
spectral peak $k_0$, while long waves always follow the linear
dispersion. To some extent, this phenomenon is correlated to the
phase shift plotted in Figure~\ref{fig:phase} and \ref{fig:phase3d},
which is also present at high wavenumbers only.

\begin{figure}
\centering
\begin{overpic}[width=0.48\textwidth]{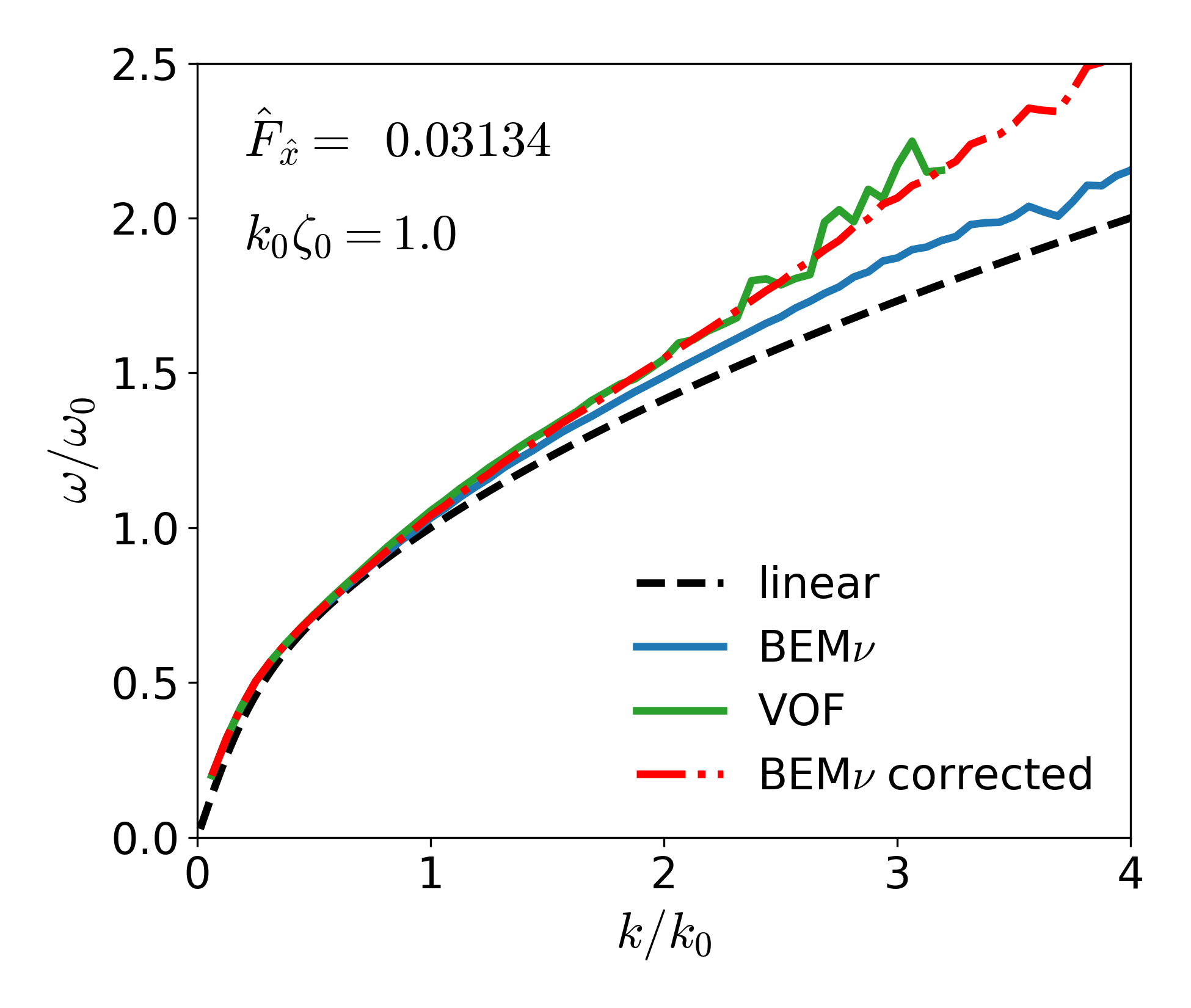}
    \put(0,75){\textbf{\color{black}(a)}}
\end{overpic}
\begin{overpic}[width=0.48\textwidth]{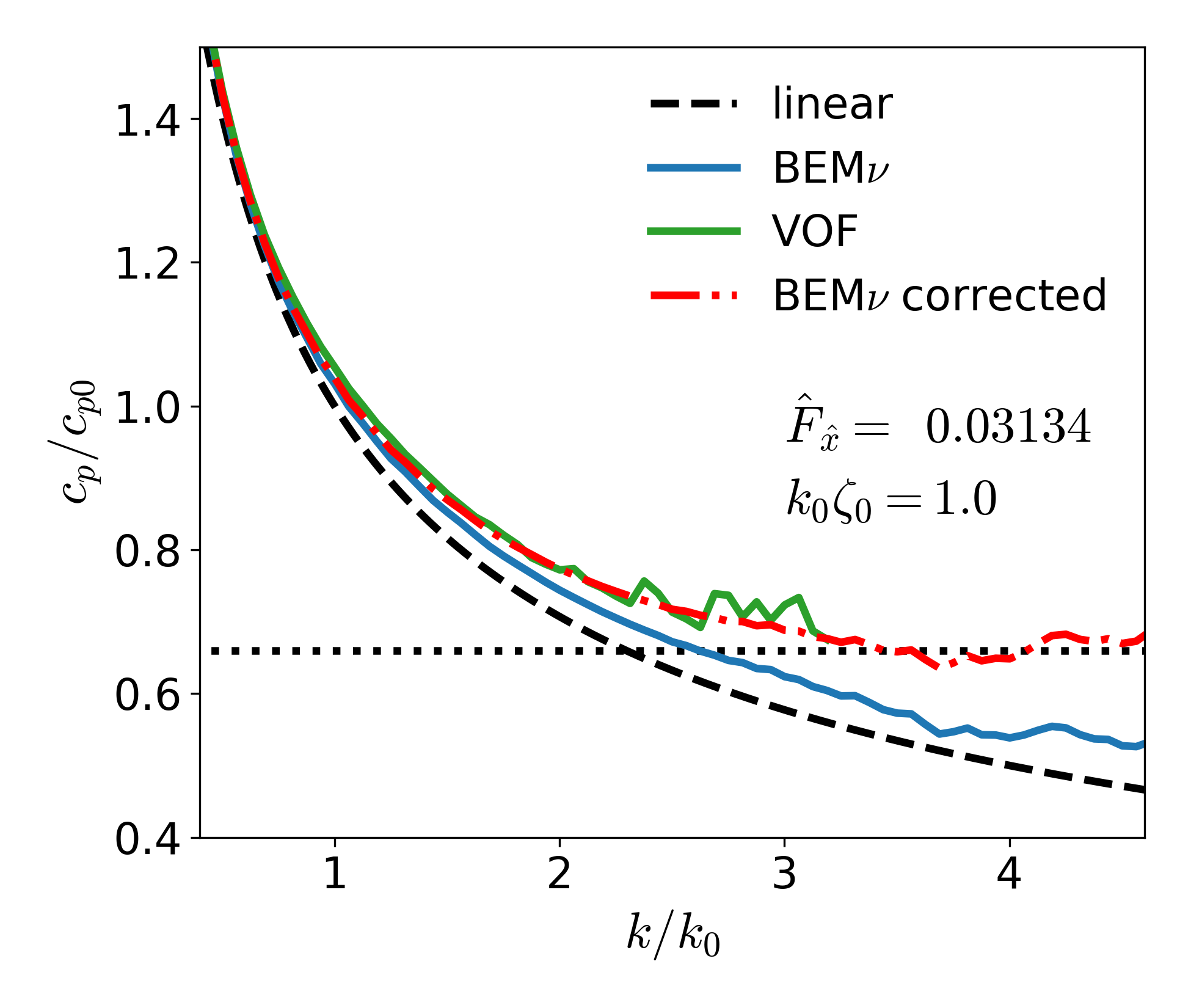}
    \put(0,75){\textbf{\color{black}(b)}}
\end{overpic}
\caption{Dependence of (a) angular frequency and (b) phase speed on the
         wavenumber obtained from the results of BEM$\nu$ and VOF
         simulations using (\ref{eq:2dfft}) and (\ref{eq:dispapprox}).
         The correction to the BEM$\nu$ curve is obtained using
         (\ref{eq:dispcorr2}).
         Wave train with the steepness parameter $k_0 \zeta_0 = 1.0$
         is considered in the plots.}
\label{fig:disp2}
\end{figure}

The dependency of frequency on the wavenumber $\omega(k)$ can be approximated from the distribution of
$|\hat{\eta} (k, \omega)|$ presented in Figure~\ref{fig:disp}
by using the following weighting:
\begin{eqnarray}
   && \omega (k_n) =
      \frac{\sum\limits_m \omega_m |\hat{\eta} (k_n, \omega_m)|^2}
           {\sum\limits_m |\hat{\eta} (k_n, \omega_m)|^2}
   \label{eq:dispapprox}
\end{eqnarray}
The approximate dispersion relations obtained from the BEM$\nu$ and VOF results
are plotted in Figure~\ref{fig:disp2} (a) for the steepest wave packet
$k_0 \zeta_0 = 1.0$.
It is clearly shown that the results of both numerical models deviate from the
linear dispersion relation. Meanwhile the deviation is less pronounced
for the BEM$\nu$ model, implying the importance of including non-potential
effects in calculations. Here we attempt to correct the dispersion curve of the
BEM$\nu$ model by involving the phase shift (\ref{eq:phaseapprox2}) and
the expression (\ref{eq:freqnl}):
%
\begin{equation}
    \omega(k) = \omega_{BEM\nu}(k) +
      \left< \frac{\partial \Delta \xi}{\partial t} \right> \approx
      \omega_{BEM\nu} \left<
      1 + \Theta_T \Xi \left[ \hat{F}_{\hat{x}} \right]^{\Theta_F}
                       \left[ \frac{k}{k_0} \right]^{\Theta_K}
                       \left[ \frac{t}{T_0} \right]^{\Theta_T-1}
      \right>
   \label{eq:dispcorr1}
\end{equation}   
%
Taking into account $(\Theta_T-1) \sim 0$, the expression \eqref{eq:dispcorr1} can be simplified as:
\begin{eqnarray}
   && \omega(k) =
      \omega_{BEM\nu} \left(
      1 + \Theta_T \Xi \left[ \hat{F}_{\hat{x}} \right]^{\Theta_F}
                       \left[ \frac{k}{k_0} \right]^{\Theta_K}
      \right)
   \label{eq:dispcorr2}
\end{eqnarray}
The corrected dispersion curve (\ref{eq:dispcorr2}) for the BEM$\nu$ model plotted in
Figure~\ref{fig:disp2} (a) is close to the 
VOF model. This suggests that the difference in wave dispersion
between the two models is caused by the phase shift
phenomenon discussed in Section \ref{sec:shift}.

The dependency of phase speed on the wavenumber $c_p(k) = \omega / k$
is studied in Figure~\ref{fig:disp2}~(b).
Linear and BEM$\nu$ dispersions show significant variation of the
value of $c_p$ within the considered range of the wavenumbers.
On the other hand, the VOF data suggests a constant value of
$c_p$ at high wavenumbers. This implies that relatively short waves propagate
at a similar speed. Note that capillarity in
(\ref{eq:disprel}) has an insignificant effect within the considered range
of wavenumbers. Contrary to the observations of
\citet{Waseda2018a, Waseda2018b},  the property
$c_p \neq c_p(k)$ is only held for high wavenumbers in the considered case.

\subsection{\label{sec:trajec} Wave train trajectory}

As demonstrated in Section \ref{sec:disper}, wave breaking 
\replaced{can cause high-frequency harmonics to propagate at a}{may lead to propagation of the high-frequency harmonics at the} speed appreciably greater than that
defined by the linear dispersion relation (\ref{eq:disprel}).
Under these circumstances, short wave components \deleted{length components} propagate together
with \deleted{the} longer ones, \added{thus} preventing the dispersive spreading of \deleted{the} wave packet.
Such \replaced{an}{wave} evolution \replaced{can lead to}{is expected to result in} the distortion of \deleted{the} wave train shape and spatio-temporal energy redistribution.
\replaced{Besides}{Alongside with that}, the propagation speed of the entire wave packet
can also be influenced by \deleted{the} wave breaking nonlinearities.
The \added{propagation} speed \replaced{of}{at which the} wave packet energy \replaced{in}{propagates through} space is
defined by the group velocity.
\myhighlight{Consider group velocity of the carrier (peak) frequency harmonic by
differentiation of (\ref{eq:dispcorr2}):}
\begin{eqnarray}
   && c_{g0} = 
      \left. \frac{\partial \omega}{\partial k} \right|_{k=k_0} \approx
      c_{g0, BEM\nu}\left(
      1 + \Theta_T \Xi \left[ \hat{F}_{\hat{x}} \right]^{\Theta_F}
      \right)
   \label{eq:cgcorr}
\end{eqnarray}
The expression \eqref{eq:cgcorr} suggests \replaced{that}{growth of} the wave train propagation velocity
\deleted{with} increases \replaced{with}{of the wave} breaking intensity defined by the energy
dissipation rate $\hat{F}_{\hat{x}}$. In this section, the dynamics of
the wave packet evolution is studied to \replaced{verify}{validate} these hypotheses.

The propagation velocity of a wave train can be evaluated by analysing
\deleted{the shape of} the surface elevation envelope \deleted{that is to  be} obtained
from \deleted{the results of the} numerical simulations.
At each instant $t$, the Hilbert transform is applied to calculate
wave train envelopes from free wave surface elevations $\eta(x,t)$
obtained \replaced{by the}{from} BEM$\nu$ and VOF \replaced{models}{computations}:
\begin{eqnarray}
   && \mathscr{H}(x) = \left| \eta(x) +
             \frac{i}{\pi} \text{PV} \int_{-\infty}^{+\infty}
             \frac{\eta (\chi)}{\chi - x} d \chi
             \right|,
   \label{eq:hilb}
\end{eqnarray}
and \replaced{the}{its} dimensionless form is given by:
\begin{eqnarray}
   && \tilde{\mathscr{H}} (x,t) = \frac{\mathscr{H} (x,t)}
      {\underset{x,t}{\max}\left(\mathscr{H} (x,t)\right)}
   \label{eq:hilbenerg}
\end{eqnarray}
In the limit of linear system, the velocity of a Gaussian wave train
is determined by the propagation speed of the envelope maximum. When waves
become highly nonlinear, the instantaneous envelope is disturbed by the fast
variations due to nonlinear interactions, which redistribute the energy
$\tilde{\mathscr{H}}^2$ within the wave train. Since these variations do
not determine the mean velocity of the energy propagation, the linear
approach is not applicable.
There is no strict method yet to determine the propagation speed
of a strongly nonlinear wave packet. In this study the wave train
trajectory in the ${x-t}$ space is evaluated by using a central
weighting of the energy distribution:
\begin{eqnarray}
   && X_H(t) = \frac{\sum\limits_{x} x \tilde{\mathscr{H}}^2 (x,t)}
                    {\sum\limits_{x} \tilde{\mathscr{H}}^2 (x,t)},
   \label{eq:hilbtraj}
\end{eqnarray}
where $X_H(t)$ is the instantaneous wave train coordinate.
The instantaneous propagation speed of the wave train is then
$v(t) = dX_H / dt$.

\begin{figure}
\centering
\begin{overpic}[width=0.45\linewidth]{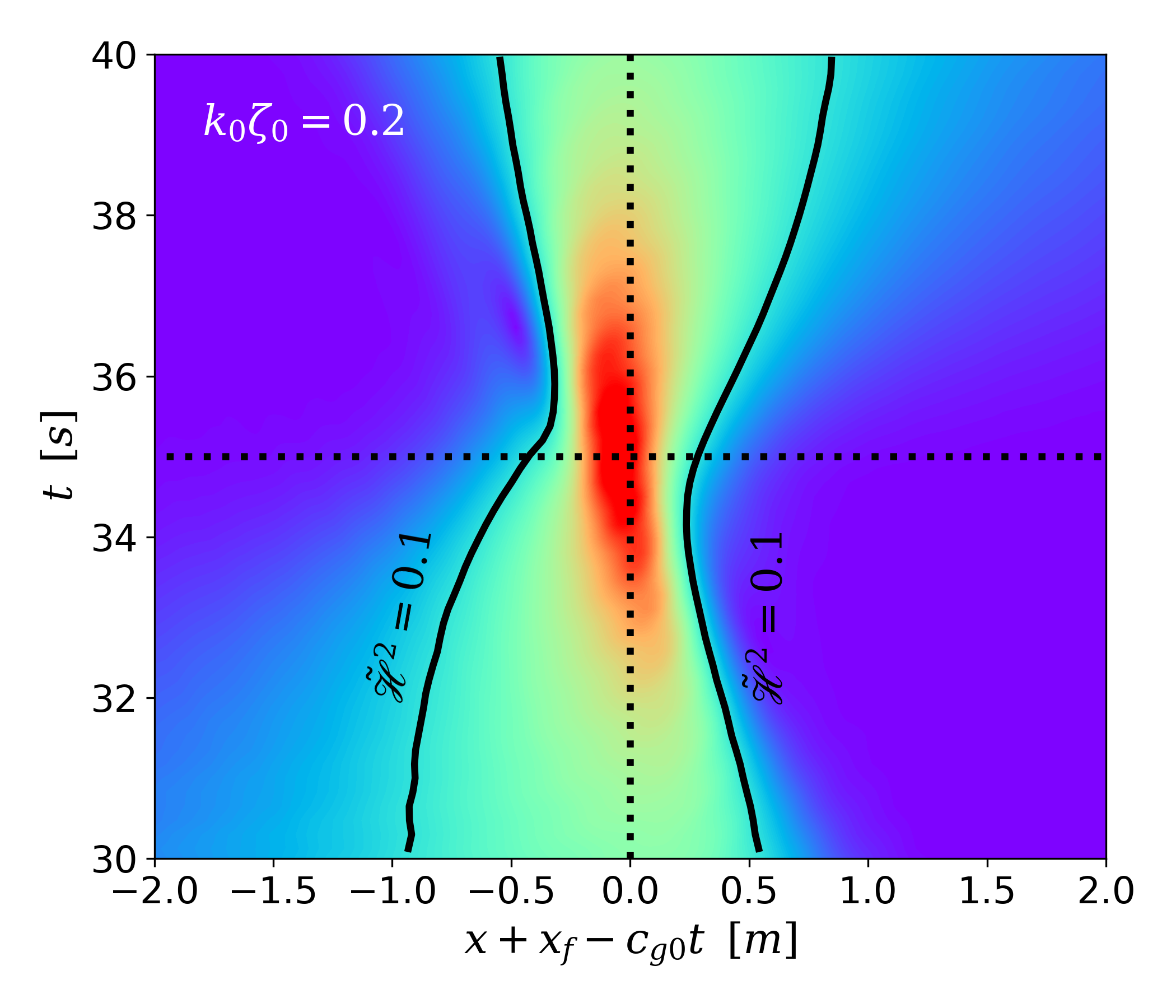}
    \put(85,18){\textbf{\color{white}(a)}}
\end{overpic}
\begin{overpic}[width=0.45\linewidth]{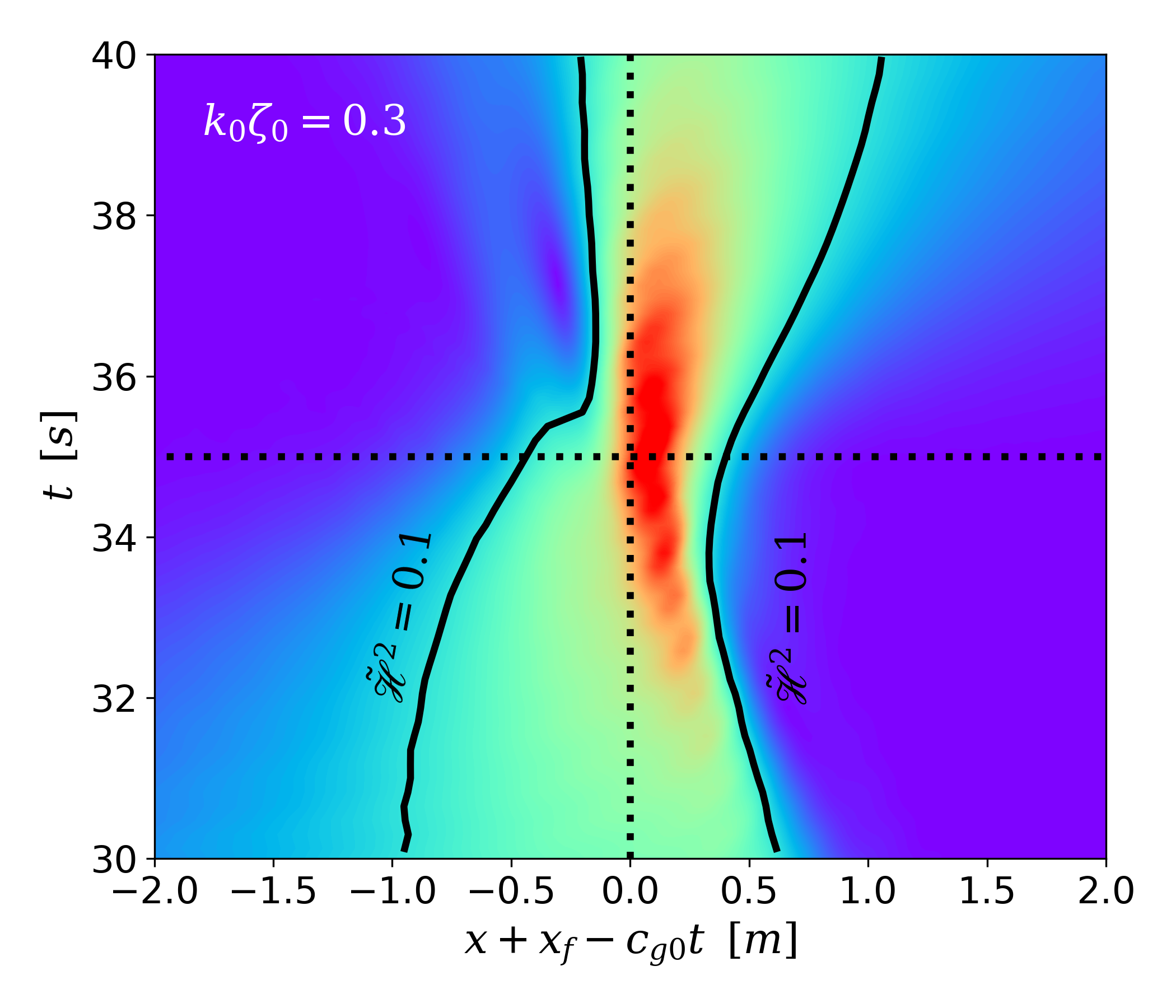}
    \put(85,18){\textbf{\color{white}(b)}}
\end{overpic}\\
\begin{overpic}[width=0.45\linewidth]{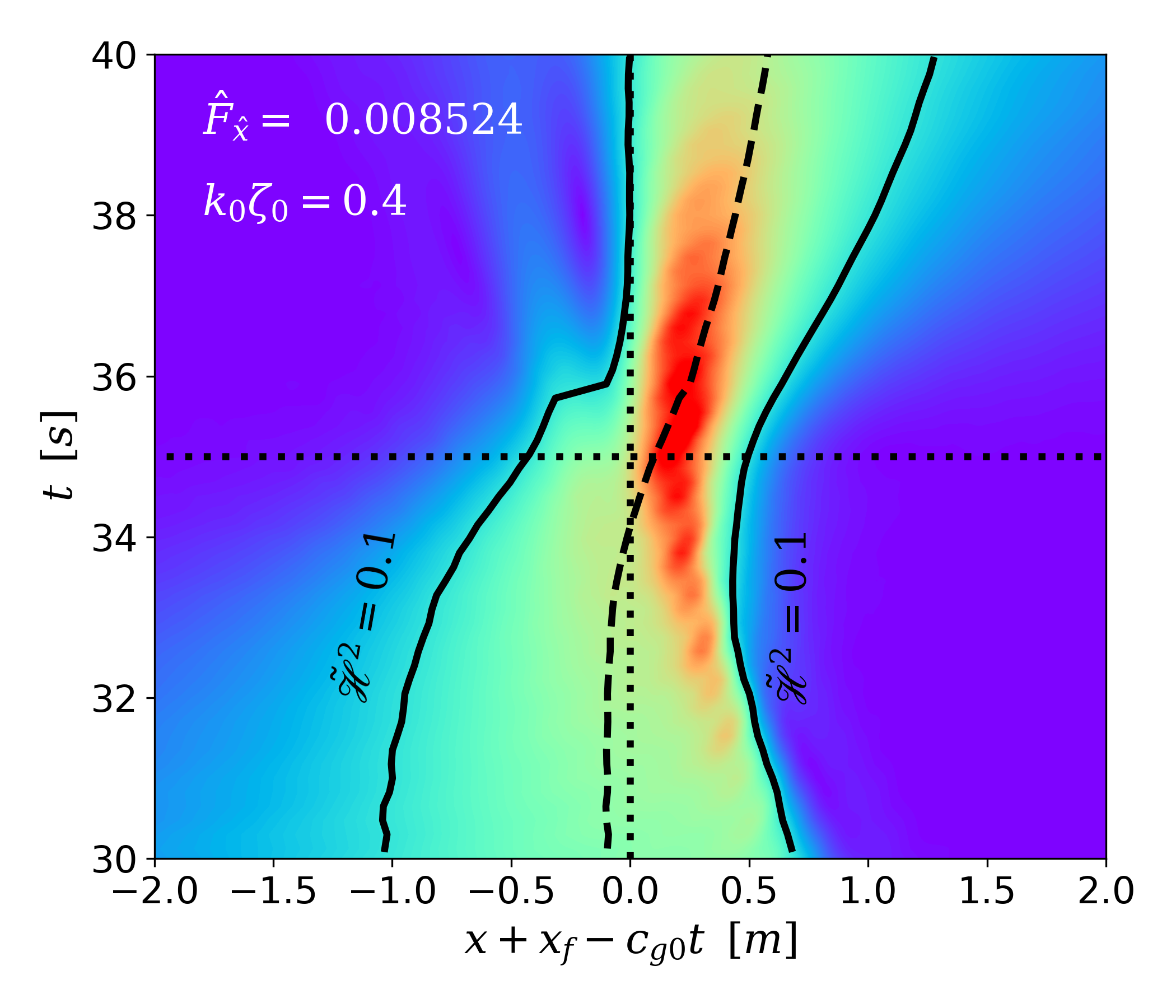}
    \put(85,18){\textbf{\color{white}(c)}}
\end{overpic}
\begin{overpic}[width=0.45\linewidth]{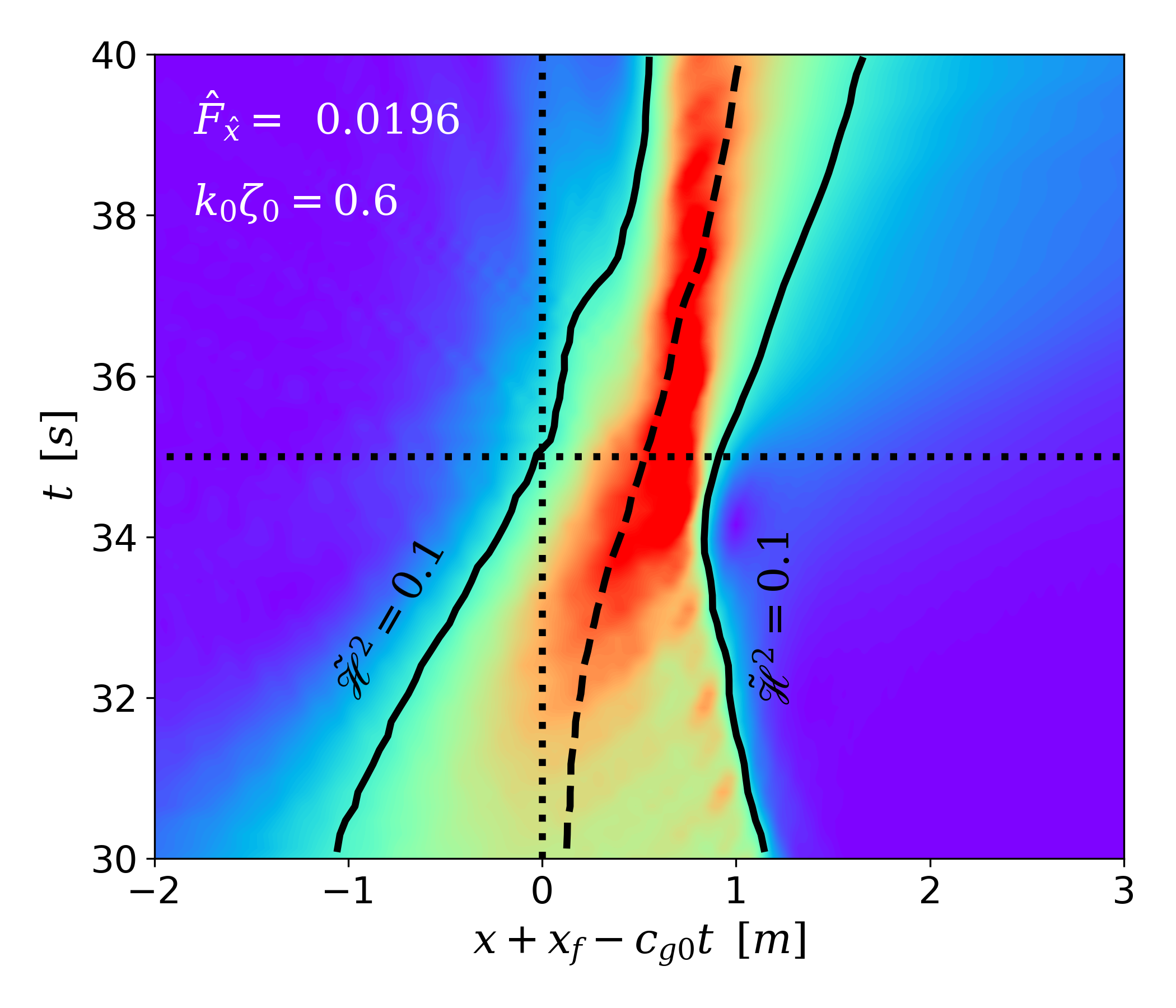}
    \put(85,18){\textbf{\color{white}(d)}}
\end{overpic}\\
\begin{overpic}[width=0.45\linewidth]{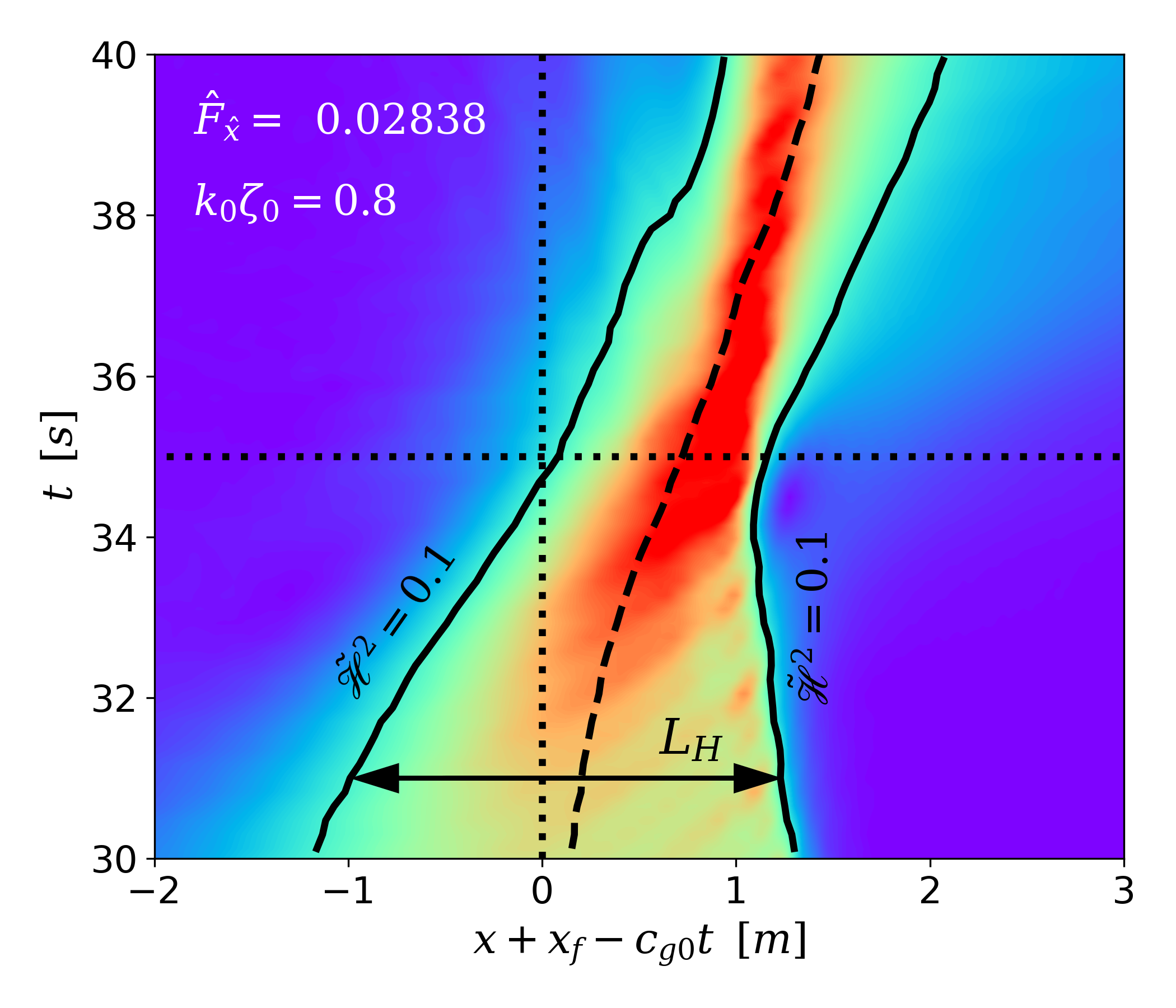}
    \put(85,18){\textbf{\color{white}(e)}}
\end{overpic}
\begin{overpic}[width=0.45\linewidth]{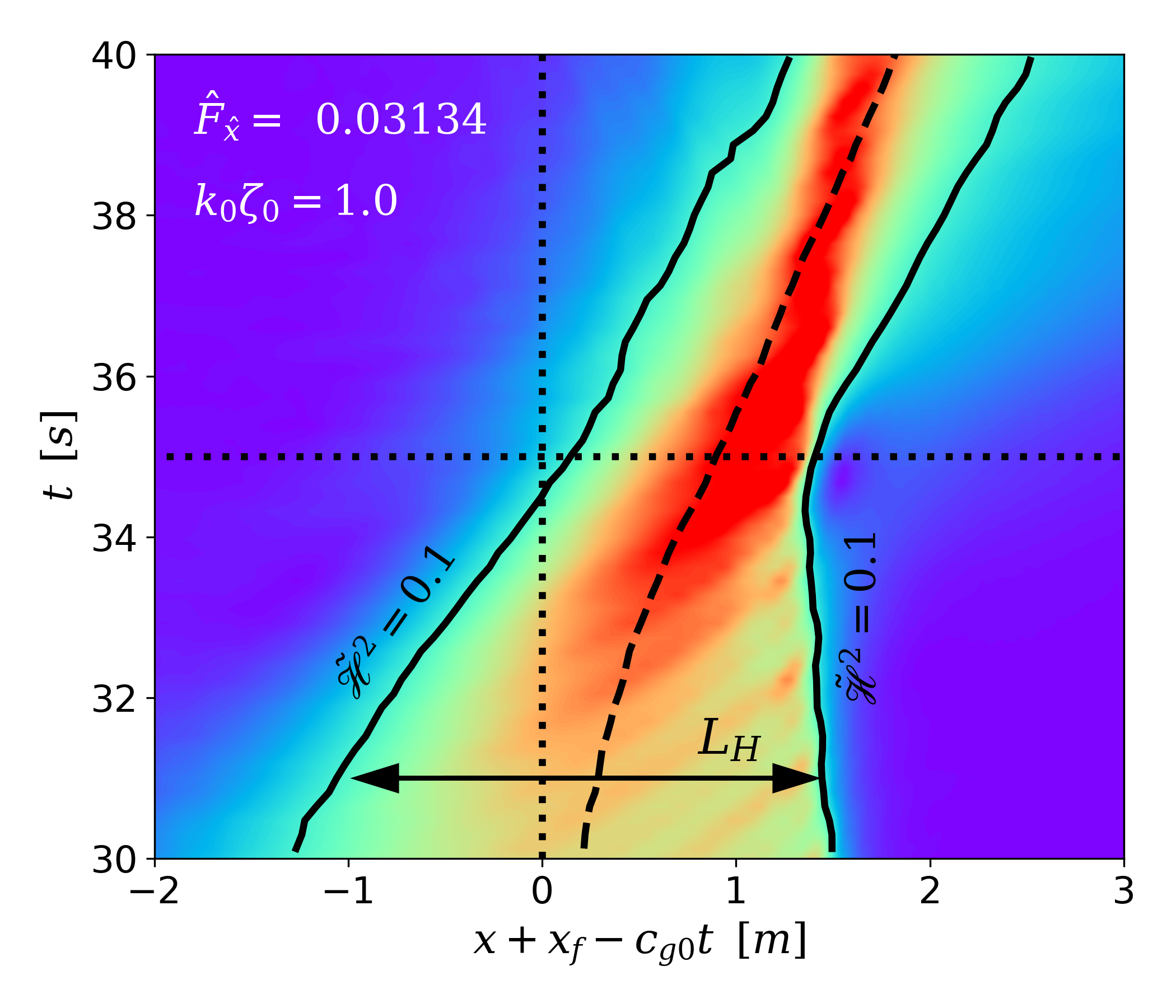}
    \put(85,18){\textbf{\color{white}(f)}}
\end{overpic}
\caption{Surface elevation envelope $\tilde{\mathscr{H}} (x,t)$
         (\ref{eq:hilbenerg}) in $x$--$t$ space obtained from the
         results of BEM$\nu$ computations: (a)-(f) present the wave
         trains of different steepness $k_0\zeta_0$.
         The color scale has the range
         $0 \leq \tilde{\mathscr{H}} \leq 0.8$; higher values
         ($\tilde{\mathscr{H}} > 0.8$) are shown by red color.
         The boundaries of the wave train corresponding to the energy
         level $\tilde{\mathscr{H}}^2 = 0.1$ are plotted by solid lines.
         Dotted lines depict the linear focal point location. The
         approximate wave train coordinates $X_H(t)$ (\ref{eq:hilbtraj})
         are plotted by dashed lines. The instantaneous wave train
         length $L_H(t)$ is indicated in panels (e) and (f).}
\label{fig:trbem}
\end{figure}

The distribution of $\tilde{\mathscr{H}} (x,t)$ obtained by the BEM$\nu$ model is plotted in Figure~\ref{fig:trbem} for wave trains with different steepness.
Note that the $x$ coordinates are displaced by
the linear focus location $x_f$ and transformed using the linear group
velocity of the wave packet. For the purpose of analysis, the range
of $x$ coordinates containing the dominant portion of the energy and
determining the wave train boundaries is chosen by using the condition
$\tilde{\mathscr{H}}^2 \geq 0.1$ as depicted by solid lines
in Figure~\ref{fig:trbem}.
The distance between these boundaries can be used as a measure \replaced{of}{for} the
wave train instantaneous length $L_H(t)$,
see panels (e) and (f).

The spectrum of weakly-nonlinear wave train varies slowly so that 
its shape is conserved within the time intervals considered in the study.
The peak values and the spectral-weighted group velocities (\ref{eq:cgs})
are thus very close to each other, i.e. $c_{g0} \approx c_{gs}$.
Consequently the value of $c_{g0}$ gives a reasonable estimation of the wave
packet propagation speed. Thus the wave train trajectory is aligned
with the vertical axis as shown in Figure~\ref{fig:trbem} (a), where
the wave train has a low steepness $k_0\zeta_0 = 0.2$.
The actual focal point for this case practically coincides with the linear
prediction shown in the figure by dotted lines.
Increasing the steepness to $k_0\zeta_0 \geq 0.3$ changes the free surface elevation envelope. The wave train trajectory $X_H(t)$
(\ref{eq:hilbtraj}) is plotted by the dashed lines in
Figures~\ref{fig:trbem} (c)-(f). It is shown that the intensification of 
wave breaking (growth of $k_0\zeta_0$ and $\hat{F}_{\hat{x}}$)
leads to the inclination of wave train  trajectory $X_H(t)$ from
the vertical line, cf. panels (a)-(f). For the cases with relatively strong
energy dissipation, i.e. $k_0\zeta_0 \geq 0.6$, the plots
of $X_H(t)$ after the focal point are almost linear. This suggests that for each of these cases, breaking
increases the wave packet propagation velocity $v(t) = dX_H / dt$ by a constant
value equal to the gradient of the straight line.

The evolution of weakly-nonlinear wave train, see Figure~\ref{fig:trbem} (a),
consists of two consecutive stages: \added{(I)} focusing within \added{the} time interval
$t = 30 - 35 \text{ s}$, and \added{(II)} defocusing (dispersive spreading) within
$t = 35 - 40 \text{ s}$. Breaking is only expected to occur 
in the vicinity of the focal point at $t = 35 \text{ s}$. Although the surface elevation
envelopes of moderately nonlinear wave packets are distorted, both the focusing and defocusing stages are still clearly present, cf. Figures~\ref{fig:trbem} (b) and (c).
However, the occurrence of multiple stronger breakers modifies significantly the envelope evolution process for the cases with high steepness $k_0\zeta_0 \geq 0.6$, for which a defocusing stage is no longer \replaced{clearly visible}{present as shown} in panels (d)-(f).

In view of similarity in the propagation regime of breaking wave trains,
\added{we} consider the strongest energy dissipation case shown in
Figure~\ref{fig:trbem} (f). \replaced{This}{Note that this} wave train exhibits appreciable
energy dissipation due to \added{the strong and nearly continuous} breaking within the full time range studied here,
as clearly \replaced{shown in}{seen from} Figures~\ref{fig:steepness}~and~\ref{fig:bemnu}.
Despite this, the focusing stage is still largely pronounced.
On the contrary, surprisingly, the spreading of wave train in the post-focusing
stage is completely suppressed, \deleted{which is kept for all breaking wave
trains,} cf. Figures~\ref{fig:trbem} (d)-(f). Therefore, the reduction
of \deleted{the} wave height due to dispersion is less pronounced than the low steepness cases presented in panels (a)-(c). \deleted{that results
in occurrence of extreme waves during longer time intervals.}
This suggests that once extreme breaking waves appear, they propagate
together without spreading until breaking dissipates the excessive
energy and \replaced{ceases}{stops}. \added{It also implies that
	such extreme waves could last longer than expected.}
\replaced{Having a close look at the red areas shown in Figure \ref{fig:trbem}, where  $\tilde{\mathscr{H}} (x,t) \geq 0.8$, we can clearly see their growth in time with wave train steepness especially for $k_0\zeta_0 \geq 0.6$. For low steepness wave trains $k_0\zeta_0 \leq 0.4$, the wave height reduces to less than 50\% of the maximum height by time $t=38$ s. But for highly nonlinear waves $k_0\zeta_0 \geq 0.6$, the red regions can extend beyond $t=40$ s. This means that increasingly intensified breaking can prolong the lifespan of extreme waves, and such an effect may not necessarily be trivial. }
{
This observation is seen as elongation of the
highest waves region defined by $\tilde{\mathscr{H}} (x,t) \geq 0.8$,
when comparing Figures~\ref{fig:trbem} (a)-(c) with Figures ~\ref{fig:trbem} (d)-(f).
Extreme waves may exist during longer time intervals in presence
of breaking as compared to the non-breaking environment.}

\begin{figure}
\centering
\begin{overpic}[width=0.48\linewidth]{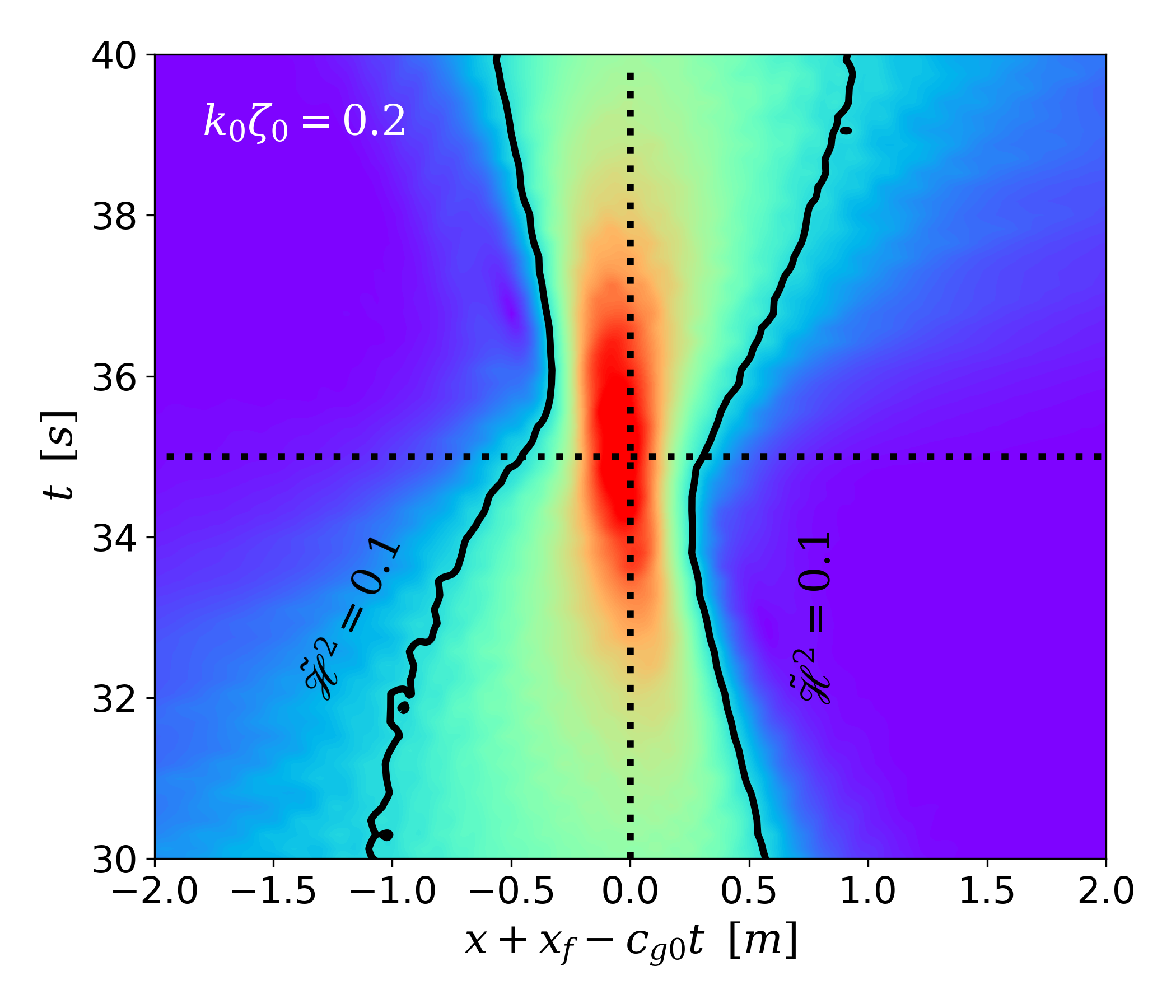}
    \put(85,18){\textbf{\color{white}(a)}}
\end{overpic}
\begin{overpic}[width=0.48\linewidth]{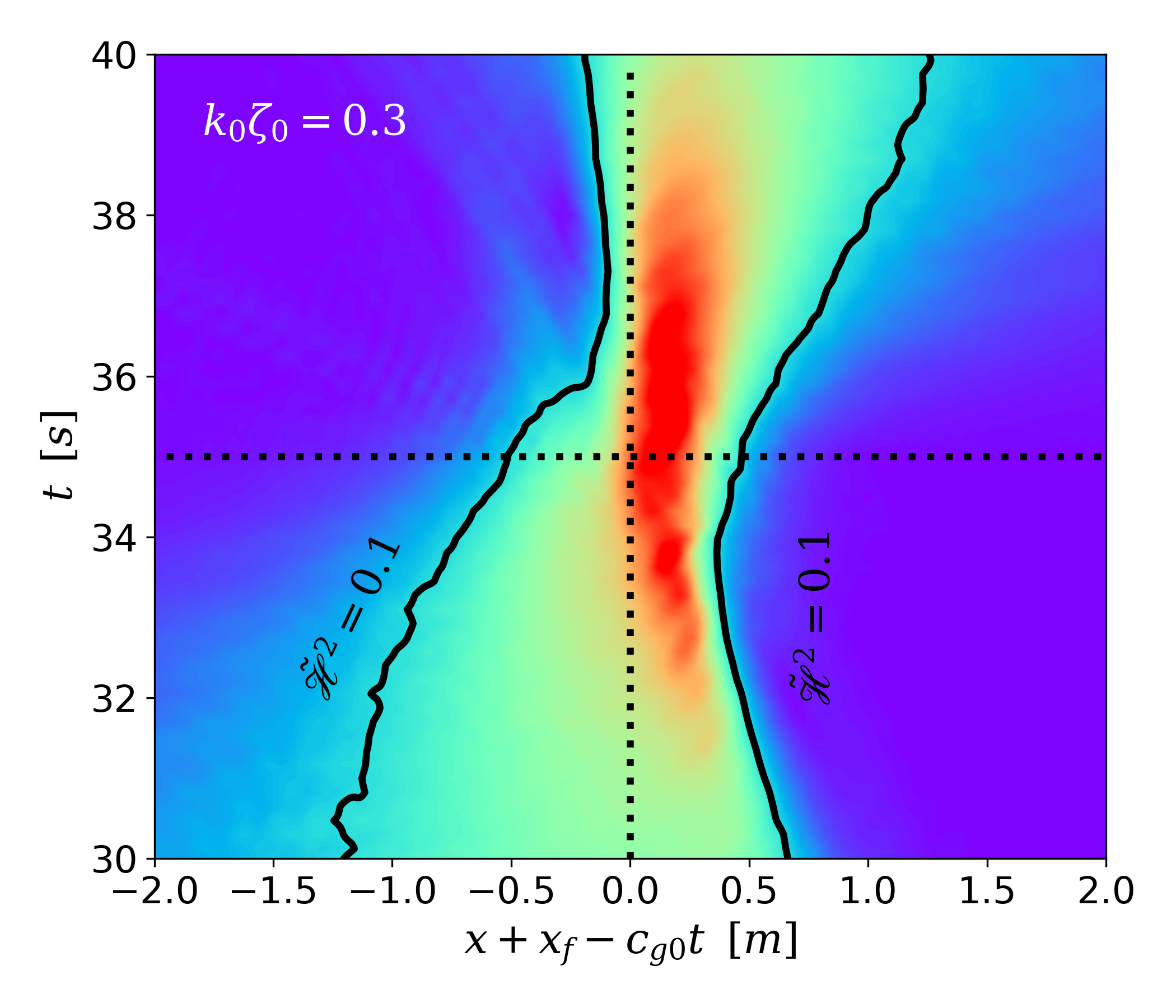}
    \put(85,18){\textbf{\color{white}(b)}}
\end{overpic}\\
\begin{overpic}[width=0.48\linewidth]{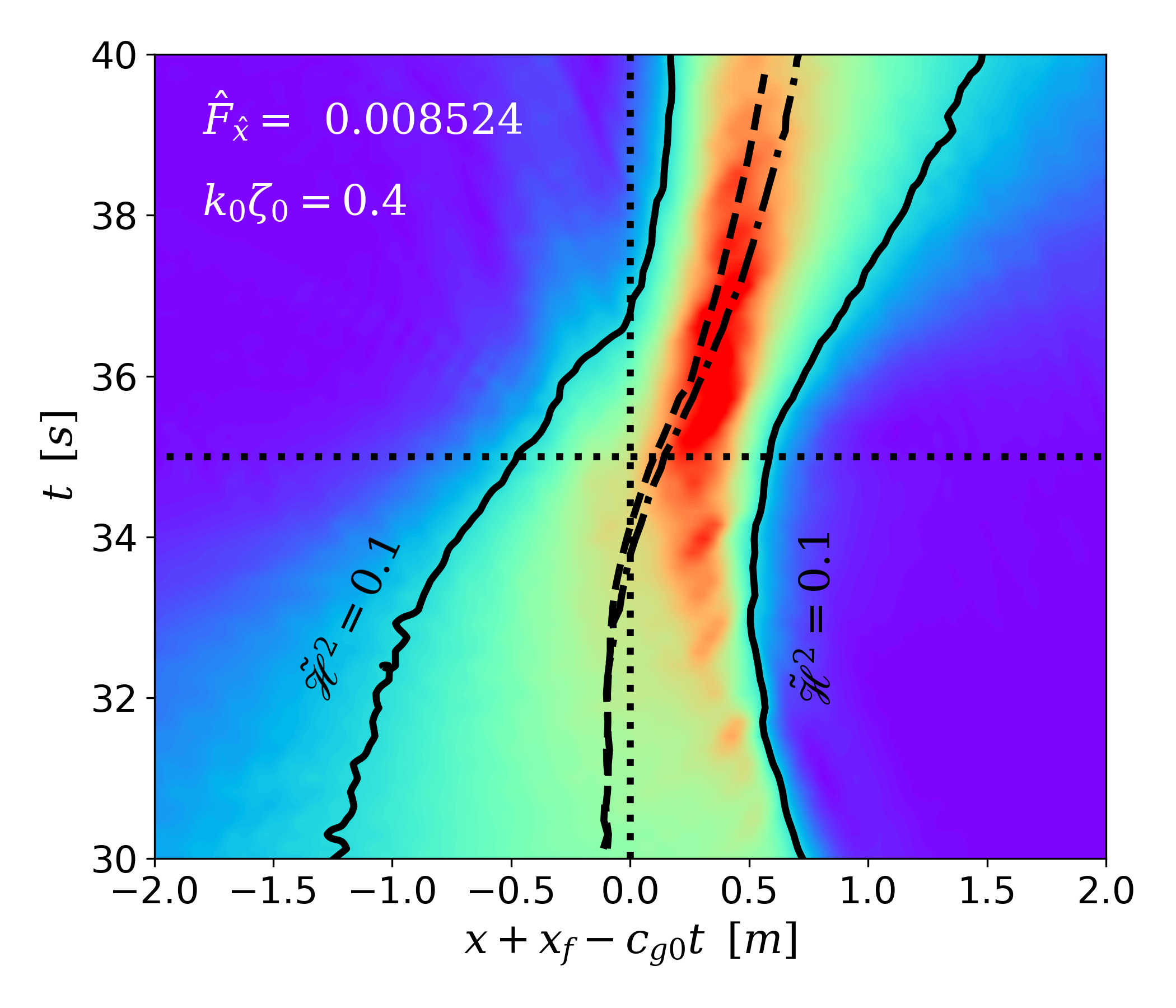}
    \put(85,18){\textbf{\color{white}(c)}}
\end{overpic}
\begin{overpic}[width=0.48\linewidth]{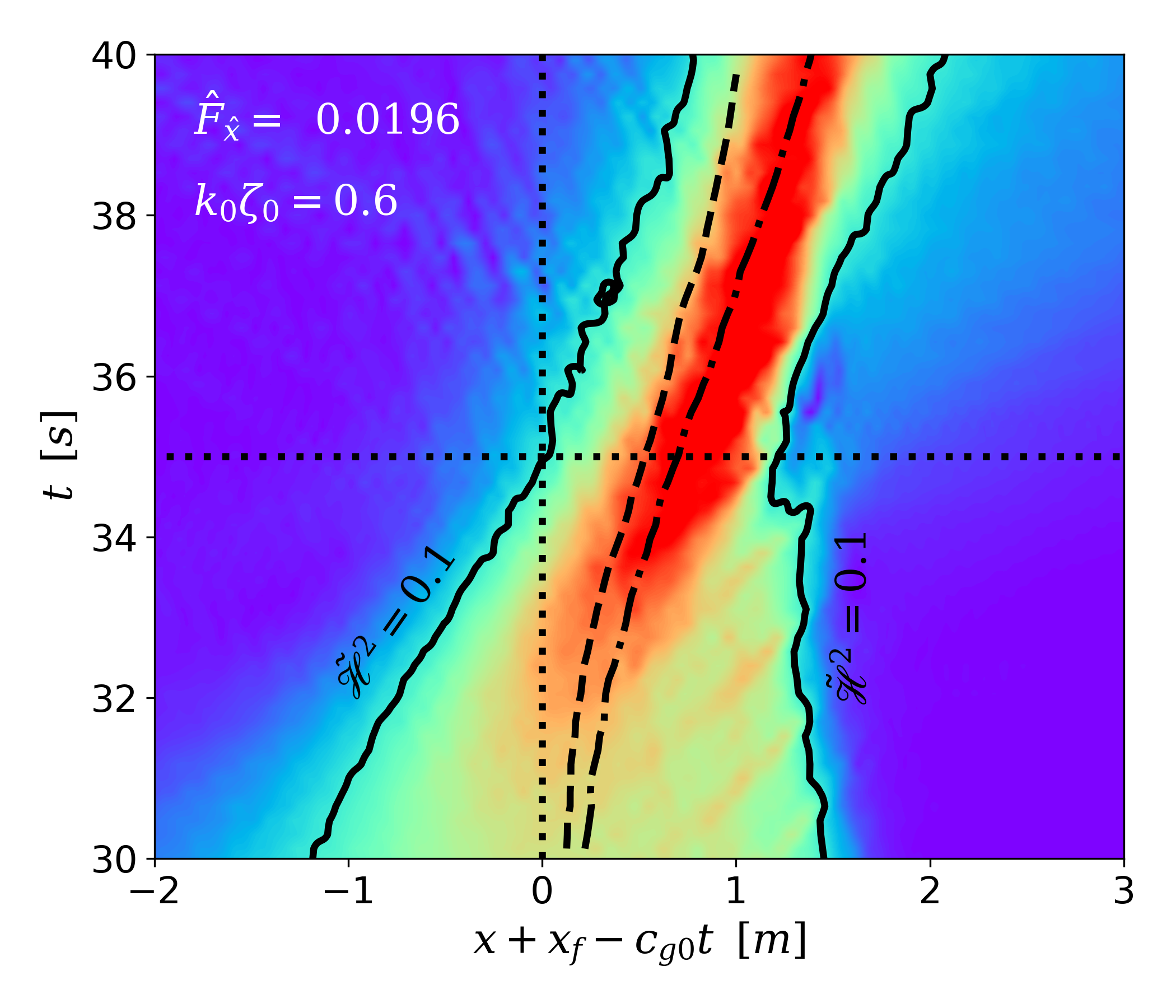}
    \put(85,18){\textbf{\color{white}(d)}}
\end{overpic}\\
\begin{overpic}[width=0.48\linewidth]{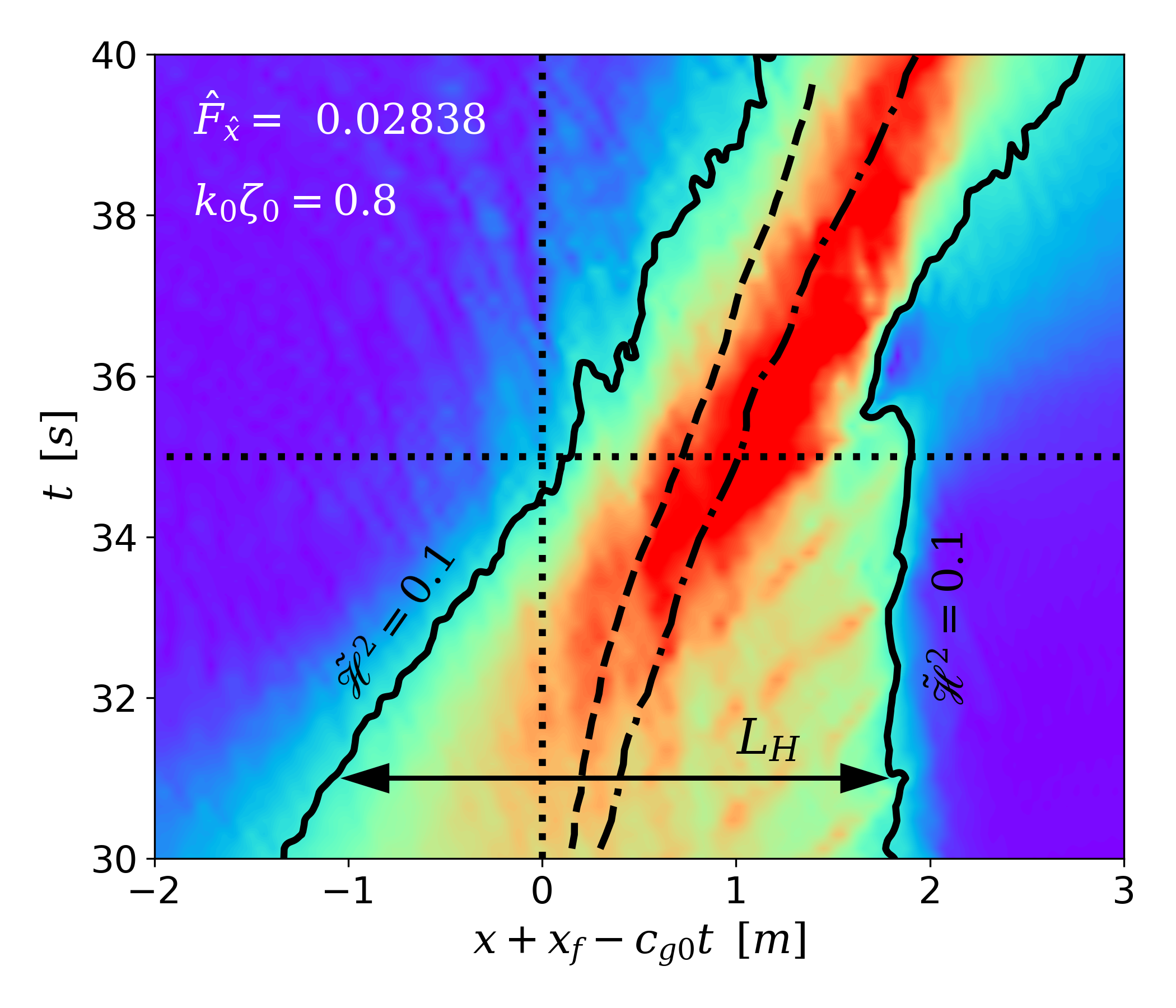}
    \put(85,18){\textbf{\color{white}(e)}}
\end{overpic}
\begin{overpic}[width=0.48\linewidth]{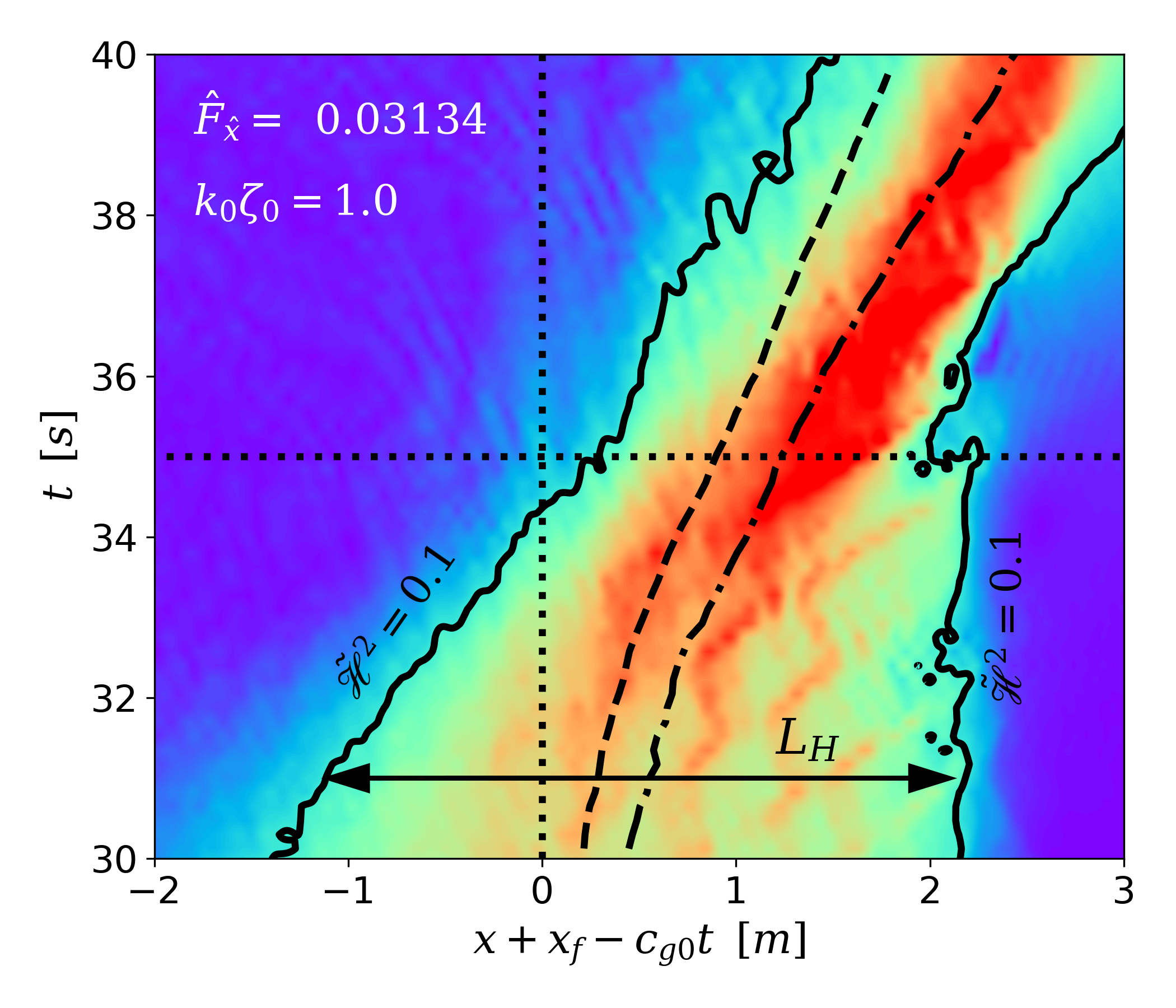}
    \put(85,18){\textbf{\color{white}(f)}}
\end{overpic}
\caption{As in Figure~\ref{fig:trbem} for VOF model. Dashed and
         dash-dotted lines show $X_H(t)$ calculated from BEM$\nu$ and
         VOF simulation results, respectively.}
\label{fig:trfoam}
\end{figure}

\replaced{The surface elevation envelopes computed by the VOF model for non-breaking wave  packets ($k_0\zeta_0 = 0.2 \text{ and } 0.3$) coincide with the BEM$\nu$ results,
cf. Figures~\ref{fig:trbem} and \ref{fig:trfoam} panels (a) and (b).}
{Evolution of the surface elevation envelopes of the steep but non-breaking
wave packets ($k_0\zeta_0 = 0.2 \text{ and } 0.3$) seen in the results
of the VOF simulations coincide with that obtained by the BEM$\nu$
model.} \deleted{and thus it is not analysed in the following.}
The distributions of $\tilde{\mathscr{H}} (x,t)$ elicited from the VOF
simulations for \replaced{breaking wave trains with}{the wave trains with} $k_0\zeta_0 \geq 0.4$ \replaced{are}{is} presented in
Figure~\ref{fig:trfoam} (c)-(f). Similar to the \deleted{results of the} BEM$\nu$
computations, \added{the} evolution of non-breaking and weakly-breaking wave trains consists of both
dispersive focusing and defocusing stages, while \replaced{there is no clear defocusing stage for strongly-breaking cases in the domain of interest}{strongly-breaking cases
have no defocusing stage}, cf. Figures~\ref{fig:trbem} and \ref{fig:trfoam}.
\replaced{At}{In} the same time, the wave train boundaries obtained \replaced{by}{in} the VOF
\replaced{model}{simulations} become \replaced{rough}{not smooth} with \added{the} intensification of \replaced{breaking}{the breaking events}
\replaced{due to the}{because of} generation of rotational flow components.

The wave train trajectories $X_H(t)$ (\ref{eq:hilbtraj}) \replaced{obtained by the VOF model are}{computed based on the VOF simulations is} plotted in
Figure~\ref{fig:trfoam} with dash-dotted lines. Corresponding plots of $X_H(t)$ taken for
reference from Figure~\ref{fig:trbem} (BEM$\nu$) are shown by dashed
lines. In the focusing stage, the trajectories of weakly-breaking wave trains computed by the BEM$\nu$ and VOF models are very close to each other as shown in Figure~\ref{fig:trfoam} (c). Once breaking is initiated in the vicinity of the focal
point, a small divergence of the trajectories appears.
\replaced{For strong breaking events as shown in Figure~\ref{fig:trfoam}}{If breaking is strong, see panels} (d)-(f), the divergence in wave trajectory between the VOF and BEM$\nu$ models becomes appreciably large. Wave trains computed by the VOF model
propagate over longer distances and thus have higher
propagation speeds compared to the BEM$\nu$ outcomes.
\replaced{These observations confirm our initial assumption that breaking can increase the propagation speed of wave trains compared to non-breaking scenarios. And this}
{These observations confirm the initial assumption on higher propagation
speed $v$ of the wave train when breaking is present as compared to
the no-breaking environment, which} is associated with the phase
shifting phenomenon as demonstrated in (\ref{eq:cgcorr}).
The non-potential flow components generated by multiple breaking events,
not taken into account in the BEM$\nu$ computations, \replaced{could}{are assumed to}
be responsible for such speed growth.
For instance, the breaking-induced sheared current varying in both time
and space \citep{Melville2017, Melville2019} may have a significant
nonlinear effect on the propagation speed.

\begin{figure}
    \centering
    \includegraphics[width=0.6\textwidth]{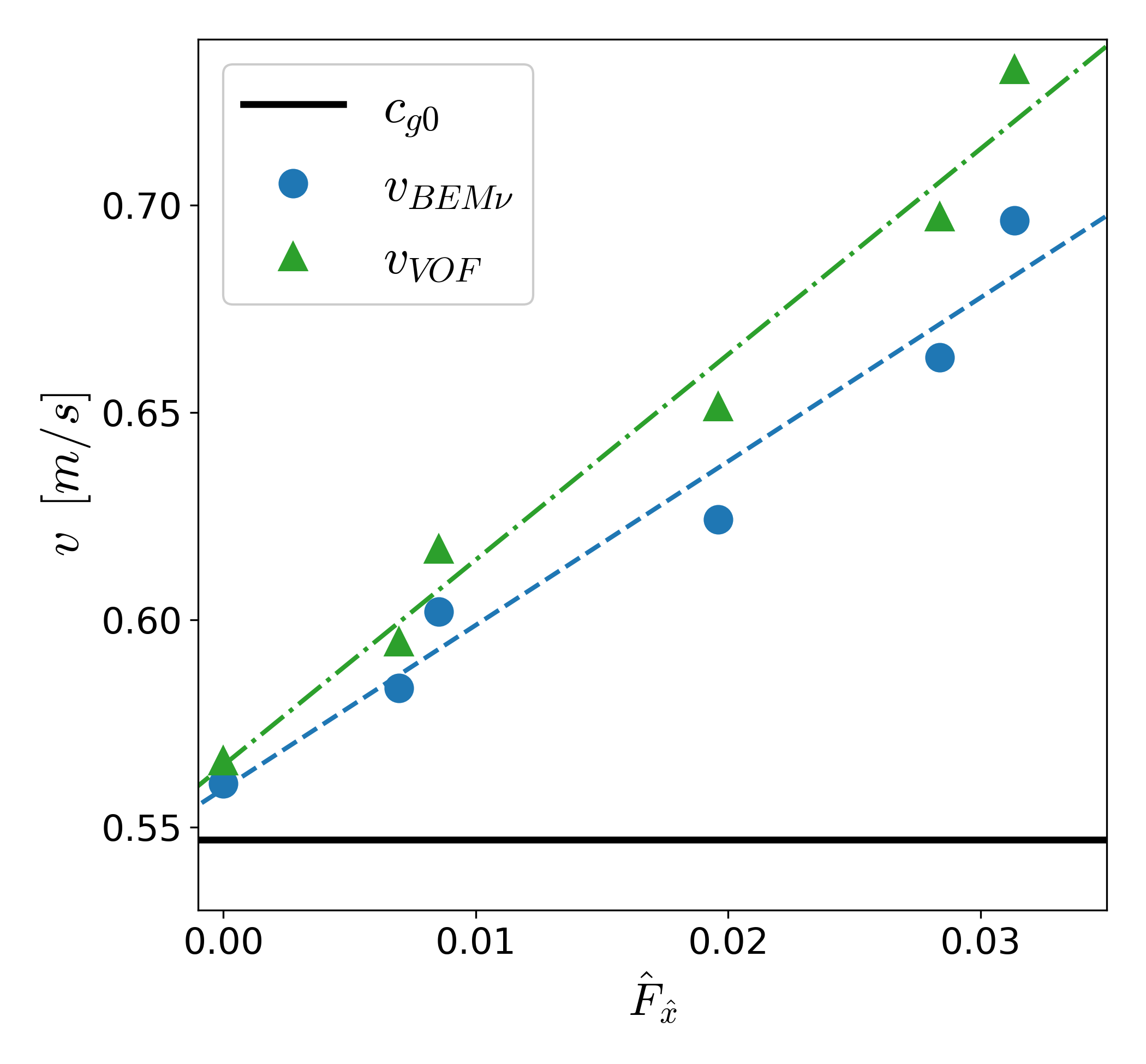}
    \caption{Dependence of the wave train propagation velocity 
             $v(t) = dX_H / dt$ on the energy dissipation rate
             $\hat{F}_{\hat{x}}$ (i.e. wave breaking strength). Dashed
             and dash-dotted lines represent the linear fit of the data
             derived from BEM$\nu$ and VOF computations.}
    \label{fig:cg}
\end{figure}

Applying linear regression to the trajectories $X_H(t)$ derived from
BEM$\nu$ and VOF computations, the wave group propagation velocity 
$v(t) = dX_H / dt$ is estimated numerically and plotted in
Figure~\ref{fig:cg}. The linear group velocity $c_{g0}$ calculated by using 
the spectral carrier frequency is included here for reference.
Strongly-nonlinear but non-breaking wave trains
(${\hat{F}_{\hat{x}} \approx 0}$) propagate at a speed moderately
higher than the linear group velocity $c_{g0}$. The growth of 
propagation speed in this case is caused by nonlinear resonant
interactions between the spectral harmonics.
A similar outcome has been obtained analytically by
\citet{Stuhlmeier2019} for nonlinear waves of
Pierson–Moskowitz spectra. Significant increases in the
propagation speed can be clearly seen for the breaking waves ($\hat{F}_{\hat{x}}>0$),
which may attain the following values: 
$v_{BEM\nu} \approx 1.27 c_{g0}$ and $v_{VOF} \approx 1.36 c_{g0}$.
While the growth observed in BEM$\nu$ computations is mostly related
to the nonlinearities in the free-surface boundary conditions
(\ref{eq:freebc1}) and (\ref{eq:freebc2}), the additional increment
of speed present in the VOF model is likely caused by  
breaking-induced rotational flows.

It is known that the balance of spectral energy in wave forecasting
models is closely dependent on \deleted{the value of} wave group velocity, which is
usually approximated \replaced{by the}{using} linear dispersion relation (\ref{eq:disprel}).
\citet{Stuhlmeier2019} \replaced{pointed out}{convinced} that \added{a} more
accurate nonlinear approximation \replaced{of}{for} the group velocity derived from
\added{the} Zakharov equation is \replaced{needed}{significant for incorporation to the wave
forecasting software}. \replaced{This means the complex effect of wave breaking on modifiying group velocity is of practical importance and needs to be taken carefully into account.}{This allows to assume that the breaking-induced enhancement of the wave propagation velocities discussed above
can have also an important effect.}

\begin{figure}
\centering
\begin{overpic}[width=0.48\linewidth]{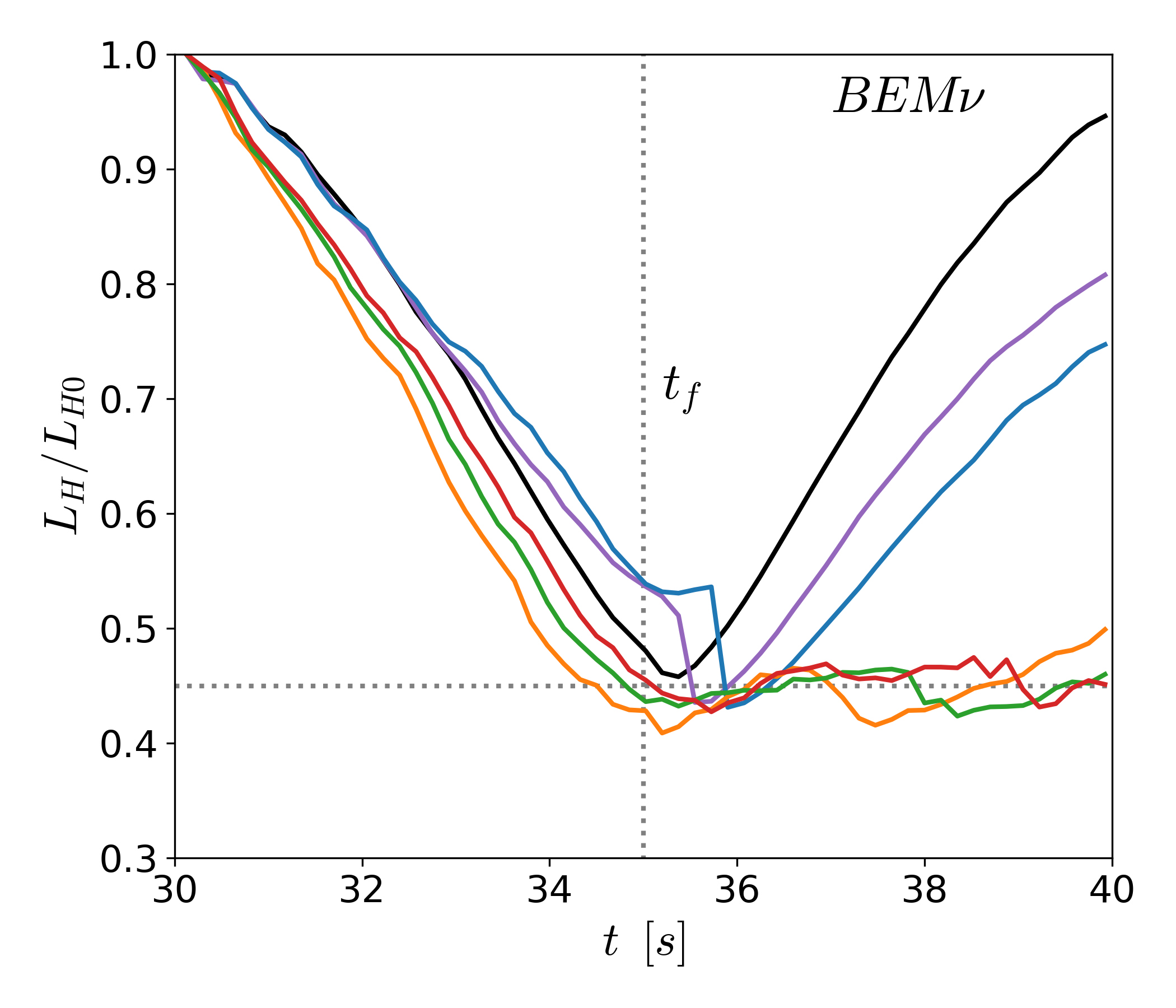}
    \put(0,76){\textbf{(a)}}
\end{overpic}
\begin{overpic}[width=0.48\linewidth]{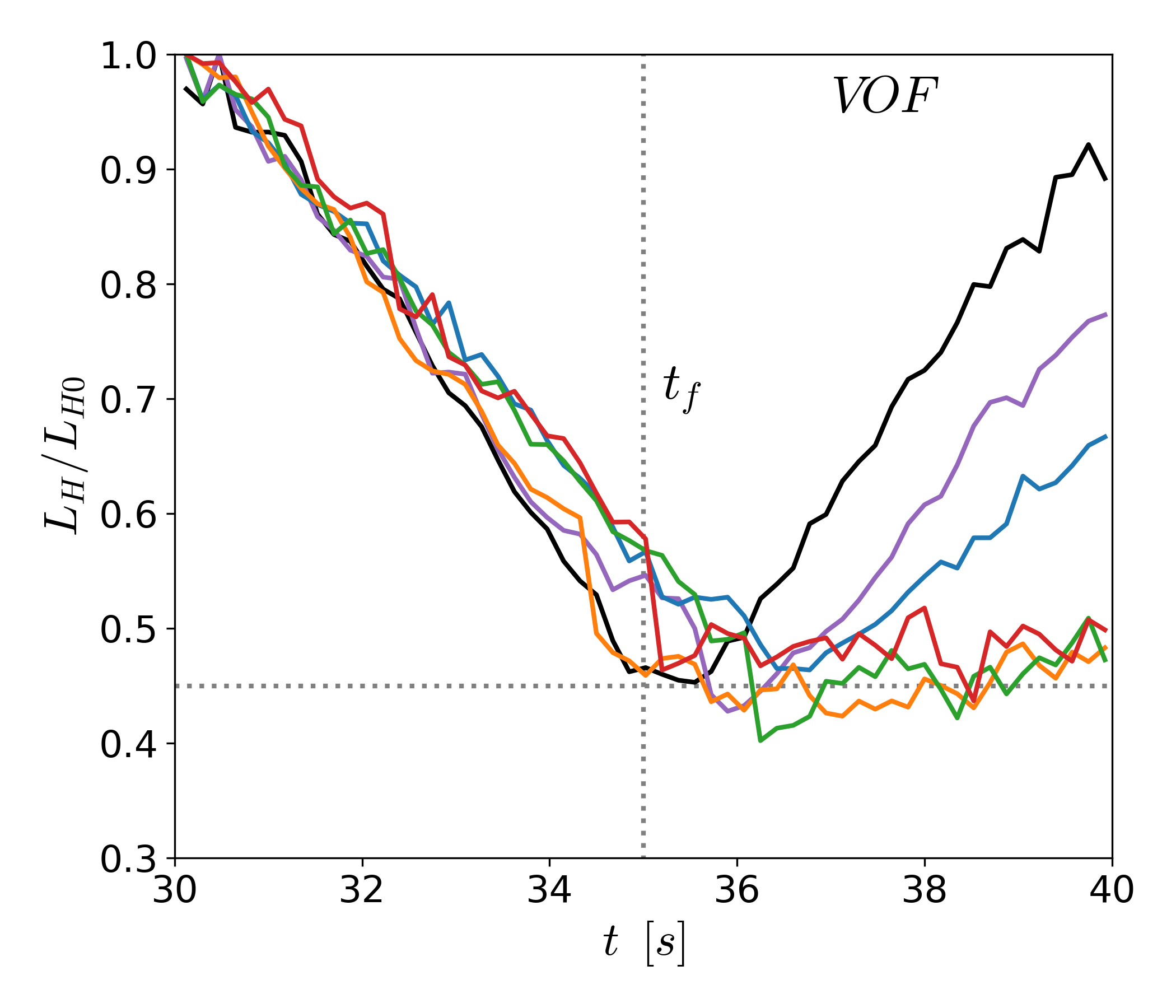}
    \put(0,76){\textbf{(b)}}
\end{overpic}
\includegraphics[width=1\textwidth]{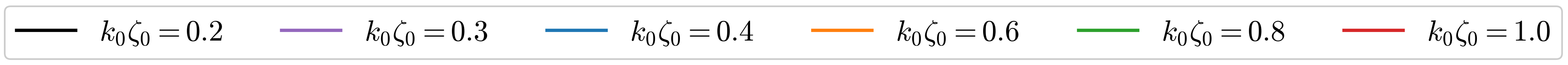}
\caption{Evolution in time of the \replaced[id=ZM]{normalised}{normalized} length of the non-breaking
         to strongly-breaking wave trains: (a) BEM$\nu$ and (b) VOF
         computations. Definition of the wave train length is shown
         schematically in figures~\ref{fig:trbem} and \ref{fig:trfoam};
         $L_{H0}$ is the initial wave train length.}
\label{fig:length}
\end{figure}

Figure~\ref{fig:length} exhibits \added{the temporal} evolution \deleted{in time} of wave group length $L_H$, which is defined as the distance between the leading and
 trailing edges of a wave train, as shown in Figures \ref{fig:trbem}
and \ref{fig:trfoam}. The wave group length is \replaced[id=ZM]{normalised}{normalized} by its \deleted{initial}
value taken at $t=30 \; \text{s}$ when wave generation is completed in
both BEM$\nu$ and VOF models.
As discussed above, all non-breaking and weakly-breaking wave trains
(${0.2 \leq k_0\zeta_0 \leq 0.4}$) are subject to consequent focusing
and defocusing stages due to \deleted{waves} dispersion. \replaced{For a strong breaking wave with}{When the breaking is
initiated} ${k_0\zeta_0 \geq 0.6}$, the defocusing stage is
suppressed, cf. Figure~\ref{fig:length} (a) and (b).
More importantly, the relative wave packet lengths after \deleted{its} focusing
are \replaced{nearly}{maintained} constant and have similar \replaced{magnitude}{value} in both BEM$\nu$ and VOF
computations: $L_H / L_{H0} = \text{const} \in [0.45, 0.5]$.
The wave packet spreading is expected to resume after the breaking process
is completed. \replaced{However}{In the same time}, the absolute values of wave group length 
\replaced{computed by the BEM$\nu$ and VOF models are different}{between BEM$\nu$ and VOF results} because of the wave evolution peculiarities during the focusing stage.

\begin{figure}
\centering
\begin{overpic}[width=0.8\linewidth]{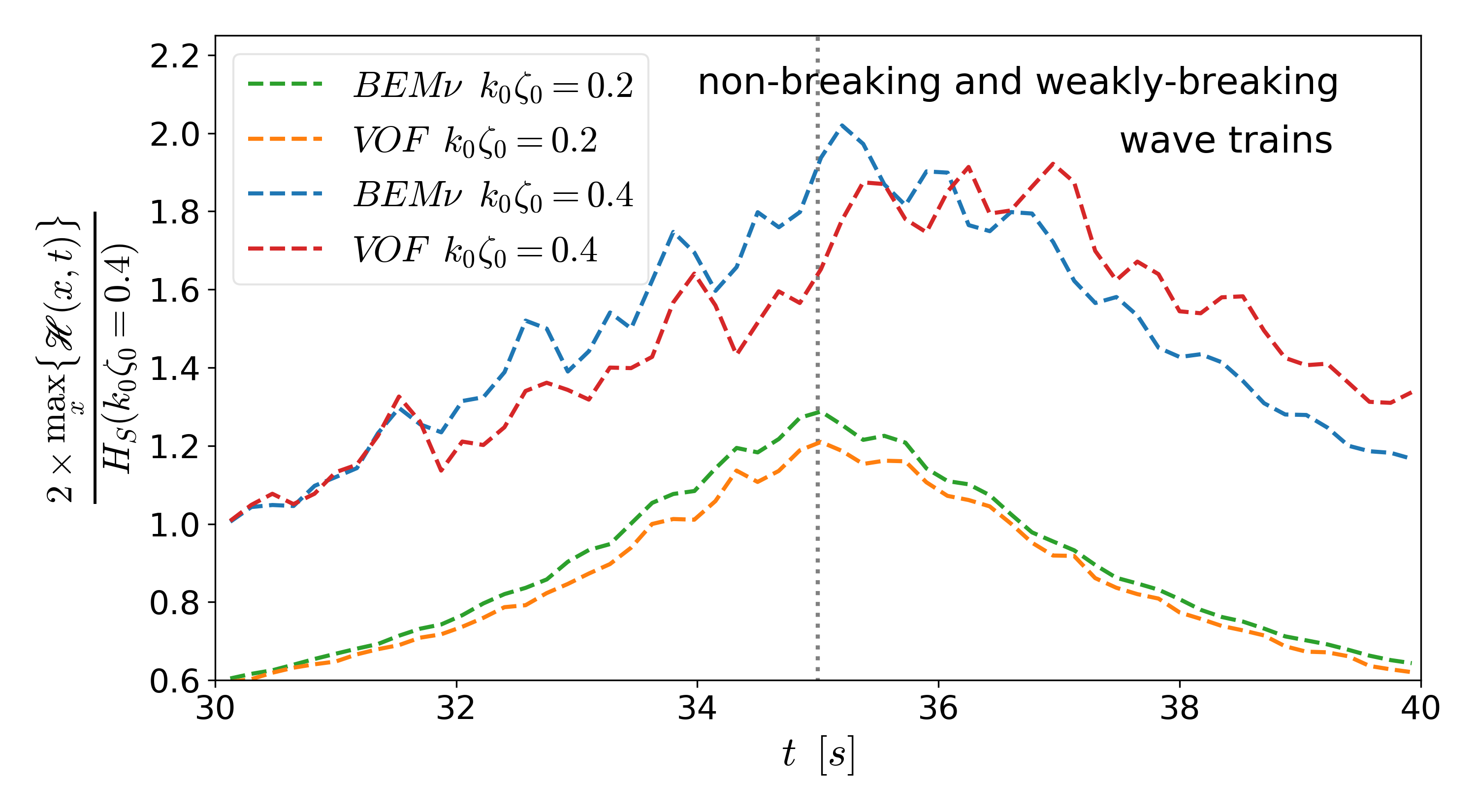}
    \put(2,50){\textbf{(a)}}
\end{overpic}\\
\begin{overpic}[width=0.8\linewidth]{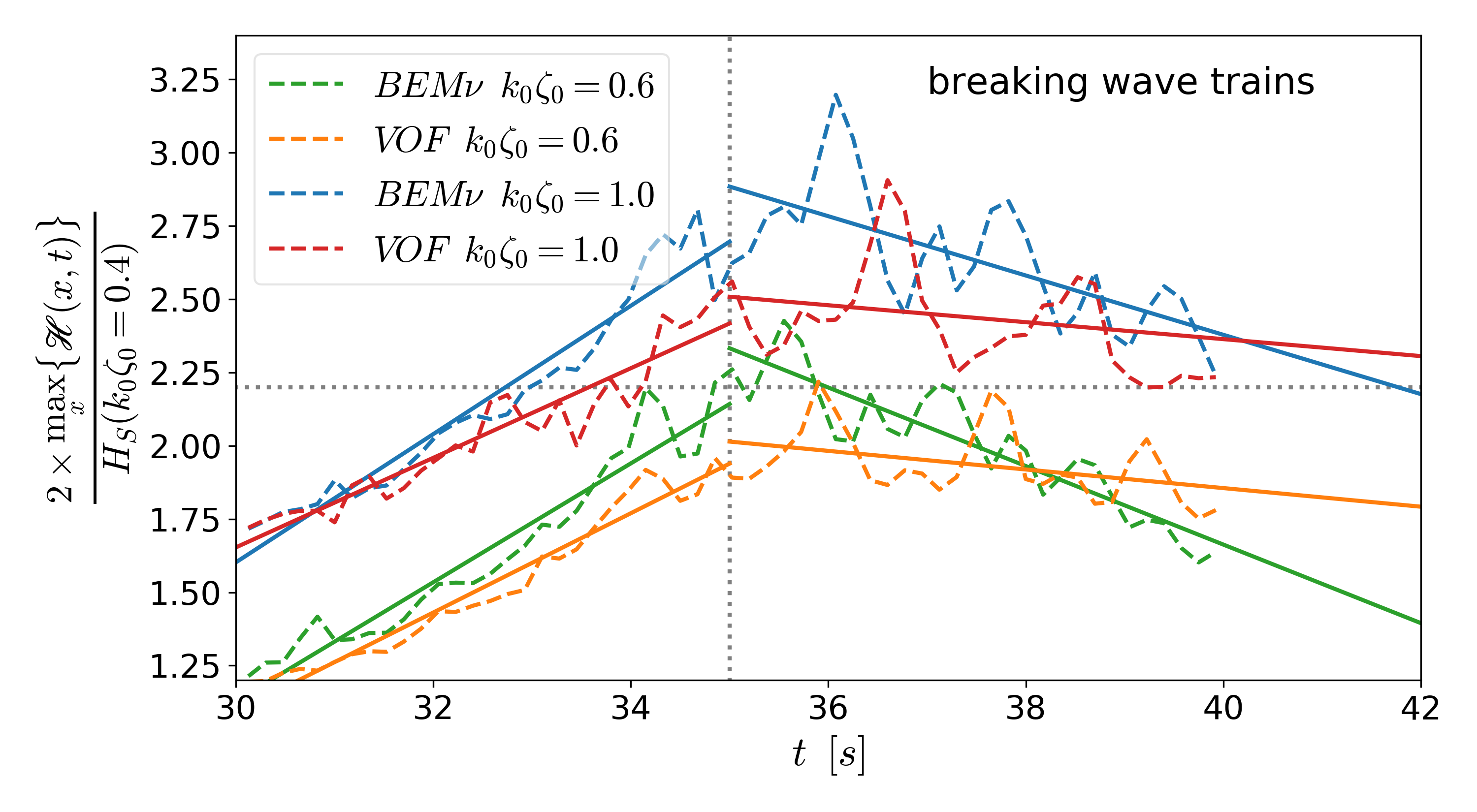}
    \put(2,50){\textbf{(b)}}
\end{overpic}
\caption{Evolution of the amplification factor $\tilde{H}$
         (\ref{eq:amplif}) observed in BEM$\nu$ and VOF computations:
         (a) non- and weakly-breaking wave trains, (b) strongly-breaking wave trains.
         Dashed lines present the raw data derived from
         the computations, while solid lines are obtained by the linear
         interpolation of the plots separately at focusing and
         defocusing stages. Vertical dotted line shows focal point
         location at $t = 35 \text{s}$.}
\label{fig:hilbmax}
\end{figure}

\replaced{It has been reported that wind has a prounced effect of amplifying wave height for broad-banded focusing wave groups in the defocusing stage \citep{Touboul2008, Kharif2010}.}
{It is known from the literature that the wind impact on the wave height
amplification in the broad-banded focusing wave groups is also pronounced
at the defocusing stage \citep{Touboul2008, Kharif2010}.}
\replaced{It}{Thus it} is \added{also} of interest \added{and importance} to analyse \replaced{the height amplification effect of wave breaking through the non-dimensional factor given by}{the wave breaking effect on the
wave height amplification}:
\begin{eqnarray}
   && \tilde{H}(t) =
      \frac{2 \times \underset{x}{\max} \{ \mathscr{H}(x,t) \}}
      {H_S(k_0\zeta_0 = 0.4)}
   \label{eq:amplif}
\end{eqnarray}
Here we introduce the \replaced{normalisation}{normalization} by using the constant significant wave
height $H_S$ calculated for the wave train at the verge of stability,
i.e. weakly-breaking group with $k_0\zeta_0 = 0.4$.
Following the standard definition \citep{Babanin2011},
significant wave height can be calculated from (\ref{eq:spec}):
\begin{eqnarray}
   && H_S  \equiv 4 \sqrt{\frac{1}{2\pi}\int_{-\infty}^{+\infty}
                       \hat{\eta}^2(\omega) d\omega} =
      \sqrt[4]{\frac{8}{\pi}} \sqrt{m \zeta_0^2 T_0 \left(
      1 + e^{-2 \pi^2 m^2} \right)}
   \label{eq:hs}
\end{eqnarray}
The factor of $2$ is introduced into the numerator of (\ref{eq:amplif}) to
evaluate the expected wave height from the envelope $\mathscr{H}$.
Figure~\ref{fig:hilbmax} shows the temporal evolution of 
wave height amplification factor defined by the expression (\ref{eq:amplif}).
Linear regression is applied in panel (b) separately
to the focusing ($t < 35 \text{s}$) and defocusing
($t > 35 \text{s}$) stages. It is clearly shown that the amplification of the
non-breaking wave group ($k_0\zeta_0 = 0.2$) is nearly symmetric about 
the focal point. Increase of the steepness parameter
$k_0\zeta_0$ causes certain asymmetry of $\tilde{H}$ in the pre- and post-focusing stages,
see panel (a).

\myhighlight{It is usually expected that strong breaking leads to instantaneous reduction of the
wave height. However, Figure~\ref{fig:hilbmax} (b) demonstrates a quite opposite and
nontrivial phenomenon. 
The height amplification factors of strongly-breaking wave groups, in particular
the solution for $k_0\zeta_0 \ge 0.6$, decay much slower than the non-breaking and
weakly-breaking ones, cf. panels (a) and (b).
This is probably due to the suppression of defocusing and the consequent conservation
of wave train length shown in Figure~\ref{fig:length}.
Both VOF and BEM$\nu$ models demonstrate asymmetry of $\tilde{H}$ plots
in the pre- and post-focusing stages with the increase of wave steepness especially
when breaking is initiated.

Under strong breaking conditions, see panel (b), the peak values of $\tilde{H}$
produced by the BEM$\nu$ model are higher than the VOF calculations for each wave train.
Unlike the BEM$\nu$ model which shows significant decay rate of $\tilde{H}$
in the post-focusing stage regardless of wave train steepness,
the VOF solutions demonstrate a very mild decay of $\tilde{H}$ over time for
$k_0\zeta_0 \ge 0.6$. It is crucial to emphasize that such conservation of
$\tilde{H}$ with time could be due to the impact of
breaking generated rotational flows and currents in the water mass, which decay
slowly and can remain in the flow for several wave periods or even longer. 
This implies that breaking induced rotational flows and currents, completely disregarded
in the BEM$\nu$ model, may play an important role in formation of extremely high waves.
Such observation may find a confirmation in \citep{Toffoli2019}.

The implication of the phenomenon discussed above is that the space (range of $x$)
where extreme waves are likely to appear might be expanded as compared to FNP predictions.
For very steep wave trains $k_0\zeta_0 \ge 0.8$, their amplification factors remain
above $2.2$ for a relatively long time in the post-focusing stage as shown in
Figure~\ref{fig:hilbmax} (b). But in the pre-focusing stage, such large values can only
be seen in the vicinity of the focal point.
In this sense, the occurrence probablity and lifespan of extreme waves might be
increased by frequent and violent breakings.
}


\section{Conclusions}
\label{sec:concl}

A suite of low-fidelity (FNP) and coupled low-/high-fidelity (FNP-NS) flow models have been  proposed to investigate the evolution of broad-banded wave trains under non-, moderate- and strong-breaking conditions. The weakly-potential approximation proposed by \citet{Ruvinsky1991} was implemented in the FNP model to take into account the energy dissipation caused by wave breaking. This approximation was closed by the eddy viscosity model proposed by \cite{Perlin2010, Perlin2012}.

The developed flow models were firstly tested with a wave train subject to modulational instability. Within the FNP model we also applied the free surface re-meshing technique, which shows a similar performance as the eddy viscosity enclosure in the stabilisation of breaking wave simulation. The computed results were compared with laboratory measurements and other published calculations in terms of surface elevation. It was found that the high-fidelity results compare well with experiments. Although the low-fidelity calculations are rather close 
to the high-order-spectral solutions reported in the literature, they deviate from laboratory measurements especially at locations far away from the wave maker.

To identify the underlying reason causing the deficiency of the eddy viscosity wave breaking model in the prediction of surface elevation, we further examined the FNP and FNP-NS models with six broad-banded focusing wave groups under non-, moderate- and strong-breaking conditions. %
A direct comparison of the low- and high-fidelity results reveals that the re-grid and eddy viscosity approaches predict accurately the energy dissipation caused by breaking.
We then applied spectral decomposition to compute the surface elevation of free waves by filtering out bound waves. Surprisingly, it was found that the amplitude spectra obtained from the FNP and NS solutions are practically identical in terms of  magnitude, regardless of wavenumber and angular frequency. This led us to speculate that the underlying reason causing the deviation of FNP solutions from high-fidelity calculations (and laboratory measurements) in terms of surface elevation is the discrepancy in phase.

To verify this hypothesis, we undertook a detailed analysis of the phase difference between the FNP and NS results. It was found that the difference in phase grows with breaking intensity, and such an effect is especially profound for high wavenumber components. We proposed an empirical formula to correlate the phase shift with wavenumber, energy dissipation rate and time in the power form by applying regression to the data. For strongly breaking wave trains it was found that the phase shift has a quadratic dependence on energy dissipation rate and wavenumber. Moreover, the shift of phase occurs at relatively high wavenumbers, but is hardly observed for long waves. It was also noticed that the growth of phase shift with time is nearly linear for strongly breaking waves.
\myhighlight{It is suggested that the observed variation in phases has similar physical origin as phase-locking effect reported by \citet{Kirby2016}. Therefore, phase-locking is considered to be the main reason for inaccuracy of FNP predictions.}

The proposed phase shift regression function was then used to study the dispersive property of breaking wave trains. It was found that weak breaking has very limited impact on the dispersion of fully-nonlinear wave packets. On the contrary, strong breaking has great influence on the dispersive property of wave packets, causing the frequencies of high wavenumber components to increase significantly. This phenomenon has been clearly demonstrated by using a 2D Fourier transform of the high-resolution spatio-temporal records of surface elevation. We also showed that the dispersion variation can be derived from the phase shift regression function. This suggests that the shift of phase could be a cause of dispersion variation.  We would like to emphasise that the complex phenomena of phase shift and dispersion change may be caused by a combined action of the nonlinear interaction of resonant waves as well as the breaking-induced rotational flows.

The change of wave dispersion caused by breaking increases the propagation speed of high wavenumber components. In the NS computation of strongly breaking waves, the phase speeds of high wavenumber components tend to be independent of the wavenumber i.e. these harmonics propagate at a similar speed. As a result, the dispersive spreading of wave train after the focal point is almost absent in all simulations of strongly breaking waves.
It was found that the evolution of wave trains involving strong breaking consists of
two distinctive stages: (1) a contraction of the wave train length and an accompanying growth of the wave height due to focusing; and (2) maintaining a nearly constant wave train length after the focal point instead of spreading out instantly as usually expected.
This unusual conservation of wave train length is of significance because it can prolong the lifespan of focused waves and expand the space for their propagation. This can raise the probability of extreme wave formation in breaking scenarios compared to non-breaking environments. Such an unexpected finding is contradictory to our general impression that strong breaking can instantly reduce wave height by destructing initially stable harmonics and dissipating their mechanical energy.

We would like to emphasise that in the current stage the conclusions drawn here are based on, and possibly limited to, the quantitative analysis of the wave trains considered in the present work. More comprehensive theoretical, numerical and experimental investigations are needed to arrive at definitive conclusions on phase shifting and other related phenomena reported in this paper. The outcomes of the current research can be beneficial to the development of more accurate theoretical models for wave breaking, which can be used in the weakly- and fully-nonlinear modelling of ocean waves for engineering and environmental science.


\section*{Acknowledgements}
This work was supported by the Engineering and Physical Sciences Research Council (EPSRC), U.K. Project:
High-fidelity Simulation of Air Entrainment in Breaking Wave Impacts, under Grant No. EP/S011862/1.
The authors would also like to express gratitude to Professor Lev Shemer from Tel Aviv University, Israel 
for his encouragement and constructive advice helping us to improve the quality of the manuscript.
\appendix

\section{Grid convergence} \label{app:A}

\begin{figure}
\centering
\includegraphics[width=1\textwidth]{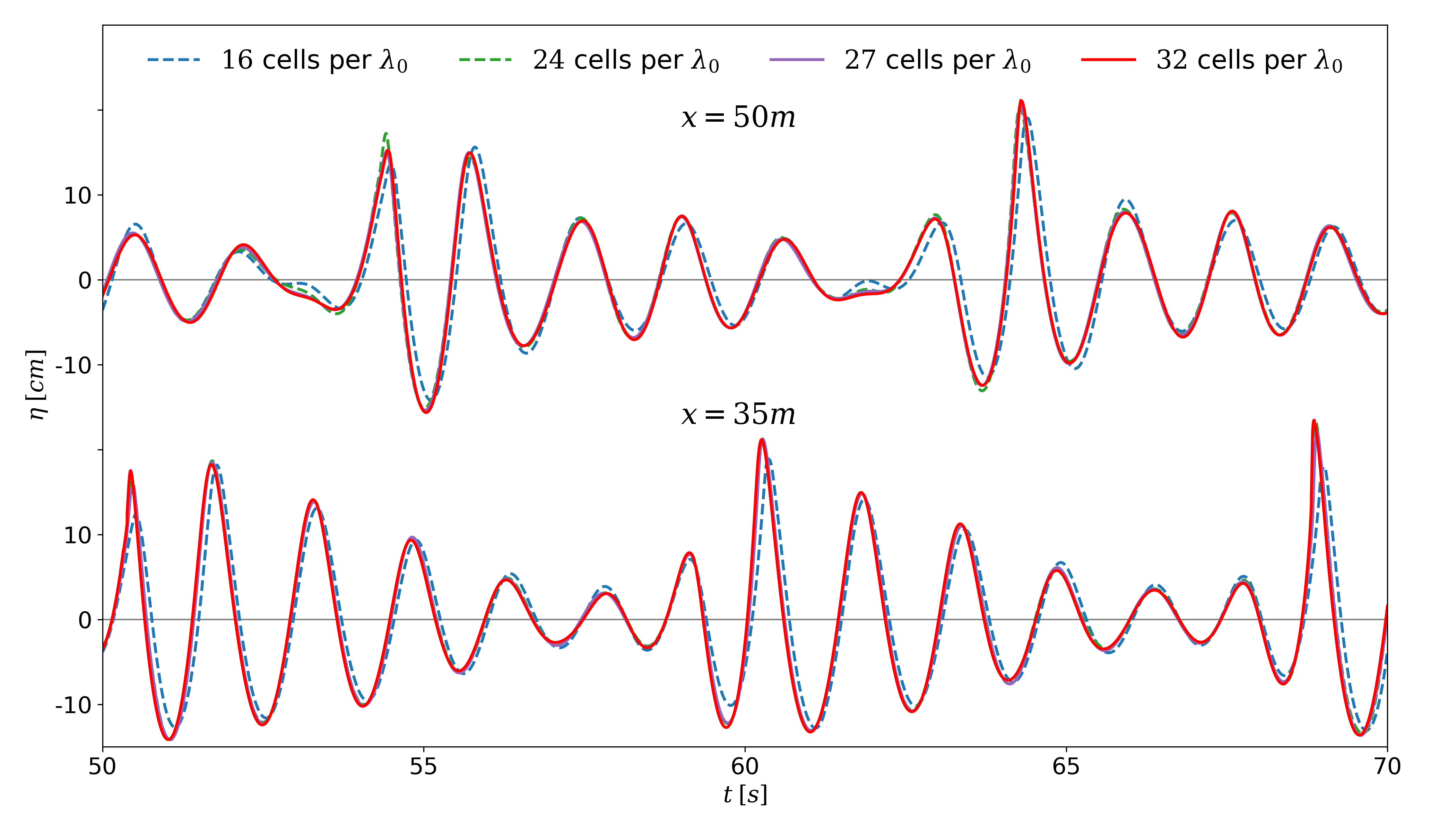}
\caption{Surface elevation obtained in BEM$\nu$ computations at two
         coordinates measured from the wavemaker location: $x = 35 m$
         and $x = 50 m$.}
\label{fig:convergbem}
\end{figure}

The grid convergence of BEM$\nu$ and VOF numerical models is studied
with respect to the modulated wave train (\ref{eq:mi}). The grid
density determined as number of cells (nodes) distributed along the carrier
wave length $\lambda_0$ is used for \replaced{characterisation}{characterization} of the grid quality.
This allows \replaced{generalisation}{generalization} of the convergence study outcomes to
other wave trains investigated in the paper. Four grids of different
free surface mesh density ranging from $16$ to $32$ cells per $\lambda_0$
are considered in \added{the} BEM$\nu$ model, see Figure~\ref{fig:convergbem}.
All meshes show identical surface elevation plots close to the wavemaker
($x = 35 m$), as well as far away from it ($x = 50 m$).
Nevertheless, the finest mesh is used in the \replaced{present work}{computations to improve
the coupling between BEM$\nu$ and VOF simulations}.

\begin{figure}
\centering
\includegraphics[width=1\textwidth]{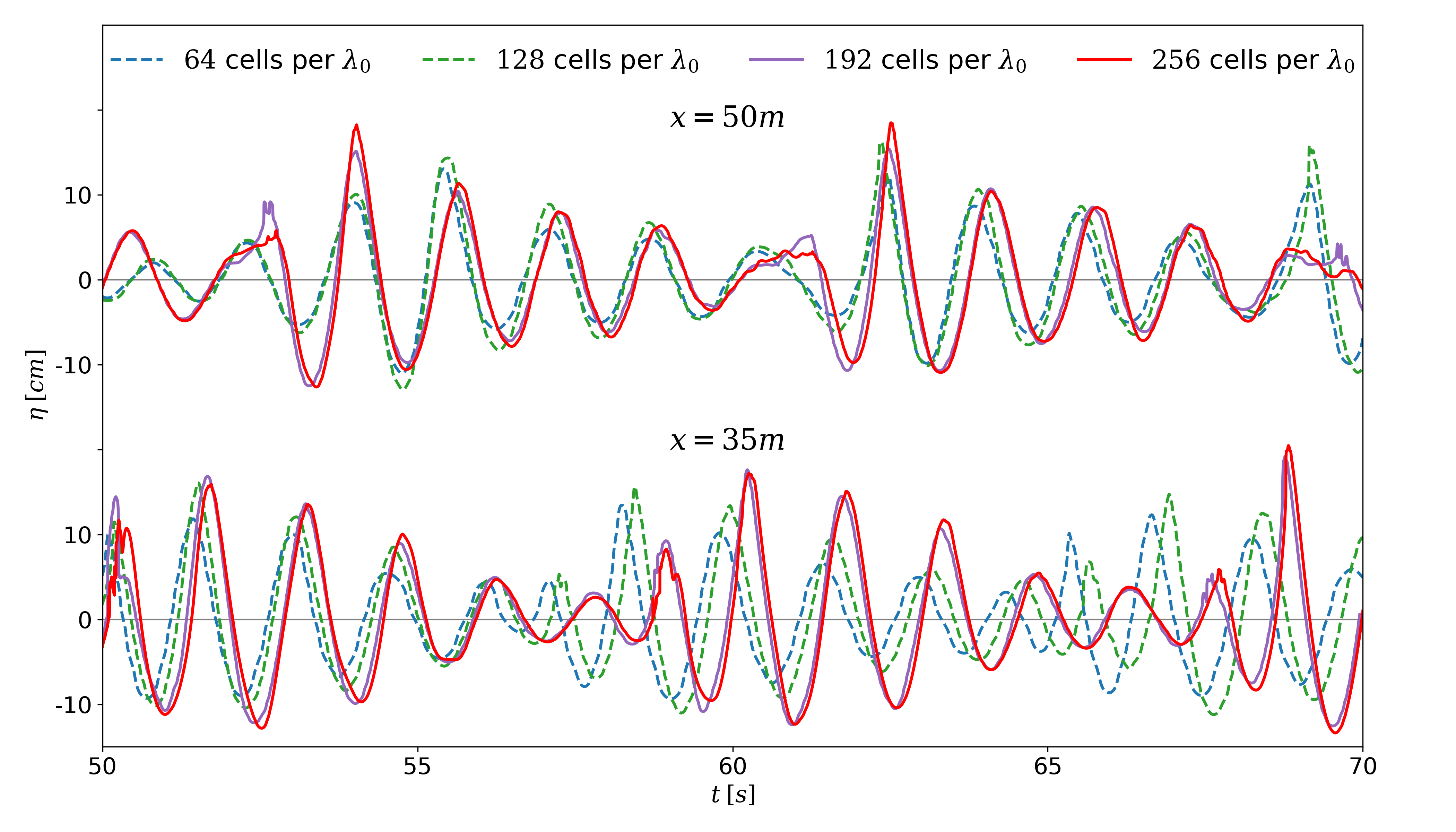}
\caption{As in Figure~\ref{fig:convergbem} for the results of VOF
         computations.}
\label{fig:convergfoam}
\end{figure}

The grid resolutions of $64$ to $256$ cells per carrier wave length
$\lambda_0$ were considered in convergence study of the VOF model,
see Figure~\ref{fig:convergfoam}.
In this case, first two grids fail in simulation of the waves dispersion.
The grid densities $192$ and $256$ cells per $\lambda_0$ showed
practically identical results at $x = 35 m$. Deviation between the
results obtained with these two grids at $x = 50 m$ is acceptably
small, thus the convergence is established with respect to the free
surface elevation. The finest grid of density $256$ cells per
$\lambda_0$ is used for the computations.

\section{Coefficients of the weakly-nonlinear Zakharov model}
\label{app:B}

Assume $A(k)$ being the discrete complex wavenumber spectrum
of the free waves only. Introduce the following wavenumbers:
\begin{eqnarray}
   && k_b = k_m + k_n \;\;\;\text{and}\;\;\;
      k_c = - k_m + k_n \;\;\;\text{and}\;\;\;
      k_d = - k_m - k_n
   \label{eq:wavenum}
\end{eqnarray}
The corresponding angular frequencies due to dispersion relation
(\ref{eq:disprel}) are:
\begin{eqnarray}
\hspace{-20pt}   && \omega_b = \sqrt{g k_b \text{tanh} (k_b h)} \;\;\;\text{and}\;\;\;
      \omega_c = \sqrt{g k_c \text{tanh} (k_c h)} \;\;\;\text{and}\;\;\;
      \omega_d = \sqrt{g k_d \text{tanh} (k_d h)}
   \label{eq:freq}
\end{eqnarray}
The complex wavenumber amplitudes in the $2^{\text{nd}}$-order bound
waves spectrum required to complete expression (\ref{eq:boundwave}) are
\citep{Shemer1987, Krasitskii1994}:
\begin{eqnarray}
   && B(k_m, k_n) = -\pi \sqrt{\frac{2 g \omega_b}{\omega_m \omega_n}}
      \frac{V(\omega_b, \omega_m, \omega_n, k_m+k_n, k_m, k_n)}
           {\omega_b - \omega_m - \omega_n}
      A_m A_n
   \label{eq:B}
\end{eqnarray}
\begin{eqnarray}
   && C(k_m, k_n) = -\pi \sqrt{\frac{2 g \omega_c}{\omega_m \omega_n}}
      \frac{W(\omega_c, \omega_m, \omega_n, -k_m+k_n, k_m, k_n)}
           {\omega_c + \omega_m - \omega_n}
      A_m^{*} A_n
   \label{eq:B}
\end{eqnarray}
\begin{eqnarray}
   && D(k_m, k_n) = -\pi \sqrt{\frac{2 g \omega_d}{\omega_m \omega_n}}
      \frac{T(\omega_d, \omega_m, \omega_n, -k_m-k_n, k_m, k_n)}
           {\omega_d + \omega_m + \omega_n}
      A_m^{*} A_n^{*}
   \label{eq:B}
\end{eqnarray}
Here $m$ and $n$ are numbers of harmonics in the discrete free waves
complex spectrum $A(k)$; the star superscript ``$*$'' stands
for the complex conjugation.
The expressions for the coefficients $V$, $W$ and $T$ are
\citep{Shemer1987}:
\begin{eqnarray}
   \begin{aligned}   
      V(\omega_0, \omega_1, \omega_2, k_0, k_1, k_2) =
      \frac{1}{4 \pi} \sqrt{\frac{g}{2}}
      \left\{
          \frac{1}{2} \sqrt{\frac{\omega_0}{\omega_1 \omega_2}}
          \left[
              \left( \frac{\omega_1 \omega_2}{g} \right)^2
              + k_1 k_2
          \right]
      \right.\\
      \left.
          - \sqrt{\frac{\omega_2}{\omega_0 \omega_1}}
          \left[
              \left( \frac{\omega_0 \omega_1}{g} \right)^2
              - k_0 k_1
          \right]
      \right\}
   \end{aligned}
   \label{eq:V}
\end{eqnarray}
\begin{eqnarray}
   \begin{aligned}   
      W(\omega_0, \omega_1, \omega_2, k_0, k_1, k_2) =
      \frac{1}{4 \pi} \sqrt{\frac{g}{2}}
      \left\{
          \sqrt{\frac{\omega_2}{\omega_0 \omega_1}}
          \left[
              \left( \frac{\omega_0 \omega_1}{g} \right)^2
              + k_0 k_1
          \right]
      \right.\\
      \left.
          - \sqrt{\frac{\omega_1}{\omega_0 \omega_2}}
          \left[
              \left( \frac{\omega_0 \omega_2}{g} \right)^2
              - k_0 k_2
          \right]
          - \sqrt{\frac{\omega_0}{\omega_1 \omega_2}}
          \left[
              \left( \frac{\omega_1 \omega_2}{g} \right)^2
              - k_1 k_2
          \right]
      \right\}
   \end{aligned}
   \label{eq:W}
\end{eqnarray}
\begin{eqnarray}
   \begin{aligned}   
      T(\omega_0, \omega_1, \omega_2, k_0, k_1, k_2) =
      \frac{1}{4 \pi} \sqrt{\frac{g}{2}}
      \left\{
          \sqrt{\frac{\omega_2}{\omega_0 \omega_1}}
          \left[
              \left( \frac{\omega_0 \omega_1}{g} \right)^2
              + k_0 k_1
          \right]
      \right.\\
      \left.
          + \frac{1}{2} \sqrt{\frac{\omega_0}{\omega_1 \omega_2}}
          \left[
              \left( \frac{\omega_1 \omega_2}{g} \right)^2
              + k_1 k_2
          \right]
      \right\}
   \end{aligned}
   \label{eq:T}
\end{eqnarray}

\bibliographystyle{jfm}
\bibliography{manuscript}

\end{document}